\documentclass[letter,11pt]{article}
\pdfoutput=1
\usepackage{graphicx,amssymb,amsmath,color,slashed}
\usepackage{jheppub}
\usepackage[T1]{fontenc}
\usepackage{multirow}
\usepackage{lscape}

\def\beq{\begin{equation}}
\def\eeq{\end{equation}}

\def\bea{\begin{eqnarray}}
\def\eea{\end{eqnarray}}
\def\bei{\begin{itemize}}
\def\eei{\end{itemize}}
\def\bmat{\begin{matrix}}
\def\emat{\end{matrix}}
\def\ble{\begin{flushleft}}
\def\ele{\end{flushleft}}
\def\={\,=\,}
\def\+{\,+\,}
\def\-{\,-\,}

\def\GeV{\,{\rm GeV}\,}
\def\TeV{\,{\rm TeV}\,}
\def\fb{\, {\rm fb} \,}
\def\pb{\, {\rm pb} \,}

\def\nn{\nonumber}
\def\op{{\mathcal O}}
\def\bu{{\bar u}}

\def\bq{{\bar q}}
\def\bQ{{\bar Q}}
\def\SU2{{\rm SU}(2)}

\def\gsubmu{\gamma_\mu}
\def\govermu{\gamma^\mu}

\newcommand{\Fig}[1]{Fig.~\ref{#1}}
\newcommand{\Eq}[1]{Eq.(\ref{#1})}
\newcommand{\Sec}[1]{Sec.\ref{#1}}

\begin{document}

\title{Renormalization group-induced phenomena of top pairs from four-quark effective operators}
\author[a,1]{Sunghoon Jung,\note{nejsh21@gmail.com}}
\author[a]{P. Ko,}
\author[a,b,2]{Yeo Woong Yoon,\note{ywyoon@kias.re.kr}}
\author[a]{Chaehyun Yu}

\affiliation[a]{School of Physics, Korea Institute for Advanced Study, Seoul 130-722, Korea}
\affiliation[b]{School of Physics, Konkuk University, Seoul 143-701, Korea}

\abstract{We study the renormalization group(RG) evolution of four-quark operators that contribute to the top pair production. In particular, we focus on the cases in which certain observables are \emph{first} induced from the one-loop RG while being absent at tree-level. From the operator mixing pattern, we classify all such RG-induced phenomena and underlying models that can induce them. We then calculate the full one-loop QCD RG evolution as the leading estimator of the effects and address the question of which RG-induced phenomena have largest and observable effects. The answer is related to the color structure of QCD. The studied topics include the RG-induction of top asymmetries, polarizations and polarization mixings as well as issues arising at this order. The RG-induction of top asymmetries is further compared with the generation of asymmetries from QCD and QED at one-loop order. We finally discuss the validity of using the RG as the proxy of one-loop effects on the top pair production. As an aside, we clarify the often-studied relations between top pair observables.
}

\preprint{KIAS-P14005}

\maketitle

\section{Introduction}

The Tevatron and the early LHC reached the percent-level precision of top pair data to test the Standard Model (SM) and to disfavor a large set of new physics models. The top quark is, however, still believed to be a sensitive probe of new physics related to the electroweak symmetry breaking. The slight discrepancy between the top asymmetry data from the Tevatron and theoretical prediction also remains unresolved needing the sub-percent level of precision measurements and theoretical calculations. The LHC13 can produce 900 pb$\times$ 20/fb $\sim$ 18 million tops a year or so and more at higher collision energies, and such a high precision will be achieved.

Along the top pair precision program, not only dedicated higher-order calculations but also qualitative and intuitive understanding of what phenomena can arise at higher-order and which underlying models can induce them will be precious knowledge. The SM higher-order calculations have been made significant progresses recently~\cite{Almeida:2008ug,Bernreuther:2012sx,Czakon:2013goa}, and more is coming. Higher-order corrections to even some new physics models have also been calculated~\cite{Xiao:2010hm,Bauer:2010iq} (but not for all models). It is also useful to study higher-order physics more model-independently and systematically using the effective operators which will aid the latter purpose.

In this paper, we assume that new physics is around the TeV scale, and we use $d=6$ four-quark effective operators to describe them. We especially focus on their effects that arise \emph{first} at one-loop order. In these cases, the RG calculation is most useful as the RG effect is the leading contribution, and its effect can be most usefully and dramatically presented. The operator mixing~\cite{oai:arXiv.org:hep-ph/9606222,oai:arXiv.org:hep-ph/9512380} is a suitable language to describe the induction of new phenomena at one-loop order because the induction of new operator leads to the new phenomena as we will see in \Sec{sec:observables}. The pattern of operator mixing which can be approximately understood from the quantum numbers of operators would indicate the possible RG-induction of phenomena and underlying models. We study them in this paper using four-quark effective operators in the top pair production; see Refs.~\cite{Bauer:2010iq,Grojean:2013kd,Elias-Miro:2013gya,oai:arXiv.org:1107.4012,oai:arXiv.org:1204.4773,Jenkins:2013zja,Alonso:2013hga,Zhang:2014rja} for other studies of higher-order effects from effective operators. One advantage of top physics in our study is that many top observables can be measured, and each operator mixing pattern can have observable impact in the future when all those observables are measured.

This paper is outlined as following. We first calculate top pair observables in terms of four-quark operators in \Sec{sec:observables}. Operator quantum numbers, relations between observables and the useful basis for the operator mixing will be discussed. Then we summarize our RG calculation in \Sec{sec:oprge}. Main qualitative discussions on the operator mixing pattern will be presented in \Sec{sec:4by4} with some emphasis on the color structure of QCD in \Sec{sec:rgafb}. Main numerical results are presented in the remaining subsections of \Sec{sec:numerical}, and we classify underlying models in \Sec{sec:matching}. Discussions on the validity of the RG calculation as a leading estimation of the one-loop effects are presented in \Sec{sec:rgvalid}. Then we conclude in \Sec{sec:conclusion}.

\section{Observables} \label{sec:observables}

We first calculate the top pair observables in terms of four-quark operators (evaluated at $m_t$ scale). Among the full set of operators that are complete and closed under QCD RG evolution (that will be introduced in \Sec{sec:oprge}), four-quark operators composed of light-quark and top-quark currents can contribute to the $q\bar{q} \to t\bar{t}$ top pair productions at leading order. They are denoted by
\beq
{\cal L}_{\rm eff} \, \ni \, \sum_{\rm A,B} \left( \frac{C_{\rm AB}^{(8)}}{\Lambda^2} {\cal O}_{\rm AB}^{(8)} \+ \frac{C_{\rm AB}^{(1)}}{\Lambda^2} {\cal O}_{\rm AB}^{(1)} \right),
\label{eq:uuttcvv} \eeq
\beq
{\cal O}_{\rm AB}^{(8)} \= \left( \bar{u} \gamma_\mu T^a \, u \right)_{\rm A} \left( \bar{t} \gamma^\mu T^a \, t \right)_{\rm B}, \qquad {\cal O}_{\rm AB}^{(1)} \= \left( \bar{u} \gamma_\mu \, u \right)_{\rm A} \left( \bar{t} \gamma^\mu \, t \right)_{\rm B}.
\eeq
We use these operators in the discussions throughout this paper except in \Sec{sec:oprge} and \ref{sec:matching}. The subscript A, B can be L(left) or R(right) chiralities in {\it L-R basis}, and V(vectorial) or A(axial-vectorial) in {\it V-A basis}. Note that, symbolically, $\rm V = R+L$ and $\rm A=R-L$ hold. The $\Lambda$ is an assumed new physics mass scale.
Although we write operators involving up quarks only, we also consider similar operators with down quarks as well.

We use operators in both L--R and V--A basis (or even linear combinations of them) throughout this paper. A caution is that the conversion formula from one basis to another do not apply to their Wilson coefficient in the same way. If the relation $\op_{i}^{\rm basis_1} = P_{ij} \op_{j}^{\rm basis_2}$ holds, the corresponding Wilson coefficients satisfy $C_{i}^{\rm basis_1} = ({P^{T}}^{-1})_{ij} C_{j}^{\rm basis_2}$  Relations between the L--R and V--A basis are collected in Appendix~\ref{app:lrva}. An example of useful relations is
\bea
{\cal O}_{\rm VV}^{(8)} &\=& {\cal O}_{\rm RR}^{(8)} + {\cal O}_{\rm LL}^{(8)} + {\cal O}_{\rm RL}^{(8)} + {\cal O}_{\rm LR}^{(8)}, \nn \\ \qquad C_{\rm VV}^{(8)} &\=& \frac{1}{4} \left( C_{\rm RR}^{(8)} + C_{\rm LL}^{(8)} + C_{\rm RL}^{(8)} + C_{\rm LR}^{(8)} \right),
\eea
which means that ${\cal O}_{\rm VV}^{(8)}$ can be decomposed into all four operators in the L--R basis with all same coefficients; in other words, if all four L--R operators have the same coefficients, the theory contains only ${\cal O}_{\rm VV}^{(8)}$ in the V--A basis.

\subsection{Independent observables of top pair} \label{sec:orthoobs}

We define top pair observables and calculate them in terms of the four-quark operators. We also show their independence. They are independent in the sense that they are contributed from independent operators. Based on the independence, we discuss relations among observables, especially the one between the top asymmetry and polarization.

In calculating top pair observables, we work up to the interference between QCD and effective operators in \Eq{eq:uuttcvv}. The square of the effective operators does not contribute to most of our discussions; the square effects are discussed in \Sec{sec:rgvssq} and Appendix~\ref{app:helamp}.

To this end, it is useful to work with helicity cross-sections. Let us denote right(left)-handed top and anti-top by $\pm$ indices. Initial spin- and color-averaged helicity cross-sections are (up to the interference)
\bea
\sigma_{++} &=& \sigma_{--} \= \frac{8\pi \alpha_s^2}{27 \hat{s}} \,\frac{m_t^2\beta_t}{\hat{s}} \left[ 1 + \frac{\hat{s}}{\Lambda^2 } \frac{2C_{\rm VV}^{(8)}}{g_s^2} \right] \,,\\
\sigma_{+-} &=& \frac{4\pi \alpha_s^2}{27 \hat{s}} \beta_t \left[ 1 + \frac{\hat{s}}{\Lambda^2 g_s^2} \Big( 2C_{\rm VV}^{(8)} + 2\beta_t C_{\rm VA}^{(8)} \Big) \right] \,,\\
\sigma_{-+} &=& \frac{4\pi \alpha_s^2}{27 \hat{s}} \beta_t \left[ 1 + \frac{\hat{s}}{\Lambda^2 g_s^2} \Big( 2C_{\rm VV}^{(8)} - 2\beta_t C_{\rm VA}^{(8)} \Big) \right] \,.
\eea
Terms not suppressed by $\Lambda^2$ are SM contributions. The forward-backward asymmetric helicity cross-sections, $a_{\lambda_1 \lambda_2}$, are defined by
\begin{equation}
a_{\lambda_1 \lambda_2} = \bigg(\int_0^1 d\cos \theta - \int_{-1}^0 d\cos \theta \bigg) \frac{d\sigma_{\lambda_1 \lambda_2}}{d\cos\theta}\,,
\label{eq:afbhelcrx}\end{equation}
where $\theta$ is the angle between incoming light quark and outgoing top quark. Then $a_{\lambda_1 \lambda_2}$ read
\bea
a_{++} &=& a_{--} \= 0 \,,\\
a_{+-} &=& \frac{2\pi \alpha_s^2}{9 \hat{s}} \,\frac{\beta_t \hat s}{\Lambda^2 g_s^2} \Big( \beta_t C_{\rm AA}^{(8)} + C_{\rm AV}^{(8)} \Big) \,,\\
a_{-+} &=& \frac{2\pi \alpha_s^2}{9 \hat{s}} \,\frac{\beta_t \hat s}{\Lambda^2 g_s^2} \Big( \beta_t C_{\rm AA}^{(8)} -C_{\rm AV}^{(8)} \Big) \,.
\eea

Now, we write various top observables in terms of these helicity cross-sections. The total top pair production and forward-backward asymmetric cross-sections are expressed as
\bea
\sigma_{\rm tot} &=& \sigma_{++}+\sigma_{--} +\sigma_{+-} + \sigma_{-+} \= \frac{8 \pi \alpha_s^2}{27 \hat{s}} \beta_t \, \left( 1+ \frac{2m_t^2}{\hat{s}} \right) \left[ 1+ \frac{\hat{s}}{\Lambda^2} \frac{2C_{\rm VV}^{(8)}}{g_s^2}\right], \\
\sigma_{\rm FB} & \propto & N(t_{\rm F}) - N(t_{\rm B}) \, \sim \, N(\cos \theta_t >0 ) - N( \cos \theta_t<0) \nonumber\\
&=& a_{++} + a_{--}+ a_{+-} + a_{-+} \=\frac{4\pi \alpha_s^2}{9 \hat{s}} \,\frac{\beta_t^2 \hat s}{\Lambda^2 g_s^2}\, C_{\rm AA}^{(8)},
\eea
where we symbolically express that the forward-backward asymmetric cross-section measures the asymmetry between the number($N$) of forward($t_{\rm F}$)- and backward-tops($t_{\rm B}$) and that it is actually measured from the direction of top quarks relative to the proton beam direction, $\cos \theta_t$. The proton direction is not unique at the LHC; either probabilistically correlated directions or the absolute rapidity difference are used instead~\cite{Bernreuther:2012sx,Antunano:2007da,Aad:2013cea}. The top polarization depends on different coupling combination
\bea
\sigma_{{\cal P}_t} &\propto& N(t_{\rm R}) - N(t_{\rm L}) \, \sim \, N(\cos \theta_{\ell} >0)-N(\cos \theta_{\bar{\ell}} <0)  \nonumber\\
&=& \sigma_{++} - \sigma_{--} + \sigma_{+-} - \sigma_{-+} \= \frac{16\pi \alpha_s^2}{27 \hat{s}} \,\frac{\beta_t^2 \hat{s}}{\Lambda^2 g_s^2}\, C_{\rm VA}^{(8)}, \\
\sigma_{{\cal P}_{\bar{t}}} &\propto& N(\bar{t}_{\rm L}) - N(\bar{t}_{\rm R}) \nonumber\\
&=& -\sigma_{++} + \sigma_{--} + \sigma_{+-} - \sigma_{-+} \= \frac{16\pi \alpha_s^2}{27 \hat{s}} \,\frac{\beta_t^2 \hat{s}}{\Lambda^2 g_s^2}\, C_{\rm VA}^{(8)}\,.
\eea
Note that the top polarization is induced by $b_\pm$ of Ref.~\cite{Degrande:2010kt} and essentially equals to $D$ of Ref.\cite{Jung:2010yn}. The spin-correlation of top pairs is expressed as
\bea
\sigma_{{\cal C}_{t\bar{t}}} &\propto&  N(t_{\rm R} \bar{t}_{\rm L}) +  N(t_{\rm L} \bar{t}_{\rm R}) - N(t_{\rm R} \bar{t}_{\rm R}) - N(t_{\rm L} \bar{t}_{\rm L}) \nonumber\\
&\sim& N(\cos \theta_{\ell} \cos \theta_{\bar{\ell}} >0 ) - N(\cos \theta_{\ell} \cos \theta_{\bar{\ell}} <0 ) \nn \\
&=& -\sigma_{++} - \sigma_{--} + \sigma_{+-} + \sigma_{-+} \=  \frac{8\pi \alpha_s^2}{27 \hat{s}} \beta_t \left( 1- \frac{2m_t^2}{\hat{s}} \right) \, \left[ 1+  \frac{\hat{s}}{\Lambda^2} \frac{2C_{\rm VV}^{(8)}}{g_s^2} \right].
\eea
Another independent observable is the forward-backward asymmetry of top polarizations
\bea
\sigma_{D_{\rm FB}} &\propto& N( t_{\rm F,R} ) - N( t_{\rm F,L} ) +N( t_{\rm B,L} ) -N( t_{\rm B,R} ) \nonumber\\
&\sim& N(\cos \theta_t \cos \theta_{\ell} >0 ) -  N( \cos \theta_t \cos \theta_{\ell} <0 ) \nonumber\\
&=& a_{++} - a_{--} + a_{+-} - a_{-+} \= a_{+-} - a_{-+} \= \frac{4\pi \alpha_s^2}{9\hat{s}} \, \frac{\beta_t \hat{s}}{\Lambda^2 g_s^2} \, C_{\rm AV}^{(8)}.
\label{eq:dfb} \eea
 The lepton direction $\cos \theta_\ell$ defined with respect to the given reference direction is used to measure the top polarization; two typical choices of the reference direction are the top direction in the $t\bar{t}$ rest frame called the helicity basis and the beam direction called the beam basis~\cite{Mahlon:1995zn}.

Observables are normalized by the total cross-section
\beq
A_{\rm FB} \= \frac{\sigma_{\rm FB}}{\sigma_{\rm tot}}, \quad {\cal P}_t \= \frac{ \sigma_{{\cal P}_t}}{ \sigma_{\rm tot}}, \quad {\cal C}_{t\bar{t}} \= \frac{ \sigma_{{\cal C}_{t\bar{t}}}}{\sigma_{\rm tot}}, \quad {\cal D}_{\rm FB} \= \frac{ \sigma_{{\cal D}_{\rm FB}}}{ \sigma_{\rm tot}}.
\eeq

In summary, all four chiral couplings $C_{\rm VV}^{(8)}, C_{\rm AA}^{(8)}, C_{\rm VA}^{(8)}, C_{\rm AV}^{(8)}$ can, in principle, be determined by four observables $\sigma_{\rm tot}, A_{\rm FB}, {\cal P}_t, D_{\rm FB}$\footnote{The spin-correlation provides new complementary information if non-four-quark operators exist~\cite{Degrande:2010kt}.}. The correspondence is summarized as (see also Refs.\cite{Degrande:2010kt,Jung:2009pi,Jung:2010yn})
\bea
&& \left. \begin{array}{l}
\cdot \quad \sigma_{\rm tot}, {\cal C}_{t\bar{t}} \, \leftrightarrow \, C_{\rm VV}^{(8)} \\
\cdot \quad A_{\rm FB} \, \leftrightarrow \, C_{\rm AA}^{(8)} \\
\end{array} \, \right\} \textrm{P-even} \nonumber\\
&& \left. \begin{array}{l}
\cdot \quad {\cal P}_t \, \leftrightarrow \, C_{\rm VA}^{(8)} \\
\cdot \quad {\cal D}_{\rm FB} \, \leftrightarrow \, C_{\rm AV}^{(8)}
\end{array} \, \right\} \textrm{P-odd}
\label{eq:corres} \eea
As advertised, each observable is contributed from single operator in the V--A basis. This basis is suitable for the operator mixing discussion because the induction of new operator in this basis directly implies the induction of new top pair observable. The parity quantum numbers of operators shown in \Eq{eq:corres} are useful to understand the pattern of operator mixing as will be discussed later.

Interestingly, the top asymmetry is induced by the parity-even operator, ${\cal O}_{\rm AA}$, although individual currents are axial and therefore parity violating. Moreover, the operator ${\cal O}_{\rm AA}$ does not induce the top polarization which is parity-odd observable. This might be viewed as counter-intuitive, at first. Consider the ${\cal O}_{\rm AR}$ operator, for example. As only right-handed tops are interacting, one may expect some degree of right-handed top polarizations. However, the top polarization is theoretically zero as we just discussed. What the ${\cal O}_{\rm AR}$ operator does is to induce the top asymmetry while leaving the total cross-section unchanged. These are understood as the production of forward (right-handed) top events and the removal of the same number of backward (right-handed) top events. Alternatively, the relation, ${\cal O}_{\rm AR}={\cal O}_{\rm RR}-{\cal O}_{\rm LR}$, shows that ${\cal O}_{\rm RR}$ produces right-handed (forward) tops while $-{\cal O}_{\rm LR}$ removes (backward) right-handed tops. Thus, the total rate and the top polarization are not modified while the top asymmetry is generated. This clarifies the commonly studied relation between top asymmetries and top polarizations; two observables have different parity quantum numbers and are completely independent, so top polarizations can still be zero even though top asymmetries are induced from new right-handed top interactions. As a corollary, if non-zero top polarizations are measured, it necessarily implies the existence of parity violating top interactions.

We comment that, in practice, non-zero top polarizations can still be measured even though zero polarizations are theoretically expected. The top polarization is measured through the angular distribution of charged lepton decay products. But, under selection cuts, different distributions of lepton rapidities from right- and left-handed tops cause certain bias among them. Moreover, top quarks themselves can be produced with different rapidity spectrum in different underlying models (even with same top polarizations). Dedicated collider studies, e.g. Ref.\cite{Krohn:2011tw}, indeed show that non-zero (and different) top polarizations will be measured from various models including the SM and the axigluon that theoretically produce zero polarizations.

Several observables proposed in literatures can be used to disentangle the subtle relation between top asymmetries, top polarizations, and chiralities of top and light quark couplings. The lepton forward-backward asymmetry depends both on the asymmetry and the chirality of tops produced~\cite{Krohn:2011tw,Bowen:2005ap,oai:arXiv.org:1201.1790,Aguilar-Saavedra:2014yea}. Furthermore, the threshold lepton asymmetry can directly measure the chirality of light quark interactions~\cite{Falkowski:2011zr}. The top polarization asymmetry, $D_{\rm FB}$, introduced in \Eq{eq:dfb} (and similarly in Refs.\cite{Jung:2010yn,Fajfer:2012si}) can also provide useful information as it is the observable induced by the ${\cal O}_{\rm AV}$ operator.

\subsection{Observables at hadron colliders}

Based on the expressions derived in previous subsection, we numerically evaluate observables at hadron colliders by convoluting with CTEQ6.6M parton distribution functions (PDF)~\cite{oai:arXiv.org:0802.0007}.

We use following SM parameters
\beq
m_t=173.0\GeV, \quad \alpha_S(m_Z) =0.1180, \quad m_Z=91.19\GeV, \quad s_W^2 = 0.2315.
\eeq
At Tevatron($\sqrt{s}=1.96\TeV$), our leading order SM prediction is
\beq
\sigma_{\rm tot}^{\rm SM} \= 5.50\pb, \qquad (gg: 0.55\pb, \quad q\bar{q} : 4.95\pb).
\eeq
We do not include any electroweak effects.

Our effective theory predictions at Tevatron($\sqrt{s}=1.96\TeV$) are (decomposed into $u\bar{u}$ and $d\bar{d}$-initiated contributions, so we use superscripts on Wilson coefficients to distinguish them)
\bea
\Delta \sigma_{\rm tot} &=& ( 37\fb^{+9.2}_{-6.6} \cdot 4C_{\rm VV}^{ut (8)} \+ 6.1\fb^{+1.6}_{-1.2} \cdot 4C_{\rm VV}^{dt (8)} ) \, \left(\frac{3\TeV}{\Lambda}\right)^2, \\
\sigma_{\rm FB} &=& ( 14\fb^{+3.6}_{-2.6} \cdot 4C_{\rm AA}^{ut(8)} \+ 2.2\fb^{+0.61}_{-0.43} \cdot 4C_{\rm AA}^{dt(8)} ) \, \left(\frac{3\TeV}{\Lambda}\right)^2, \\
\sigma_{{\cal P}_t} &=& ( 19\fb^{+4.9}_{-3.5} \cdot 4C_{\rm VA}^{ut(8)} \+ 2.9\fb^{+0.62}_{-0.58} \cdot 4C_{\rm VA}^{dt(8)} ) \, \left(\frac{3\TeV}{\Lambda}\right)^2, \\
\sigma_{C_{t\bar{t}}} &=& ( 21\fb^{+5.4}_{-3.9} \cdot 4C_{\rm VV}^{ut(8)} \+ 3.4\fb^{+0.92}_{-0.65} \cdot 4C_{\rm VV}^{dt(8)} ) \, \left(\frac{3\TeV}{\Lambda}\right)^2,\\
\sigma_{D_{\rm FB}} &=& ( 22\fb^{+5.5}_{-3.9} \cdot 4C_{\rm AV}^{ut(8)} \+ 3.6\fb^{+0.96}_{-0.68} \cdot 4C_{\rm AV}^{dt(8)} ) \, \left(\frac{3\TeV}{\Lambda}\right)^2,
\eea
where uncertainties are obtained by varying the renormalization and factorization scales $\mu_{\rm R,F}$ by factor of 2 around $m_t$. $\Delta \sigma_{\rm tot}$ is defined as $\Delta \sigma_{\rm tot} = \sigma_{\rm tot} - \sigma_{\rm tot}^{\rm SM}$. Additional renormalization scale uncertainty arises from RG evolution of Wilson coefficients and this will be separately calculated later. We keep the factor 4 in front of the Wilson coefficients for our convenience. In the region of high mass $m_{t\bar{t}} \geq 800\GeV$, SM is $ \sigma_{\rm tot}^{\rm SM} = 38\fb $ and
\bea
\Delta \sigma_{\rm tot} &=& ( 1.1\fb^{+0.40}_{-0.29} \cdot 4C_{\rm VV}^{ut(8)} \+ 0.081\fb^{+0.030}_{-0.021} \cdot 4C_{\rm VV}^{dt(8)} ) \, \left(\frac{3\TeV}{\Lambda}\right)^2, \\
\sigma_{\rm FB} &=& ( 0.69\fb^{+0.25}_{-0.17} \cdot 4C_{\rm AA}^{ut(8)} \+ 0.052\fb^{+0.020}_{-0.013} \cdot 4C_{\rm AA}^{dt(8)} ) \, \left(\frac{3\TeV}{\Lambda}\right)^2, \\
\sigma_{{\cal P}_t} &=& ( 0.91\fb^{+0.34}_{-0.22} \cdot 4C_{\rm VA}^{ut(8)} \+ 0.069\fb^{+0.027}_{-0.018} \cdot 4C_{\rm VA}^{dt(8)} ) \, \left(\frac{3\TeV}{\Lambda}\right)^2, \\
\sigma_{C_{t\bar{t}}} &=& ( 0.92\fb^{+0.33}_{-0.23} \cdot 4C_{\rm VV}^{ut(8)} \+ 0.069\fb^{+0.027}_{-0.017} \cdot 4C_{\rm VV}^{dt(8)} ) \, \left(\frac{3\TeV}{\Lambda}\right)^2, \\
\sigma_{D_{\rm FB}} &=& ( 0.74\fb^{+0.28}_{-0.18} \cdot 4C_{\rm AV}^{ut(8)} \+ 0.056\fb^{+0.022}_{-0.014} \cdot 4C_{\rm AV}^{dt(8)} ) \, \left(\frac{3\TeV}{\Lambda}\right)^2.
\eea
For $m_{t\bar{t}} \geq 650\GeV$, the highest mass bin where top asymmetry is measured ($ \sigma_{\rm tot}^{\rm SM} = 218\fb$)
\bea
\Delta \sigma_{\rm tot} &=& ( 4.1\fb \cdot 4C_{\rm VV}^{ut(8)} \+ 0.41\fb \cdot 4C_{\rm VV}^{dt(8)} ) \, \left(\frac{3\TeV}{\Lambda}\right)^2, \\
\sigma_{\rm FB} &=& ( 2.5\fb \cdot 4C_{\rm AA}^{ut(8)} \+ 0.24\fb \cdot 4C_{\rm AA}^{dt(8)} ) \, \left(\frac{3\TeV}{\Lambda}\right)^2.
\eea

The results at LHC8 are ($\sigma_{\rm tot}^{\rm SM} \= 142\pb$)
\bea
\Delta \sigma_{\rm tot} &=& (188\fb \cdot 4C_{\rm VV}^{ut(8)} \+ 111\fb \cdot 4 C_{\rm VV}^{dt(8)} ) \left( \frac{ 3\TeV}{\Lambda} \right)^2, \\
\sigma_{{\cal P}_t} &=& ( 116\fb \cdot 4C_{\rm VA}^{ut(8)} \+ 67\fb \cdot 4C_{\rm VA}^{dt(8)} ) \, \left(\frac{3\TeV}{\Lambda}\right)^2,
\eea
and for $m_{t\bar{t}} \geq 1\TeV$ for which top resonances were searched ($\sigma_{\rm tot}^{\rm SM} \= 2.54\pb$)
\bea
\Delta \sigma_{\rm tot} &=& (19\fb \cdot 4C_{\rm VV}^{ut(8)} \+ 9.5\fb \cdot 4 C_{\rm VV}^{dt(8)} ) \left( \frac{ 3\TeV}{\Lambda} \right)^2,\\
\sigma_{{\cal P}_t} &=& ( 17\fb \cdot 4C_{\rm VA}^{ut(8)} \+ 8.8\fb \cdot 4C_{\rm VA}^{dt(8)} ) \, \left(\frac{3\TeV}{\Lambda}\right)^2.
\eea

Most constraining data are the Tevatron total cross-section and total and high-mass top asymmetries and the LHC8 total cross-section and heavy resonance searches. Those data are collected in Table.\ref{tab:datapred} in comparison with our benchmark model predictions that will be discussed in \Sec{sec:numerical}. At LHC with higher collision energy, $gg$-initiated production becomes more important, so we do not consider them in this paper.

\section{Operator renormalization} \label{sec:oprge}

We now calculate the RG evolution of four-quark operators. We introduce our full operator basis used for the renormalization in \Sec{sec:opbasis} and then summarize the procedure to solve RG equations in \Sec{sec:rge}. Main physics of the operator renormalization underlying our study is qualitatively discussed in \Sec{sec:4by4} before we move on to a full numerical study in the next section.

\subsection{Operator basis closed under QCD RG evolution} \label{sec:opbasis}

Once the SM equations of motion are consistently used, a set of four-quark operators form a complete basis closed under QCD RG evolution~\cite{Jenkins:2013zja}. We denote the four-quark effective operators as
\bea
{\cal L}_{\rm eff} &=& \sum_{i,j =\{q_A\}} \frac{C_{ij}^{(8)}(\mu)}{\Lambda^2} {\cal O}_{ij}^{(8)} \+ \sum_{i,j=\{q_A\}} \frac{C_{ij}^{(1)}(\mu)}{\Lambda^2} {\cal O}_{ij}^{(1)}, \label{eq:fullopbasis1}\\
&& {\cal O}_{ij}^{(1)} \= j_i^\mu j_{\mu j}, \quad {\cal O}_{ij}^{(8)} \= j_i^{a \mu} j_{\mu j}^a.
\label{eq:fullopbasis}
\eea
Operators are bilinears of the current $j_{q_A}^{\mu} = (\bar q \, \gamma^\mu \, q)_{\rm A}$ or $j_{q_A}^{a \mu} = (\bar q \, T^a \, \gamma^\mu \, q)_{\rm A}$. Indices $i,j$ denote the flavor and chirality of each current. We assume that new physics is flavor-conserving, so are the currents.

We emphasize that the operators ${\cal O}_{\rm AB}$ used to calculate the top pair production at $m_t$ scale in \Sec{sec:observables} are subsets of full set of operators ${\cal O}_{ij}$ here; the operator subset in \Sec{sec:observables} is the one that contributes to the $q{\bar q} \to t\bar{t}$ at tree-level. These operators can be induced from the full set of operators introduced here through the operator mixing during QCD RG evolution. The phenomenology of this RG effect is our subject and is studied in \Sec{sec:numerical}.

We do not consider non-four-quark operators that can still affect top pair production such as chromo-magnetic penguin operator ${\cal O}_{gh} = H \bar Q_3 \sigma^{\mu\nu} T^a t G_{\mu\nu}^a$ and triple gluon field strength operator ${\cal O}_{G} = f_{abc} G_\mu^{a \nu} G_\nu^{b\rho} G_\rho^{c\mu}$ which were discussed in Refs.~\cite{Degrande:2010kt,Zhang:2010dr}. Obviously those operators are generated at $\Lambda$ through loop correction which is a higher-order effect to the leading RG contribution. On top of that, the operator mixing from four-quark operators to ${\cal O}_{gh}$ begins with two-loop diagram analogous to the case of weak Hamiltonian in $B$-physics~\cite{Ciuchini:1993ks,Ciuchini:1993fk}. Apparently, this is also the case for the mixing effect from four-quark operators to ${\cal O}_{G}$. Those two-loop RG effects may be minor and the treatment of them is beyond the scope of this work. We will consider only models generating four-quark operators at tree-level matching and ignore all non-four-quark operators.

We introduce two operator bases that we use to calculate QCD RG evolution in \Sec{sec:tpop} and \Sec{sec:expop}. They are closed under QCD RG evolution and complete even though they are subsets of full dimension-6 effective operators categorized in Ref.~\cite{Grzadkowski:2010es}. They are complete because we are considering only models generating four-quark operators. We cross-checked our RG equations with those from Ref~\cite{Alonso:2013hga} by converting the operator basis properly.

\subsubsection{Tree-penguin classified basis} \label{sec:tpop}

Our first basis is intended to clearly distinguish the origin of operators as tree-level, one-loop penguin and two-loop double penguin diagrams. This classification is analogous to that of the weak Hamiltonian in $B$ physics~\cite{oai:arXiv.org:hep-ph/9512380}. Since we consider a tree-level matching in this paper, only the tree-level operators are generated at the matching scale; but the other two are generated through the QCD RG evolution with penguin coefficients. The coefficients of penguin operators are the convenient measures of penguin effects.

The tree-operators are defined by
\begin{eqnarray}
\label{eq:opt}
{\rm LL}:~~ &&~~~~\op_{Q_i Q_j}^{(1)} =\lambda_{ij} \left(\bQ_i \gsubmu Q_i\right)\left(\bQ_j \govermu Q_j\right) ,\nn \\
 && ~~~~\op_{Q_i Q_j}^{(8)} =\lambda_{ij} \left(\bQ_i \gsubmu T^a Q_i\right)\left(\bQ_j \govermu T^a Q_j\right)
,\nn \\
{\rm RR}:~~&&~~~~\op_{qq^\prime}^{(1)} = \left(\bq \gsubmu q \right) \left( \bq^\prime \govermu q^\prime\right),
~~~~~\op_{qq^\prime}^{(8)} = \left(\bq \gsubmu T^a q \right) \left( \bq^\prime \govermu T^a q^\prime\right)\,,~~~~~ (q\neq q^\prime) \nn \\
&&~~~~\op_{qq}^{(8)} =\frac{1}{2} \left(\bq \gsubmu T^a q \right) \left( \bq \govermu T^a q\right), \nn\\
{\rm LR}:~~&&~~~~\op_{Q_i q}^{(1)} = \left(\bQ_i \gsubmu Q_i \right) \left( \bq \govermu q \right),
~~~\op_{Q_i q}^{(8)} = \left(\bQ_i \gsubmu T^a Q_i \right) \left( \bq \govermu T^a q \right)\,.
\label{eq:treeop} \end{eqnarray}
We keep the $\SU2_L$ gauge symmetry in the operator basis.
$Q_i$ are $\SU2$ doublet left-handed quarks with generation index $i=1,2,3$. $q$'s are $\SU2$ singlet right-handed quarks which can be $u, d, s, c, b, t$. Color-singlet $\rm RR$-type operators with four identical quarks are reduced to its color-octet operators using color identities and Fierz transformations: for example $\op_{uu}^{(1)}= \frac{1}{2}\left(\bu \gsubmu u \right) \left( \bu \govermu u\right) = 3\op_{uu}^{(8)}$. It is convenient to introduce a symmetry factor $1/2$ or  $\lambda_{ij} = 1/(1+\delta_{ij})$ for such operators with four identical quarks.
These tree operators form a complete basis closed under QCD RG evolution by themselves.

The penguin operators are defined by
\begin{eqnarray}
\label{eq:opp}
{\rm LL}:~~~ &&\op_{Q_i \Sigma_Q}^{(1)} = \left(\bQ_i \gsubmu Q_i\right) \sum_k \lambda_{ik}\left(\bQ_k \govermu Q_k\right),
\nn \\
~~~ &&\op_{Q_i \Sigma_Q}^{ (8)} = \left(\bQ_i \gsubmu T^a Q_i\right) \sum_k \lambda_{ik}\left(\bQ_k \govermu T^a Q_k\right),
\nn \\
{\rm RR}:~~~&&\op_{q^\prime \Sigma_q}^{(1)} = \left(\bq^\prime \gsubmu q^\prime\right) \sum_q \lambda_{q^\prime q}\left(\bq \govermu q\right),
~ \op_{q^\prime \Sigma_q}^{ (8)} = \left(\bq^\prime \gsubmu T^a q^\prime\right) \sum_q \lambda_{q^\prime q}\left(\bq \govermu T^a q\right), \nn \\
{\rm LR}:~~~&&\op_{Q_i \Sigma_q}^{(1)} = \left(\bQ_i \gsubmu Q_i\right) \sum_q\left(\bq \govermu q\right),
~~~\op_{Q_i \Sigma_q}^{(8)} = \left(\bQ_i \gsubmu T^a Q_i\right) \sum_q\left(\bq \govermu T^a q\right), \nn \\
{\rm RL}:~~~&&\op_{q \Sigma_Q}^{(1)} = \left(\bq \gsubmu q\right)  \sum_i \left(\bQ_i \govermu Q_i\right),
~~~~\op_{q \Sigma_Q}^{(8)} = \left(\bq \gsubmu T^a q\right)  \sum_i \left(\bQ_i \govermu T^a  Q_i\right)\,.
\end{eqnarray}

Likewise, double penguin operators contain effects from double penguin diagrams, and are defined by
\begin{eqnarray}
\label{eq:opdp}
{\rm LL}:~~\op_{\Sigma_Q \Sigma_Q}^{(1)} &=& \sum_i\left(\bQ_i \gsubmu Q_i\right) \sum_{j} \lambda_{ij} \left(\bQ_{j} \govermu Q_{j}\right), \nn \\
\op_{\Sigma_Q \Sigma_Q}^{(8)} &=& \sum_i\left(\bQ_i \gsubmu T^a Q_i\right) \sum_{j} \lambda_{ij} \left(\bQ_{j} \govermu T^a Q_{j}\right),\nn \\
{\rm RR}:~~\op_{\Sigma_q \Sigma_q}^{(1)} &=& \sum_q \left(\bq \gsubmu q \right) \sum_{q^\prime} \lambda_{qq\prime}\left( \bq^\prime \govermu q^\prime\right), \nn \\
 \op_{\Sigma_q \Sigma_q}^{(8)} &=& \sum_q \left(\bq \gsubmu T^a q \right) \sum_{q^\prime} \lambda_{qq\prime}\left( \bq^\prime \govermu T^a q^\prime\right)\,.
~~~~~~~~~ \nn \\
{\rm LR}:~~\op_{\Sigma_Q \Sigma_q}^{(1)} &=& \sum_i\left(\bQ_i \gsubmu Q_i\right) \sum_q\left(\bq \govermu q\right), \nn \\
\quad \op_{\Sigma_Q \Sigma_q}^{(8)} &=& \sum_i\left(\bQ_i \gsubmu T^a Q_i\right) \sum_q\left(\bq \govermu T^a q\right)
\,.
\end{eqnarray}
Even though the penguin and double penguin operators are redundantly defined, they are useful since all the penguin effect in RG evolution is separately absorbed into penguin operators so that one can easily distinguish the origin of the RG effect. A complete list of anomalous dimension matrix(ADM) is provided in Appendix~\ref{sec:app-adm68}.

\subsubsection{Fully expanded basis without $\rm SU(2)_L$} \label{sec:expop}

Our second basis does not assume the $\rm SU(2)_L$ gauge symmetry. The basis is intended for most straightforward calculation, and can perhaps be used to study the evolution of small $\rm SU(2)_L$-breaking effects from new physics. But the $\rm SU(2)_L$ can still be imposed by boundary conditions; see Appendix~\ref{sec:app-op144}. Both left- and right-handed quarks are denoted by $q_{\rm A} $ for the quark flavor $q$ and the chirality $\rm A$. For any given $q_{\rm A}$, we have the currents $j^\mu_{q_A}$ and $j^{a\mu}_{q_A}$ in \Eq{eq:fullopbasis}. This basis really contains all current-current combinations (see \Eq{eq:fullopbasis1})
\beq
{\cal O}_{q_A q^\prime_B}^{(1)} \= j_{q_A}^\mu j_{q^\prime_B \mu},
\qquad  {\cal O}_{q_A q^\prime_B}^{(8)} \= j_{q_A}^{a \mu} j_{q^\prime_B \mu}^{a}.
\eeq
As mentioned, when $q_{\rm A}=q^\prime_{\rm B}$, color-singlet operators are reduced to their color-octet counterparts. No symmetry factors are factored out. The basis spans 144 operators including all six flavors. A complete list of $144 \times 144$ ADM is provided in Appendix~\ref{sec:app-op144}.

\subsection{Renormalization group equations} \label{sec:rge}

\begin{figure}[t]
\begin{center}
\includegraphics[width=12cm]{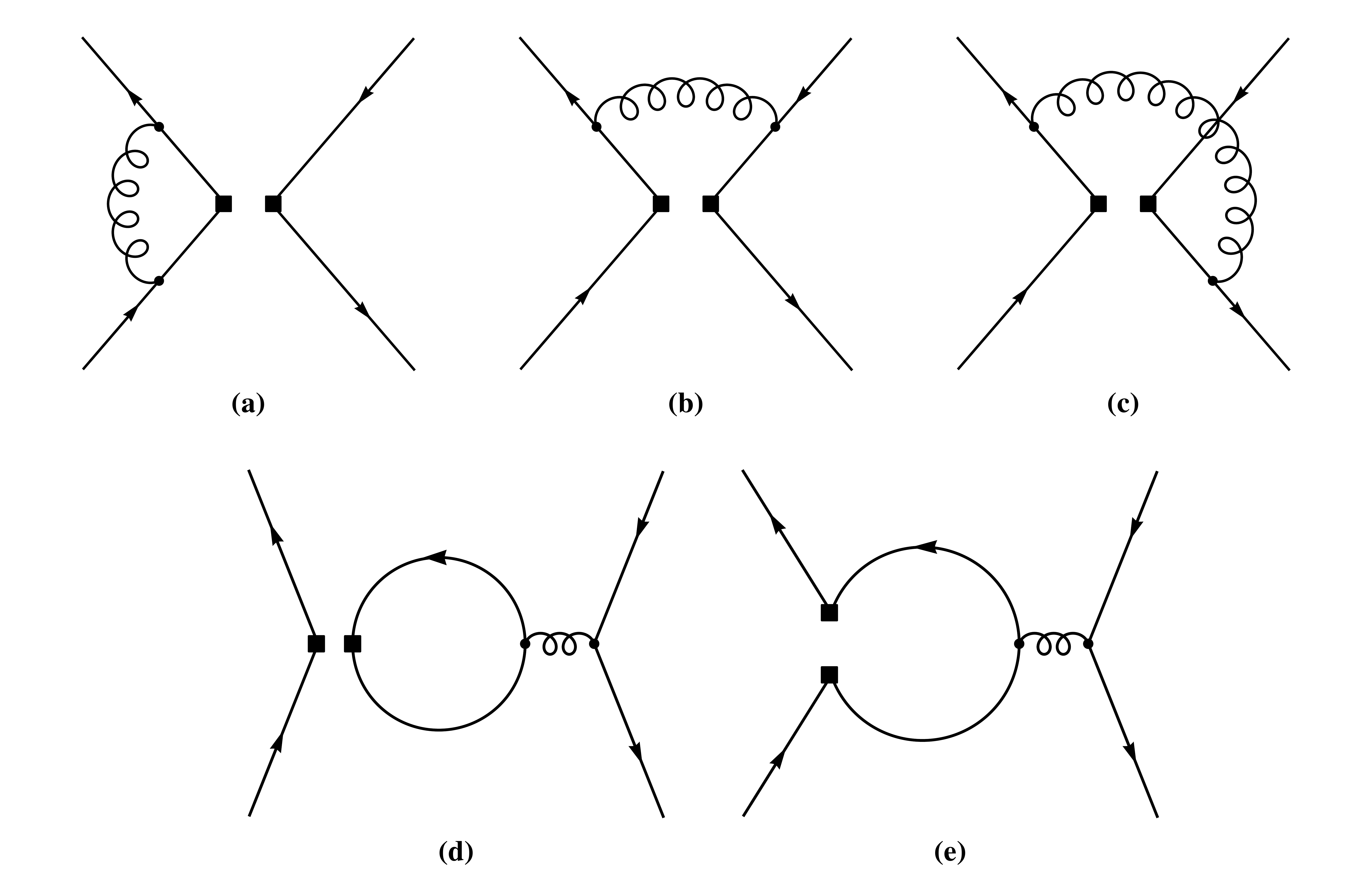}
\end{center}
\vspace{-0.6cm}
\caption{
\label{fig:ADMdiagrams}
Feynman diagrams for the QCD ADM of effective four-quarks operators. Square dots denote the operator insertion. Diagrams (a), (b) and (c) generate tree-operators and diagrams (d) and (e) generate penguin-operators in the tree-penguin basis.}
\end{figure}

For the completeness and introduction of our notation, we summarize how to solve RG equations.
Ultraviolet divergences in the one-loop calculation of effective operators are absorbed by operator renormalization constants
\begin{equation}
\label{eq:OpRen}
\op_i^{\rm bare} = Z_{\op\, ij} \op_j\,.
\end{equation}
The one-loop diagrams of effective four-quarks operators are shown in \Fig{fig:ADMdiagrams}.
If we expand  $Z_{\op\,ij}$ in $1/\varepsilon$ within the dimensional regularization with $d=4-2\,\varepsilon$
\begin{eqnarray}
Z_{\op\,ij} = \delta_{ij} + \sum_{k=1} \frac{Z_{\op,ij}^{(k)}}{\varepsilon^k},
\end{eqnarray}
one can derive the ADM $\gamma_{ij}$ can be obtained by
\begin{equation}
\gamma_{ij} \= Z_{\op\,ik}^{-1} \frac{dZ_{\op\,kj}}{d\ln \mu} \=  -2 \alpha_s \frac{d Z_{\op,\,ij}^{(1)}}{d\alpha_s}\,.
\end{equation}

Wilson coefficients satisfy the following RG equation which can be obtained by demanding the effective theory amplitude is scale-independent.
\begin{equation}
\frac{d C_i(\mu) }{d \ln \mu} \= C_j(\mu) \gamma_{ji} = \gamma^T_{ij} C_j(\mu).
\end{equation}
By diagonalizing the ADM with a matrix $V$
\beq
\widehat{\gamma} \, \equiv \, V^{-1} \gamma^T V, \qquad \widehat{C}_i(\mu) \= V^{-1}_{ij} C_j(\mu)\,,
\eeq
where $\hat \gamma$ is diagonalized matrix,
one can decouple the RG equation
\begin{equation}
\label{eq:dRGE}
\frac{d \widehat{C}_i(\mu)} {d \ln \mu} \= \widehat{\gamma}_{i} \widehat{C}_i (\mu).
\end{equation}
It is straightforward to solve Eq. (\ref{eq:dRGE}) at one-loop order; each Wilson coefficient now runs individually according to its ADM eigenvalue
\beq
\widehat{C}_i(\mu) \= \left( \frac{\alpha_s(\Lambda)}{\alpha_s(\mu)} \right)^{\frac{\widehat{\gamma}_i^{0}}{2\beta_0}}
\widehat{C}_i(\Lambda),
\eeq
where $\beta_0 = \frac{11 N_c}{3} - \frac{2}{3}n_f = 7$ with $n_f=6$ for $\mu>m_t$. At the renormalization scale $\mu_R$, we transform coefficients back to the original basis
\begin{equation}
C_i(\mu_R) \= V_{ij} \widehat{C}_j(\mu_R)  \= V_{ij}  \left( \frac{\alpha_s(\Lambda)}{\alpha_s(\mu_R)} \right)^{\frac{\widehat{\gamma}_j^{0}}{2\beta_0}} V_{jk}^{-1} C_k(\Lambda).
\end{equation}

\section{Leading RG effects absent at tree-level} \label{sec:numerical}

This section contains our main numerical results. We first qualitatively discuss the pattern of operator mixing by simplifying the structure of the full ADM. This allows us to classify possible RG-induced phenomena and underlying models. We then numerically compute each RG-induced phenomena in turn.

\subsection{Operator mixing pattern} \label{sec:4by4}

Most important features of QCD RG evolution that will be the basis of our study can be read from the following $4\times 4$ subset of the full ADM:
\beq
\{ \, {\cal O}^{(1)}_{\rm LL}, \, {\cal O}_{\rm LL}^{(8)}, \, {\cal O}^{(1)}_{\rm LR}, \, {\cal O}_{\rm LR}^{(8)} \, \},
\label{eq:4x4basis}\eeq
where we use the notation of \Sec{sec:observables}: for example, ${\cal O}_{\rm LL}^{(8)} = (\bar{u}_L \gamma^\mu T^a u_L) (\bar{t}_L \gamma_\mu T^a t_L)$ for the up-quark contribution. Penguin effects are numerically subdominant and therefore we ignore them in this approximate discussion. Under this approximation, the subset is closed under QCD RG evolution and flavors of operators are not mixed. Penguin effects are, however, included in our full numerical studies and discussed in relevant places.

The corresponding $4\times 4$ ADM is given by (ignoring penguin contributions)
\beq
\gamma^{0 } \=
\left(
\begin{array}{cccc}
0 & 12  & 0 & 0 \\
\frac{8}{3} & -4  & 0 & 0 \\
0 & 0 & 0 & -12 \\
0 & 0 & -\frac{8}{3} & -14
\end{array}
\right),
\qquad \gamma \= \frac{\alpha_s}{4\pi} \gamma^0.
\label{eq:4by4adm} \eeq
The eigenvalues are $\widehat{\gamma}^0_i = -8,\,4$ and $-16,\,2$, respectively for each block-diagonal $2\times 2$ matrix. The same ADM is obtained for the remaining four operators of ${\cal O}_{\rm RR}$ and ${\cal O}_{\rm RL}$ types.

We extract two main features of QCD RGE:
\bei
\item ${\cal O}_{\rm LL}^{(8)}$ and ${\cal O}_{\rm LR}^{(8)}$ color-octet operators run differently (while ${\cal O}_{\rm LL}^{(8)}$ and ${\cal O}_{\rm RR}^{(8)}$ and, separately, ${\cal O}_{\rm LR}^{(8)}$ and ${\cal O}_{\rm RL}^{(8)}$ run in the same way). In other words, ${\cal O}_{\rm VV}^{(8)}$ and ${\cal O}_{\rm AA}^{(8)}$ mix with each other and, separately, ${\cal O}_{\rm VA}^{(8)}$ and ${\cal O}_{\rm AV}^{(8)}$ mix, but two sets do not mix. This mixing pattern is consistent with the parity quantum numbers of operators; QCD is parity-conserving and parity-odd operators are not induced from parity-even operators\footnote{This parity argument does not prohibit the RG-induction of parity-even operators from parity-odd operators. Indeed, penguin diagrams induce such mixing.}. The mixing between ${\cal O}_{\rm VV}^{(8)}$ and ${\cal O}_{\rm AA}^{(8)}$ provides a useful insight on how QCD generates top asymmetries at one-loop order as will be discussed in \Sec{sec:rgafb}.

\item Color-singlet and -octet operators with same chiralities, e.g. ${\cal O}_{\rm LL}^{(8)}$ and ${\cal O}_{\rm LL}^{(1)}$, mix with each other. This mixing pattern implies that color-singlet models can interfere with the SM at one-loop order although they do not at tree-level.  Various one-loop effects of color-singlet models are indeed relevant and can be studied from the QCD RG evolution. It also provides a useful way to understand how QED generates top asymmetries as will be discussed in \Sec{sec:rgafb}. Furthermore, the ADM is not a symmetric matrix. Thus, the mixings of octet and singlet operators into each other are different.
\eei

All these features are reflected in the following approximate solutions of the RG equations using the ADM in \Eq{eq:4by4adm}
\bea
C_{\rm VV}^{(8)}(m_t) &\, \simeq \,& C_{\rm VV}^{(8)}(\Lambda) \+ \frac{\alpha_s}{4\pi}\ln\frac{m_t}{\Lambda} \, \left( 12 C_{\rm AA}^{(1)}(\Lambda) - 9 C_{\rm VV}^{(8)}(\Lambda) + 5 C_{\rm AA}^{(8)}(\Lambda) \right) \label{eq:sol1-4by4}\,,\\
C_{\rm AA}^{(8)}(m_t) &\, \simeq \,& C_{\rm AA}^{(8)}(\Lambda) \+ \frac{\alpha_s}{4\pi}\ln\frac{m_t}{\Lambda} \, \left( 12 C_{\rm VV}^{(1)}(\Lambda) - 9 C_{\rm AA}^{(8)}(\Lambda) + 5 C_{\rm VV}^{(8)}(\Lambda) \right) \label{eq:sol2-4by4}\,,\\
C_{\rm VA}^{(8)}(m_t) &\, \simeq \,& C_{\rm VA}^{(8)}(\Lambda) \+ \frac{\alpha_s}{4\pi}\ln\frac{m_t}{\Lambda} \, \left( 12 C_{\rm AV}^{(1)}(\Lambda) - 9 C_{\rm VA}^{(8)}(\Lambda) + 5 C_{\rm AV}^{(8)}(\Lambda) \right) \label{eq:sol3-4by4}\,,\\
C_{\rm AV}^{(8)}(m_t) &\, \simeq \,& C_{\rm AV}^{(8)}(\Lambda) \+ \frac{\alpha_s}{4\pi}\ln\frac{m_t}{\Lambda} \, \left( 12 C_{\rm VA}^{(1)}(\Lambda) - 9 C_{\rm AV}^{(8)}(\Lambda) + 5 C_{\rm VA}^{(8)}(\Lambda) \right) \label{eq:sol4-4by4}\,,
\eea
where we expand the leading RG effect up to order $\alpha_s$. RG-induced terms are proportional to large logarithmic term $\frac{\alpha_s}{4\pi}\ln\frac{m_t}{\Lambda}$.

\subsection{Models and matching} \label{sec:matching}

We select benchmark models to illustrate RG-induced phenomena. Most dramatic and important illustrative phenomena are ones absent at tree-level but induced first at one-loop order. Here, calculations of one-loop RG equations are most useful and important because one-loop RG effects are leading contributions. At this order, the relevant observables receive only RG-induced terms with large logarithmic terms in \Eq{eq:sol1-4by4}-\Eq{eq:sol4-4by4}.

The benchmark models are listed in Table.~\ref{tab:initial}. Models are chosen to have only one effective operator in the V--A basis at $\Lambda$ so that the operator mixing effect is more clearly separated. In this case, the RG-induction of other operators at low-energy becomes the leading contribution to the phenomena (see \Sec{sec:observables} for the correspondence between observables and operators in the V--A basis and \Sec{sec:rgvssq} for another reason why the V--A basis is suitable for us).  The last model in the Table is considered for another purpose.

\begin{table}[t] \centering
\begin{tabular}{c||c|c|c}
\hline \hline
Initial condition & $s$-channel resonance & $g_u$ & $g_t$ \\
\hline \hline
VV(1), VV(8)     & color-singlet/octet vector & $g_{u_{\rm R}}=g_{u_{\rm L}}=g_{d_{\rm R}}$ & $g_{t_{\rm R}}=g_{t_{\rm L}}=g_{b_{\rm R}}$ \\
AA (1), AA(8)    & color-singlet/octet vector & $g_{u_{\rm R}}=-g_{u_{\rm L}}=g_{d_{\rm R}}$ & $g_{t_{\rm R}}=-g_{t_{\rm L}}=g_{b_{\rm R}}$ \\
AV(1), AV(8) & color-singlet/octet vector & $g_{u_{\rm R}}=-g_{u_{\rm L}}=g_{d_{\rm R}}$ & $g_{t_{\rm R}}=g_{t_{\rm L}}=g_{b_{\rm R}}$ \\
VA(1), VA(8)   & color-singlet/octet vector & $g_{u_{\rm R}}=g_{u_{\rm L}}=g_{d_{\rm R}}$ & $g_{t_{\rm R}}=-g_{t_{\rm L}}=g_{b_{\rm R}}$ \\
VR(8)   & color-octet vector & $g_{u_{\rm R}}=g_{u_{\rm L}}=g_{d_{\rm R}}$ & $g_{t_{\rm R}}=g_{b_{\rm R}},\, g_{t_{\rm L}}=0$ \\
\hline\hline
\end{tabular}
\caption{Models used to describe the phenomena first induced by RG at one-loop. Two parameters, $g_u$ and $g_t$, specify all couplings; $g_u(g_t)$ denotes couplings to the first two (third) generations. We do not use different notations for octet and singlet couplings for simplicity. The matching is given by \Eq{eq:matching1} and \Eq{eq:matching2}.}
\label{tab:initial} \end{table}

The models containing a heavy $s$-channel resonance $X_\mu$,
\bea
{\cal L}_{\rm full} \, &\ni& \, g_{q_A} \, j_{q_A}^\mu X_\mu\, \quad \textrm{or} \quad  g_{q_A} \, j_{q_A}^{a\mu} X^a_\mu
\eea
is matched to four-quark effective operators in \Eq{eq:fullopbasis} at $\Lambda$. The $g_{q_A}$ with the flavor index $q$ and the chiral index $A$ denotes the coupling strength of the new particle $X^\mu$ to the current $j_{q_A}^\mu$. We comment that $t$-channel models~\cite{Jung:2009jz}, after the Fierz transformations, do not satisfy any of the initial conditions in Table~\ref{tab:initial}, thus we do not consider them.

If a new physics respects the $\rm SU(2)_L$ symmetry, the tree-level matching coefficients in the tree-penguin classified basis of \Sec{sec:tpop} are as follows
\bea
C_{Q_i Q_j} \= - g_{q_L} g_{q^\prime_L} \,,~~
C_{Q_i q^\prime} \= - g_{q_L} g_{q^\prime_R} \,,~~
C_{q q^\prime} \= - g_{q_R} g_{q^\prime_R}\,,
\label{eq:matching1}\eea
where generation index $i(j)$ corresponds to that of $q(q^\prime)$. Penguin and double penguin operators vanish at $\Lambda$. The Wilson coefficients can be either color-singlet or -octet. For the fully expanded basis introduced in \Sec{sec:expop},
\beq
C_{q_{\rm A} q^\prime_{\rm B}} \= \left\{ \bmat - g_{q_{\rm A}} g_{q^\prime_{\rm B}} \qquad \textrm{for } q_{\rm A} \ne q^\prime_{\rm B} \\ - g_{q_{\rm A}}^2 / 2 ~~\qquad \textrm{for } q_{\rm A} = q^\prime_{\rm B} \emat \right..
\label{eq:matching2} \eeq
The factor of 1/2 in \Eq{eq:matching2} is absorbed into the definition of operators with $i=j$ in the tree-penguin basis; e.g. see ${\cal O}_{qq}^{(8)}$ in \Eq{eq:treeop}. The factor of 1/2 arises due to the symmetry factors that arise from the contraction of 4 identical quarks with bilinear operator. We simply normalize the scale of effective operators to $\Lambda$.\footnote{The actual effective operator scale $\Lambda$ is determined not only by the mass of new particles, $M_X$, but also by operator power counting schemes. For example, the naive dimensional analysis implies $\Lambda \sim M_X / 4\pi$~\cite{Jenkins:2013zja,Jenkins:2013sda} while a direct matching would give $\Lambda \sim M_X$. Since we consider only four-quark operators, the operator power counting scheme is not so important. We simply use a general notation $\Lambda$ which can be properly interpreted later.}

\subsection{RG-induction of top asymmetries} \label{sec:rgafb}

\begin{figure}[t] \centering
\includegraphics[width=0.49\textwidth]{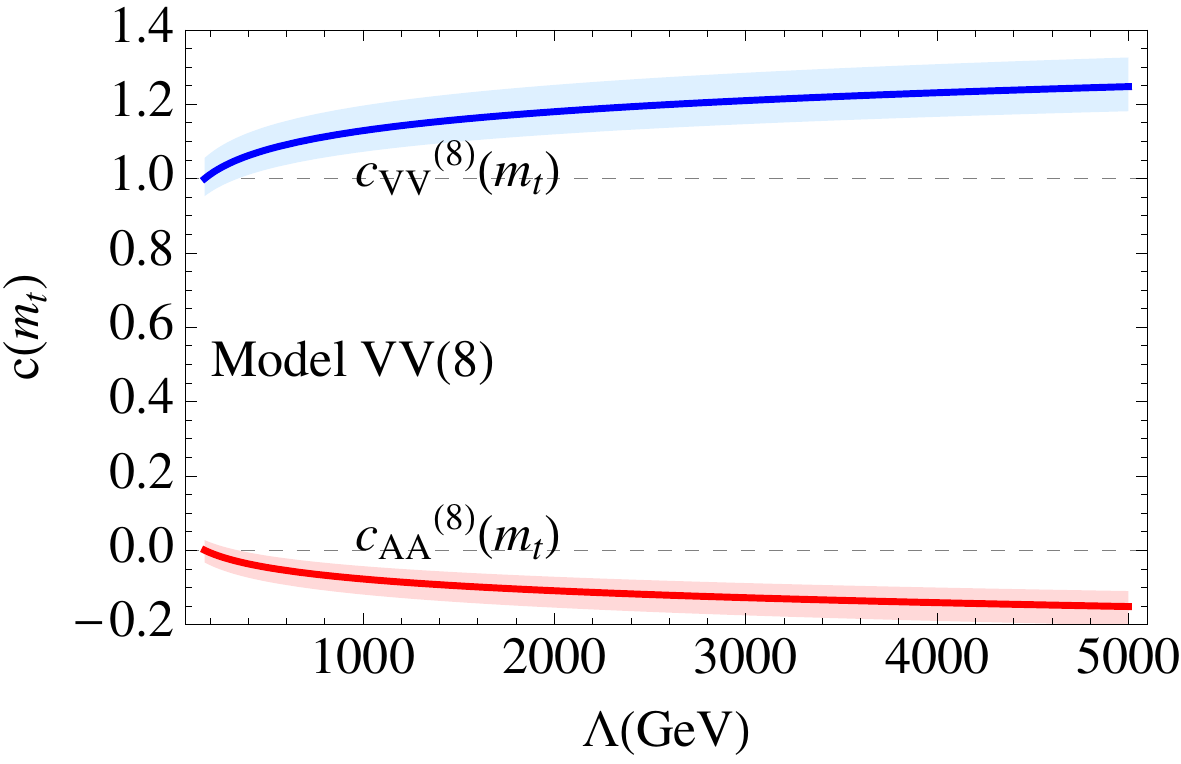}
\includegraphics[width=0.49\textwidth]{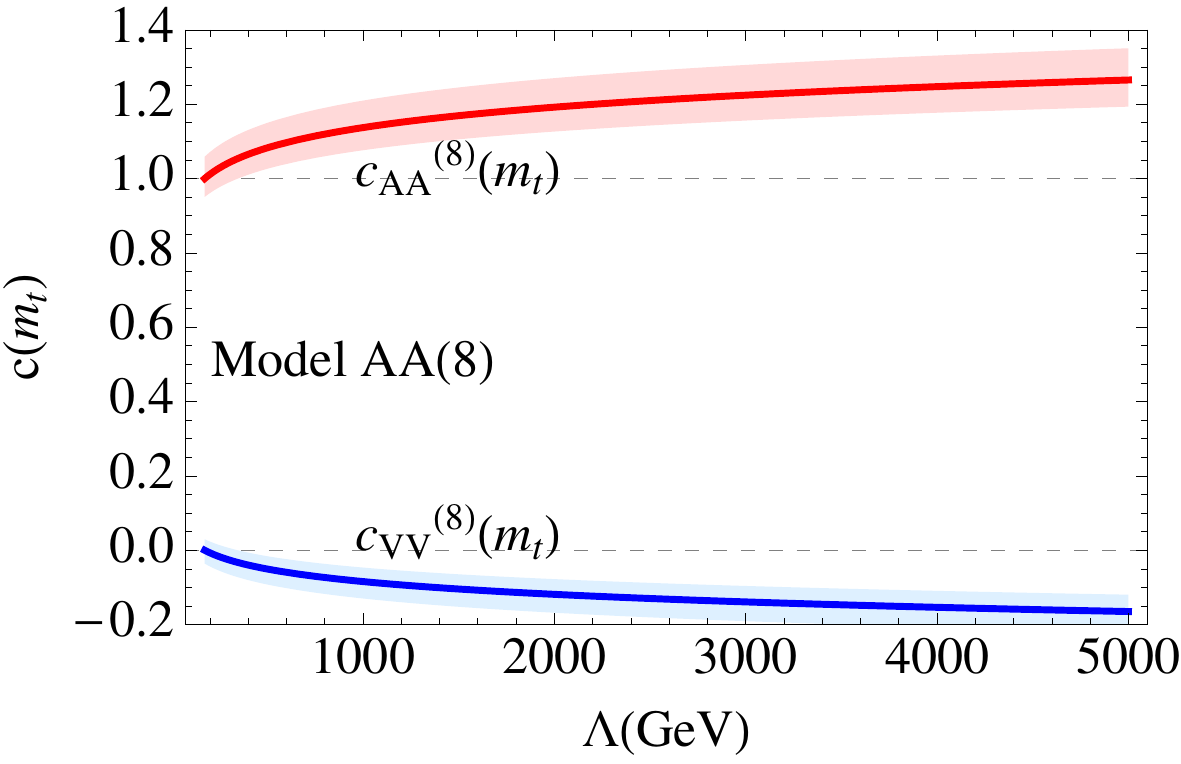}\\
\includegraphics[width=0.49\textwidth]{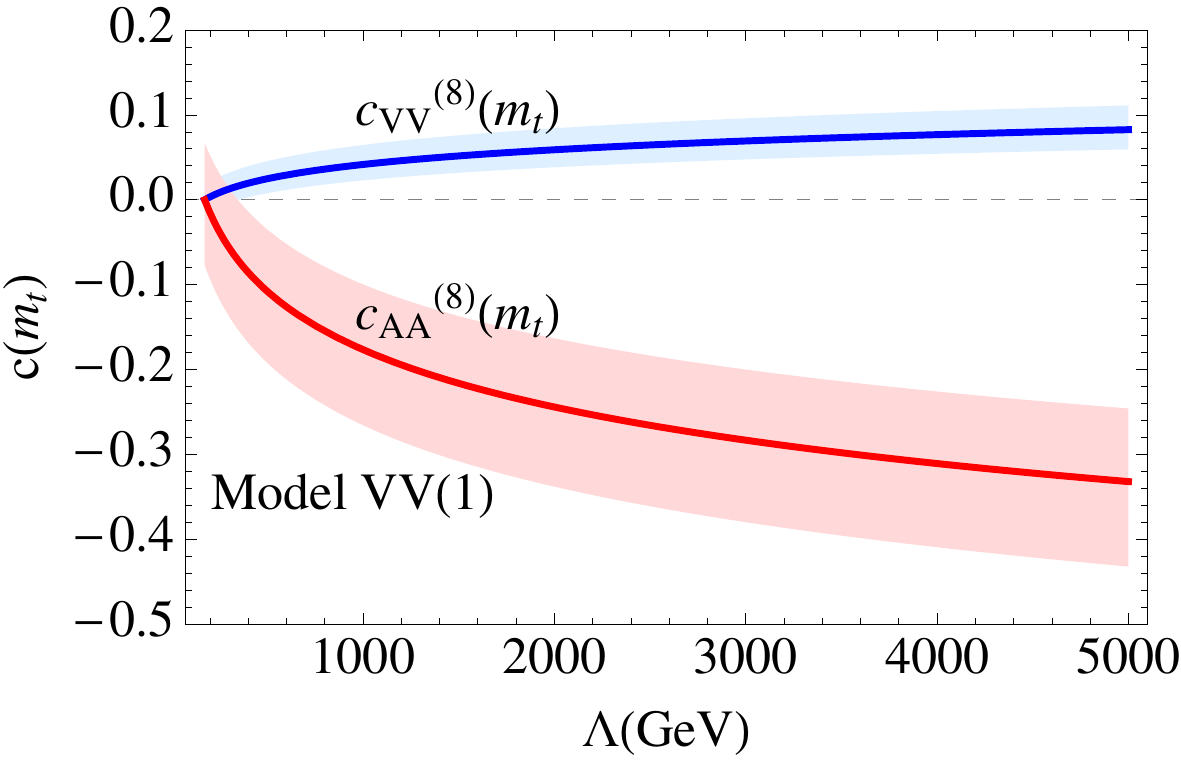}
\includegraphics[width=0.49\textwidth]{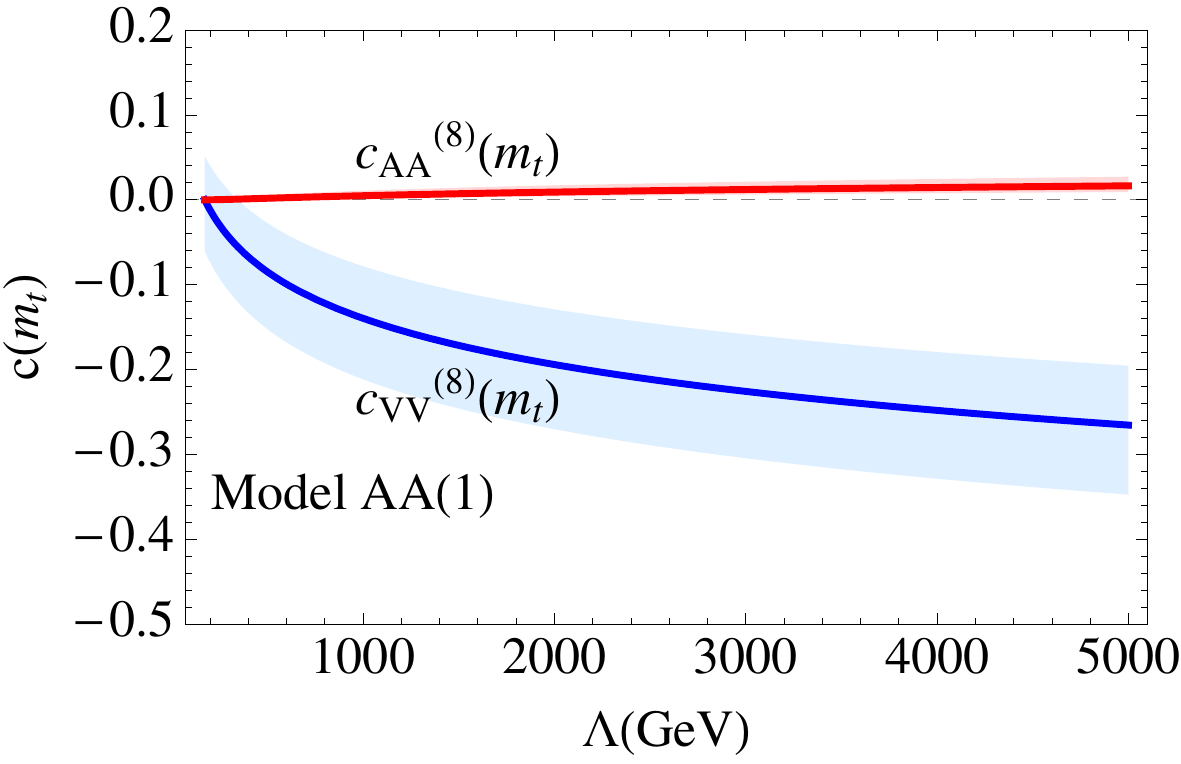}
\caption{Wilson coefficients of models with parity-even operators evaluated at the $m_t$ scale. They are RG evolved down from $\Lambda$ where models with $g_u = -g_t =1$ are matched to operators. Shaded bands are scale uncertainties.}
\label{fig:coeff-afb}
\end{figure}

\begin{figure}[t] \centering
\includegraphics[width=0.49\textwidth]{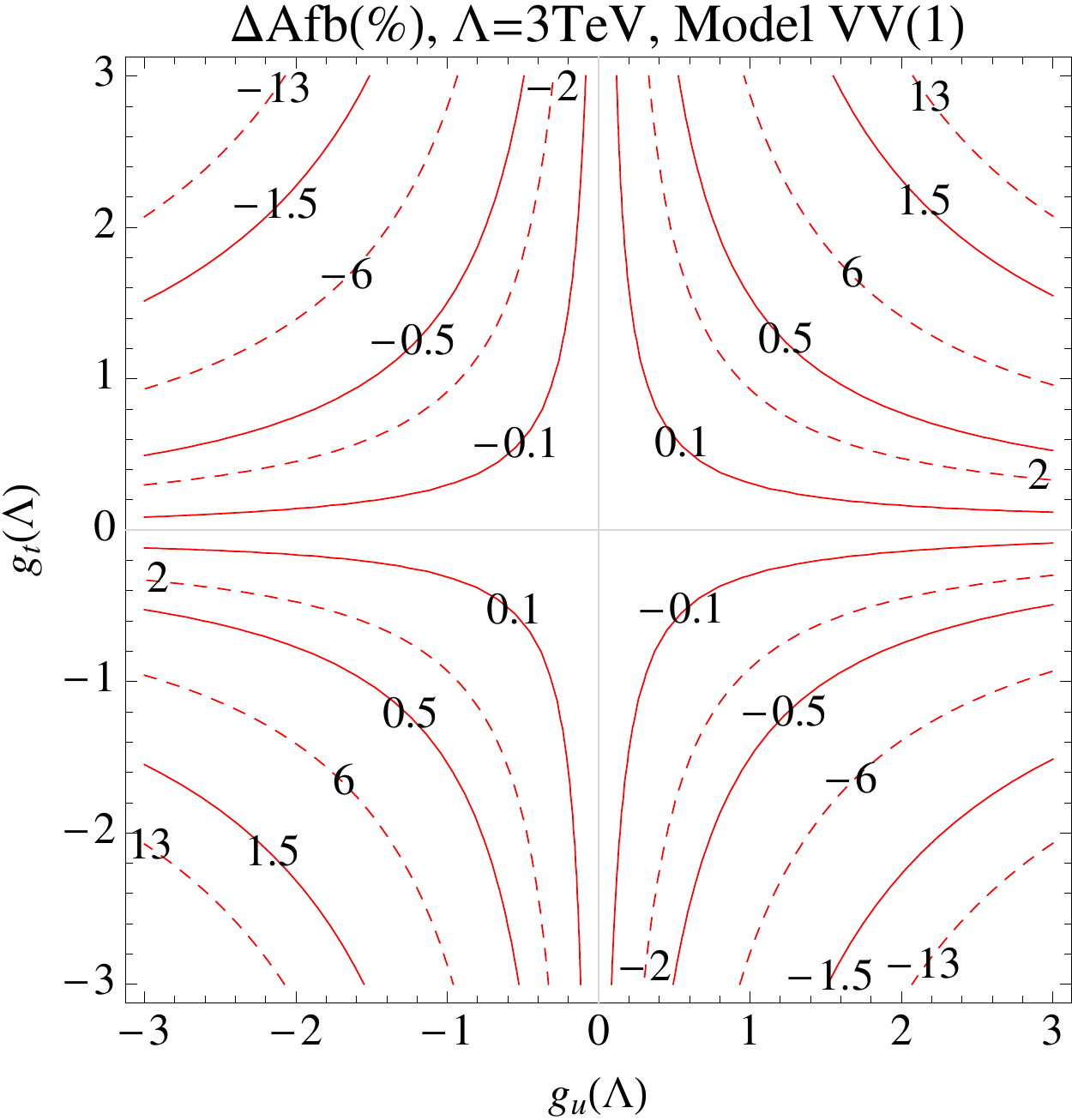}
\includegraphics[width=0.49\textwidth]{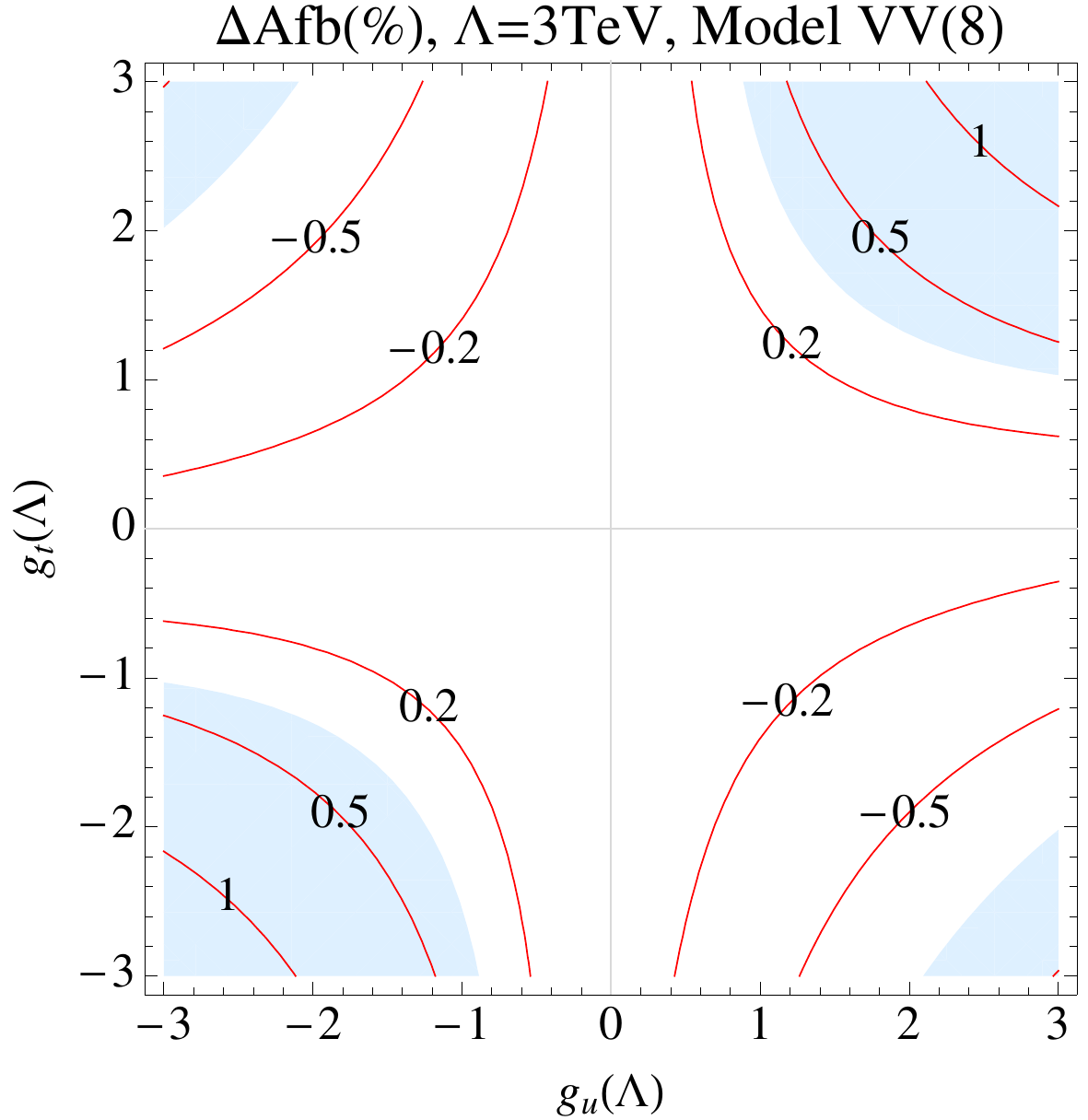}
\caption{RG-induced total top asymmetries in units of \%. Dashed lines are the top asymmetry in high-mass region $m_{t\bar{t}}>800
\GeV$. Models are matched at $\Lambda$ and RG evolved to the $m_t$ scale where top asymmetries are calculated. Shaded regions are disfavored by top pair cross-section measurements at $2\sigma$ level. $\Lambda = 3\TeV$.}
\label{fig:span1}
\end{figure}

At tree-level, the top asymmetry is induced by ${\cal O}_{\rm AA}^{(8)}$ operator. All possible RG-induction of top asymmetries is thus based on one of the following operator mixing patterns: ${\cal O}_{\rm VV}^{(8)}, {\cal O}_{\rm VV}^{(1)}, {\cal O}_{\rm AA}^{(1)} \, \to \, {\cal O}_{\rm AA}^{(8)}$. Four models, VV(8), VV(1) and AA(8), AA(1) are considered to illustrate each pattern. They span a whole set of models that can induce top asymmetries first at one-loop order.

RG-induced Wilson coefficients of ${\cal O}_{\rm VV}$ and ${\cal O}_{\rm AA}$ are evaluated in \Fig{fig:coeff-afb}, the top asymmetry at Tevatron is calculated in the plane of model parameters in \Fig{fig:span1} and detailed model predictions are compared with current data in Table~\ref{tab:datapred}.

The model VV(1) can have largest RG effects on the asymmetry. With $g_{u,t} \sim 2.5$ at $\Lambda =3$TeV, the VV(1) model induces the observable size of top asymmetries in the high-mass region ($m_{t\bar{t}} > 650\GeV$); see Table~\ref{tab:datapred} that about 9\% asymmetry is predicted while current theoretical uncertainty is only about 4.3\%. With a weaker coupling about QCD coupling strength, however, the effect falls below the current theoretical uncertainty as shown in \Fig{fig:span1}.

\begin{figure}[t] \centering
\includegraphics[width=0.55\textwidth]{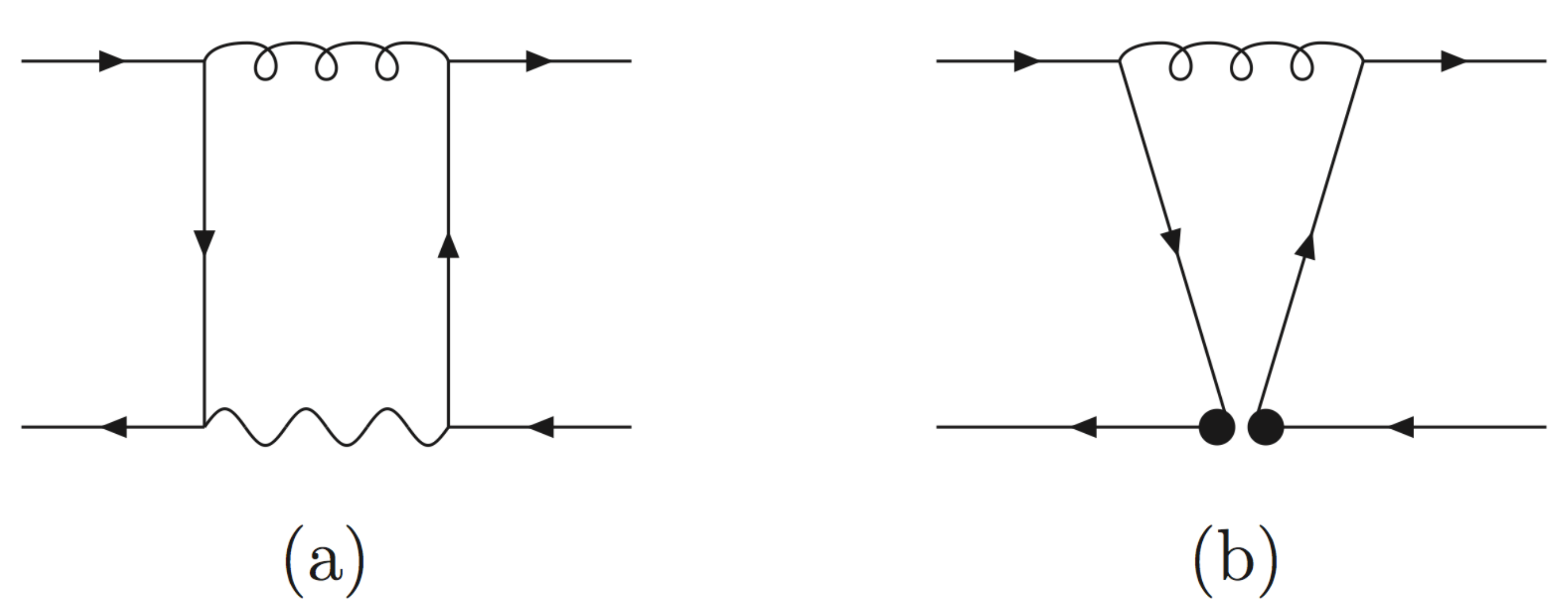}
\caption{(a) QCD-QED box diagrams generating the top asymmetry. (b) QCD RG evolution of four-quark operators. If photons were heavy in (a), integrating out heavy photons approximately reproduces the diagram (b).}
\label{fig:qcdandop}
\end{figure}

The VV(1)'s RG-induction of top asymmetries is analogous to the QED's generation of top asymmetries at one-loop order. QED does not interfere with QCD at tree-level, thus no asymmetry is generated. At one-loop order, however, it interferes with QCD through box diagrams and generates non-zero asymmetries~\cite{Berends:QED}. This underlying physics is captured by QCD RG evolutions here. Consider a heavier version of photons so called heavy photons. The box diagram inducing the top asymmetry is drawn as \Fig{fig:qcdandop}(a) in the full theory side. If heavy photons are integrated out to form ${\cal O}_{\rm VV}^{(1)}$ effective operators at $\Lambda$, the subsequent QCD RG evolution of the operators inducing the top asymmetry are triggered by diagram \Fig{fig:qcdandop}(b). Notably and clearly, two diagrams in \Fig{fig:qcdandop} are originated from the same physics. One can effectively think of the mechanism of QED's generation of top asymmetries in terms of the relevant operator mixing and induction.

In exactly the same way, the QCD's generation of top asymmetries through the one-loop box diagrams~\cite{oai:arXiv.org:hep-ph/9802268} are analogous to the RG-induction of top asymmetries from the VV(8) model -- note that ${\cal O}_{\rm AA}^{(8)}$ is RG-induced from ${\cal O}_{\rm VV}^{(8)}$. However, the VV(8) model interferes with the SM at tree-level modifying the top pair production rate sizably, thus this type of models is strongly constrained as depicted in \Fig{fig:span1}.

Another notable result in \Fig{fig:coeff-afb} is that the AA(1) model induces a small ${\cal O}_{\rm AA}^{(8)}$ operator. It is actually related to the fact that the VV(1) model has the largest RG effects on the top asymmetry. There are several ways to understand this. First of all, the approximate solution in \Eq{eq:sol2-4by4} shows that the $C_{\rm VV}^{(1)}(\Lambda)$ is a main source of the low-energy $C_{\rm AA}^{(8)}(m_t)$. Equivalently, one can also find its origin from the $4\times 4$ ADM in \Eq{eq:4by4adm}. Off-diagonal elements of each $2 \times 2$ sub-matrix have the same magnitude but just opposite signs; one is $12$ and the other is $-12$. For the VV(1) model, the same $C_{\rm LL}^{(1)} = C_{\rm LR}^{(1)}$ color-singlet Wilson coefficients would induce approximately opposite $C_{\rm LL}^{(8)} = - C_{\rm LR}^{(8)} $ color-octet coefficients, thus a maximal $C_{\rm AA}^{(8)} \propto C_{\rm LL}^{(8)} - C_{\rm LR}^{(8)} + \cdots$. By exactly the same argument, on the other hand, the AA(1) model with $C_{\rm LL}^{(1)} = - C_{\rm LR}^{(1)}$ would induce large $C_{\rm VV}^{(8)} \propto C_{\rm LL}^{(8)} + C_{\rm LR}^{(8)} + \cdots$ but small $C_{\rm AA}^{(8)}$.

In Ref.~\cite{Jung:2014gfa}, based on the QCD eikonal approximation and its color structure, it was shown that soft real correction contributions to top asymmetries are very small for the AA(1) model (but not small for the VV(1)), and the soft virtual correction would have similar suppression because it is inherently related to the soft real correction to cancel soft singularities in inclusive processes. The arguments based on the QCD eikonal approximation and on the ADM should be related with each other via QCD color factors; see also Refs.~\cite{Zhu:2012ts,Li:2013mia,Skands:2012mm,Gripaios:2013rda} for how QCD color factors are related with top asymmetries.

We make a useful but warning remark on \Fig{fig:span1}; the figure shows that the VV(8) and VV(1) with $g_u g_t >0$ , i.e., same-sign couplings, can induce positive top asymmetries. This feature may interestingly imply that a model may not need any flavor structure to induce a positive asymmetry -- in other words, flavor-independent couplings are good enough. Although QCD and QED can do so, it is known that the majority of new physics models need some flavor structure for the positive asymmetry. The only known possible new physics model without flavor structure is light axigluons~\cite{lightaxi1,Krnjaic:2011ub,AguilarSaavedra:2011ci}. Thus, the possibility of flavorless model building is thought exciting. However, the full one-loop study of the VV(1) model in Ref.~\cite{Jung:2014gfa} showed that the actual top asymmetry induced at one-loop order has an opposite sign from that predicted solely based on the RG calculation here. As will be discussed in \Sec{sec:rgvalid}, this is due to full one loop contributions that are not large logarithmic and not resummed by our RG calculation. Nevertheless, we emphasize that the useful qualitative discussions on the operator mixing pattern (hence, the classification of models for the loop-induced asymmetries) remain true.

\subsection{RG-induction and -mixing of top polarizations} \label{sec:polmix}

\begin{figure}[t] \centering
\includegraphics[width=0.49\textwidth]{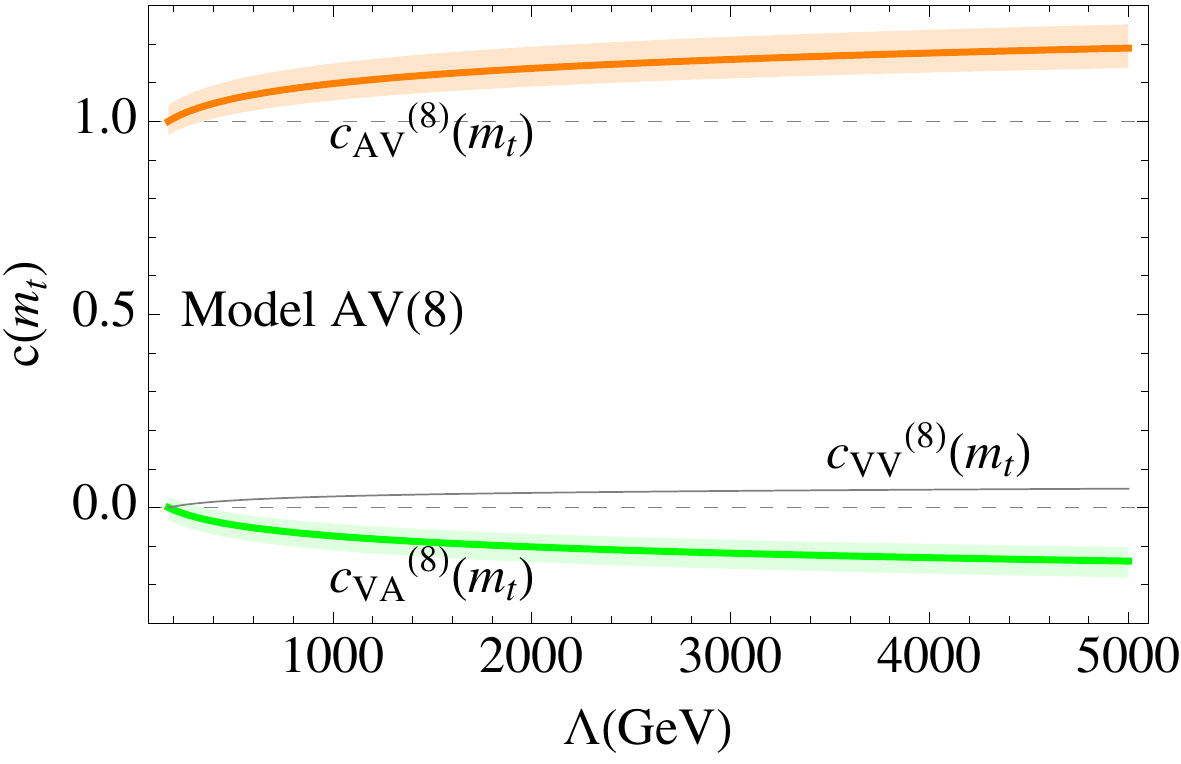}
\includegraphics[width=0.49\textwidth]{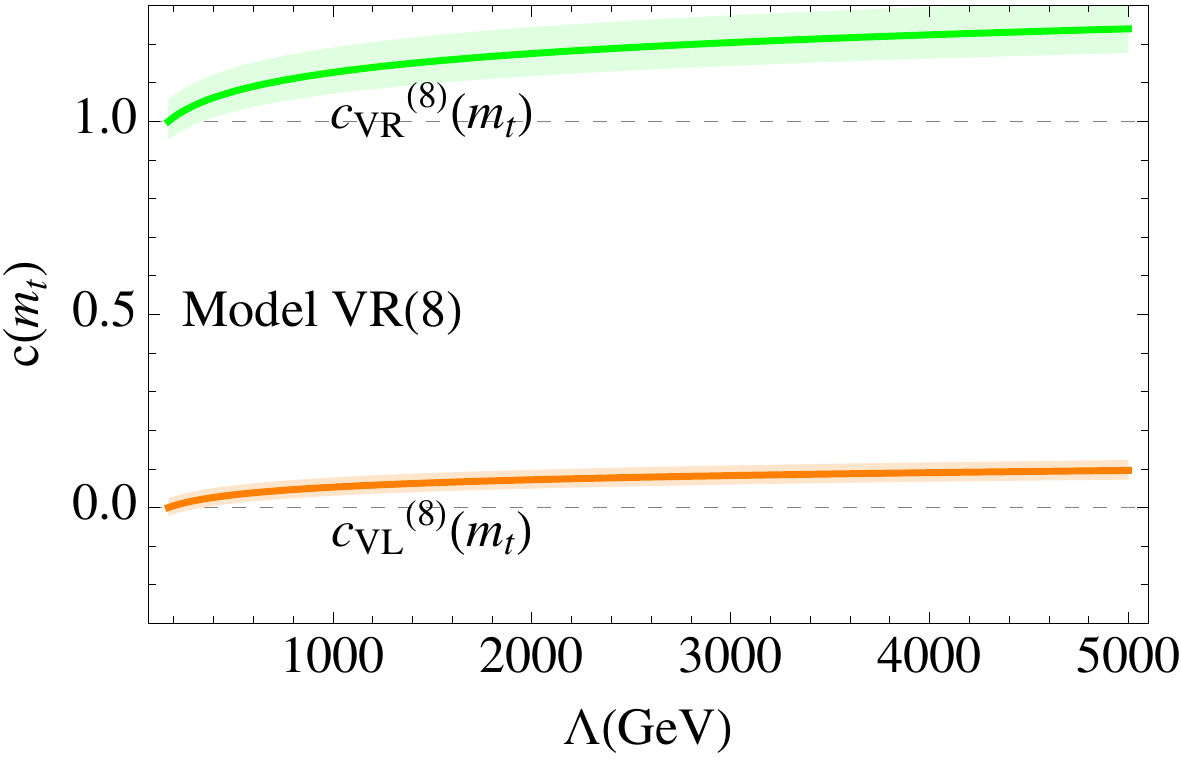}\\
\includegraphics[width=0.49\textwidth]{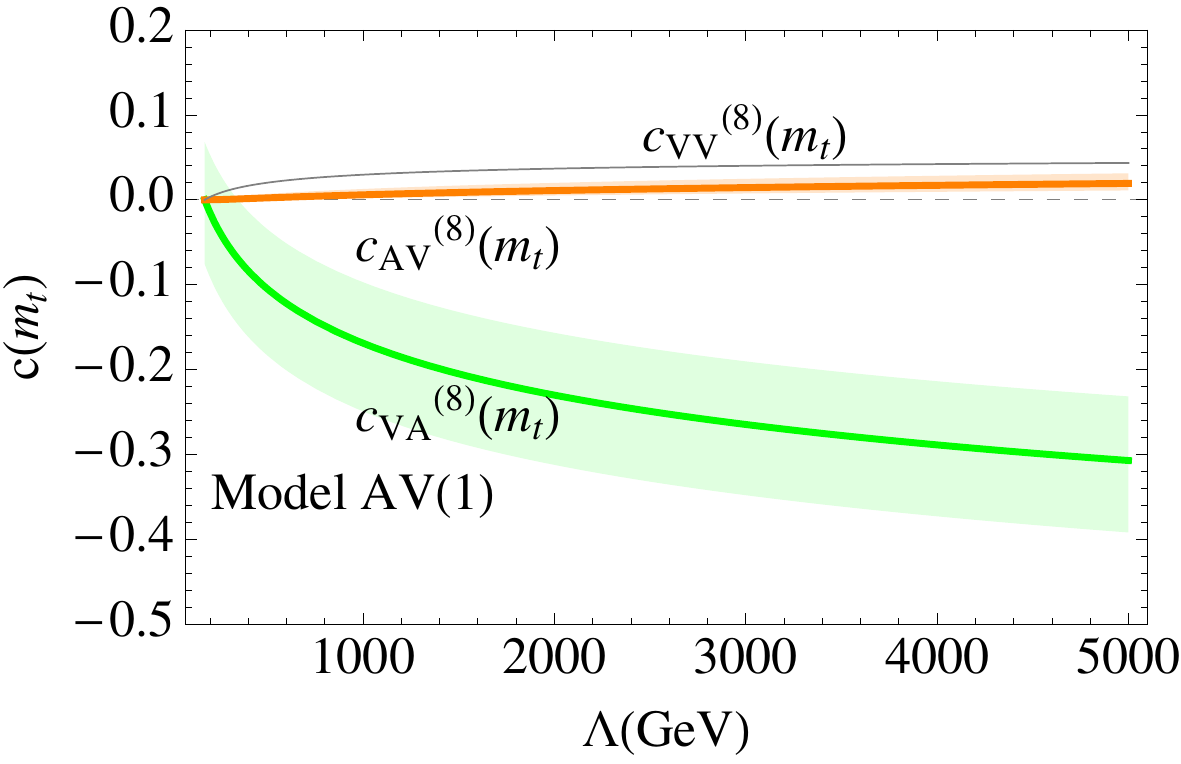}
\includegraphics[width=0.49\textwidth]{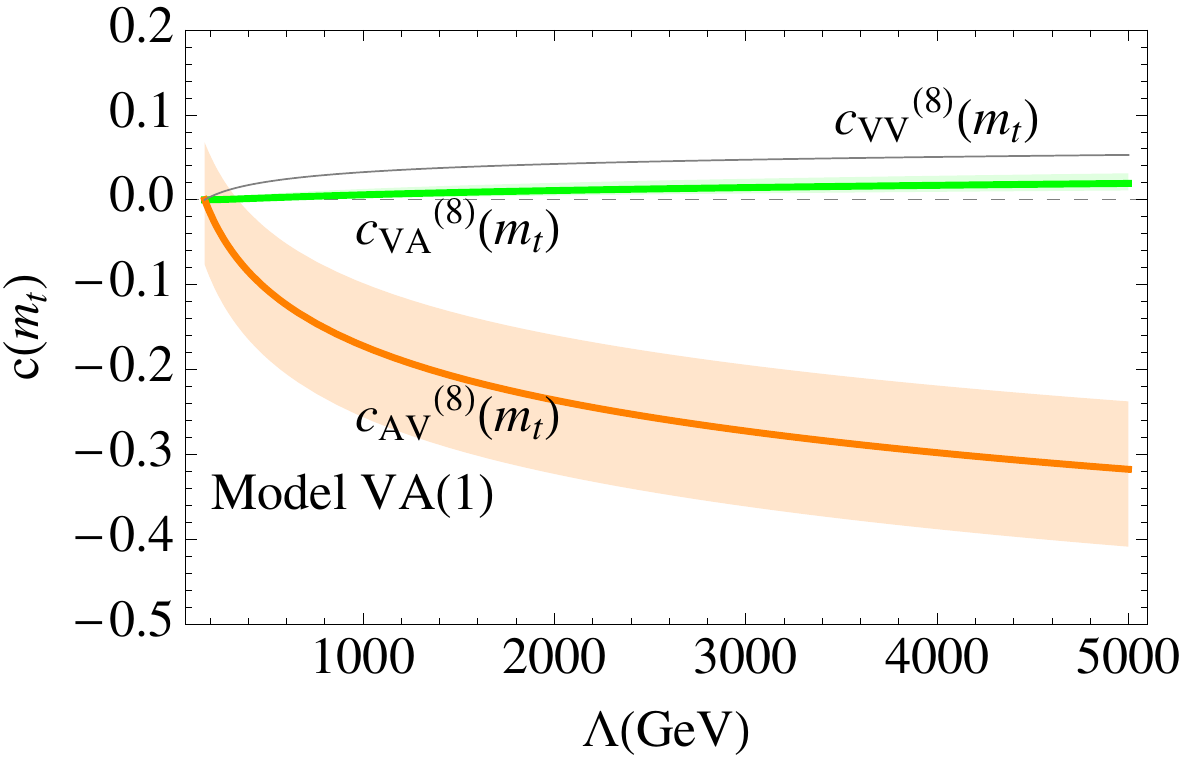}
\caption{Same as in \Fig{fig:coeff-afb} but showing other Wilson coefficients for models with parity-odd operators.}
\label{fig:coeff-pol}
\end{figure}

\begin{figure}[t] \centering
\includegraphics[width=0.49\textwidth]{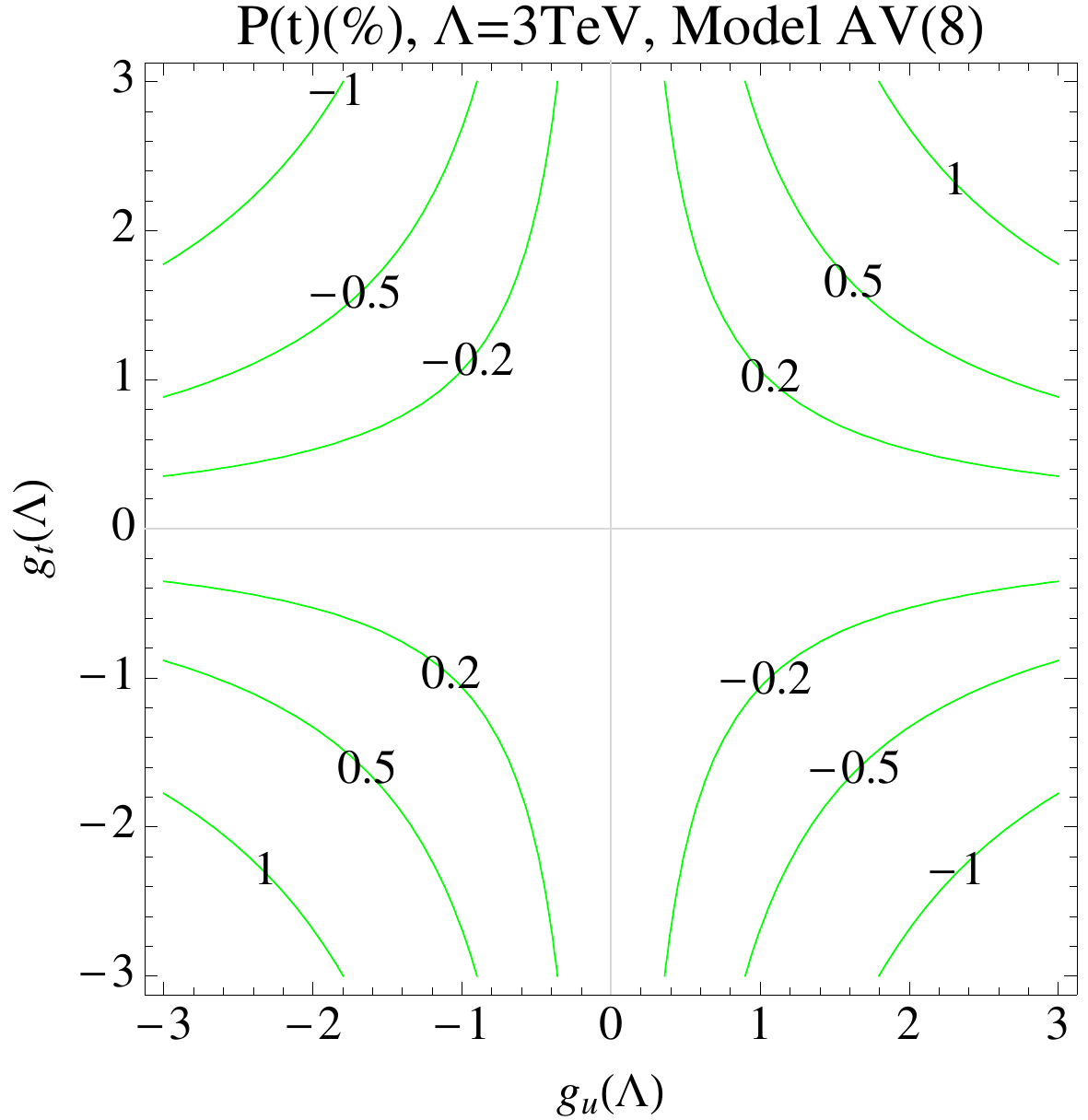}
\includegraphics[width=0.49\textwidth]{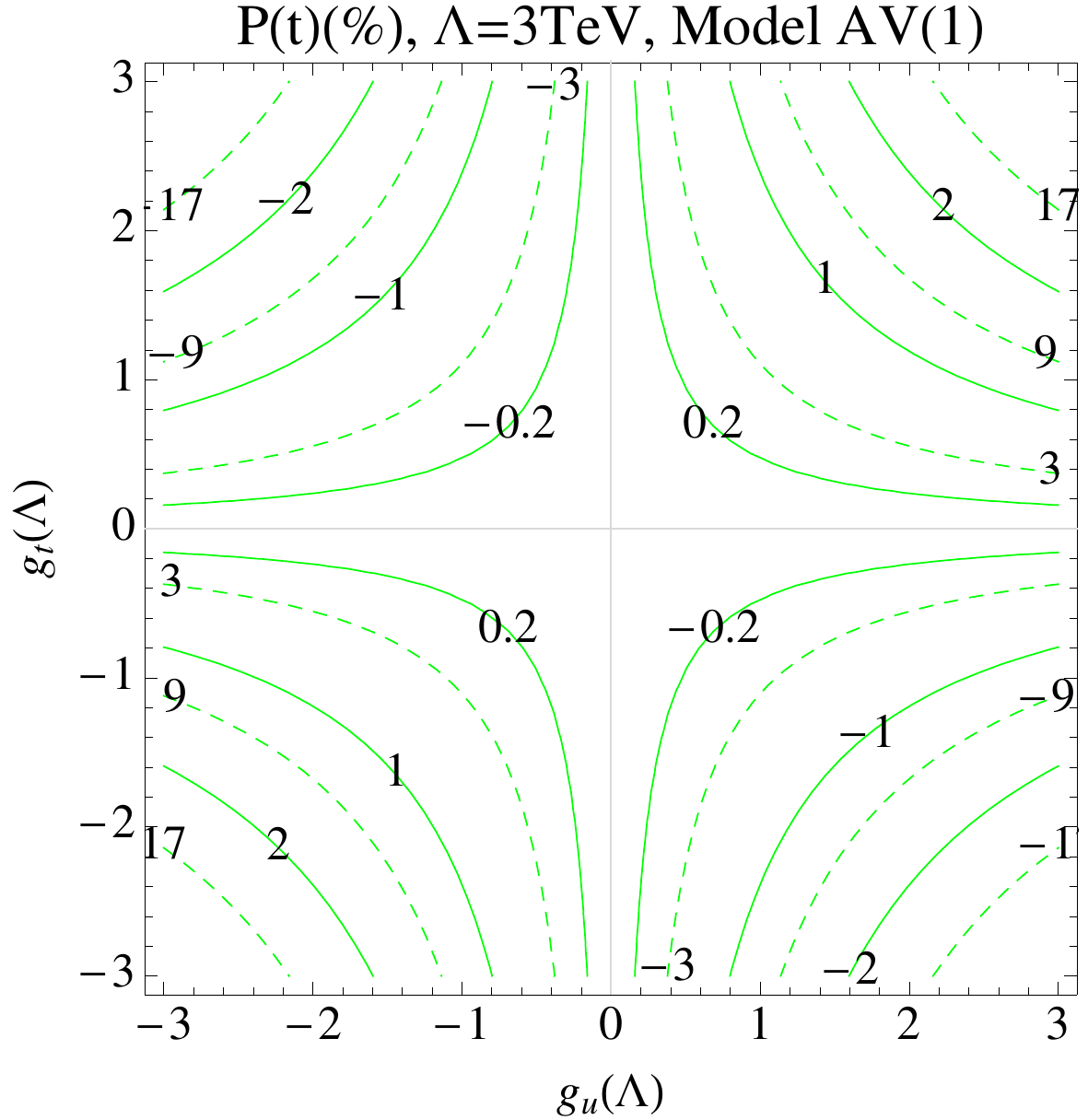}
\includegraphics[width=0.49\textwidth]{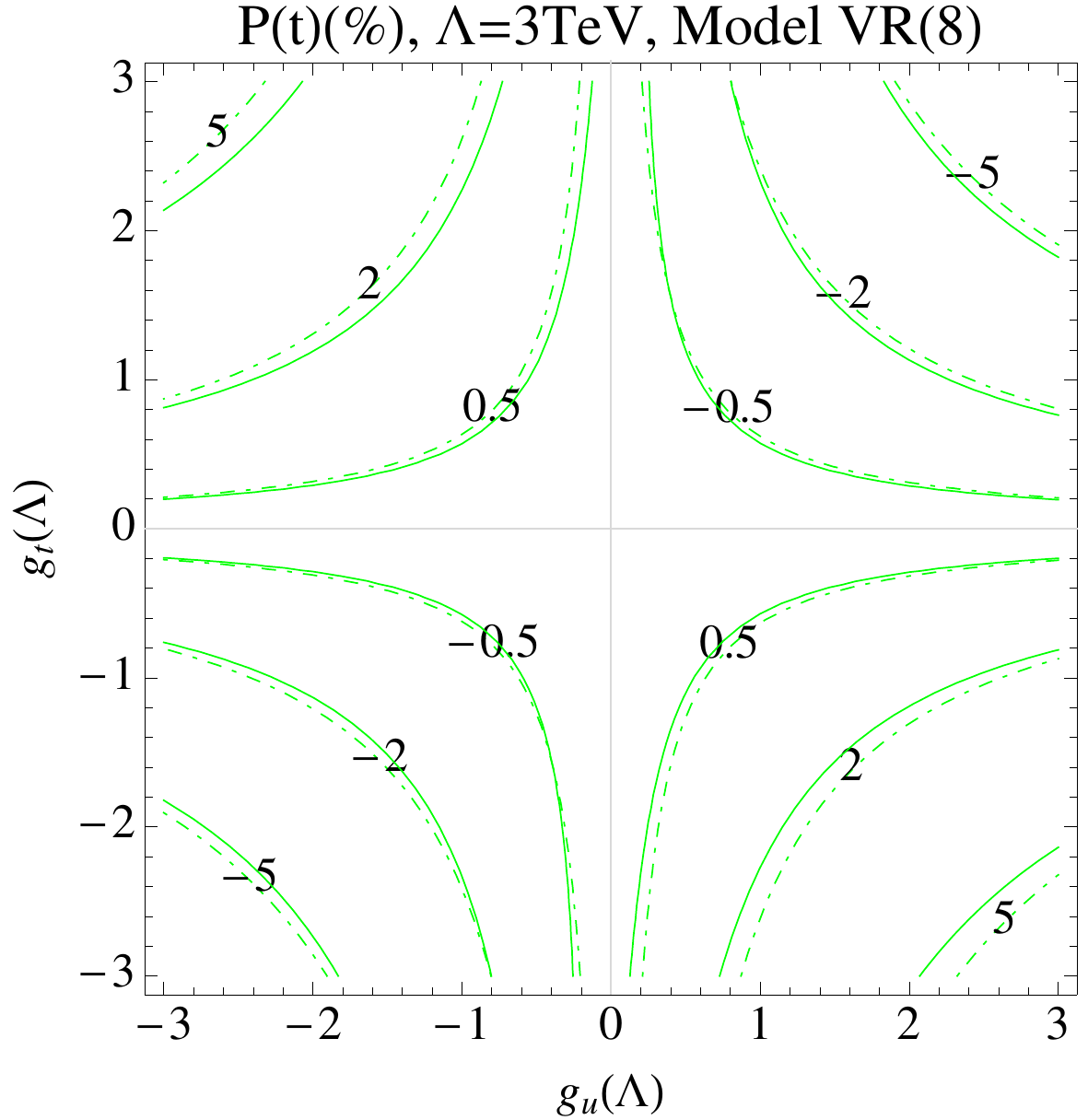}
\includegraphics[width=0.49\textwidth]{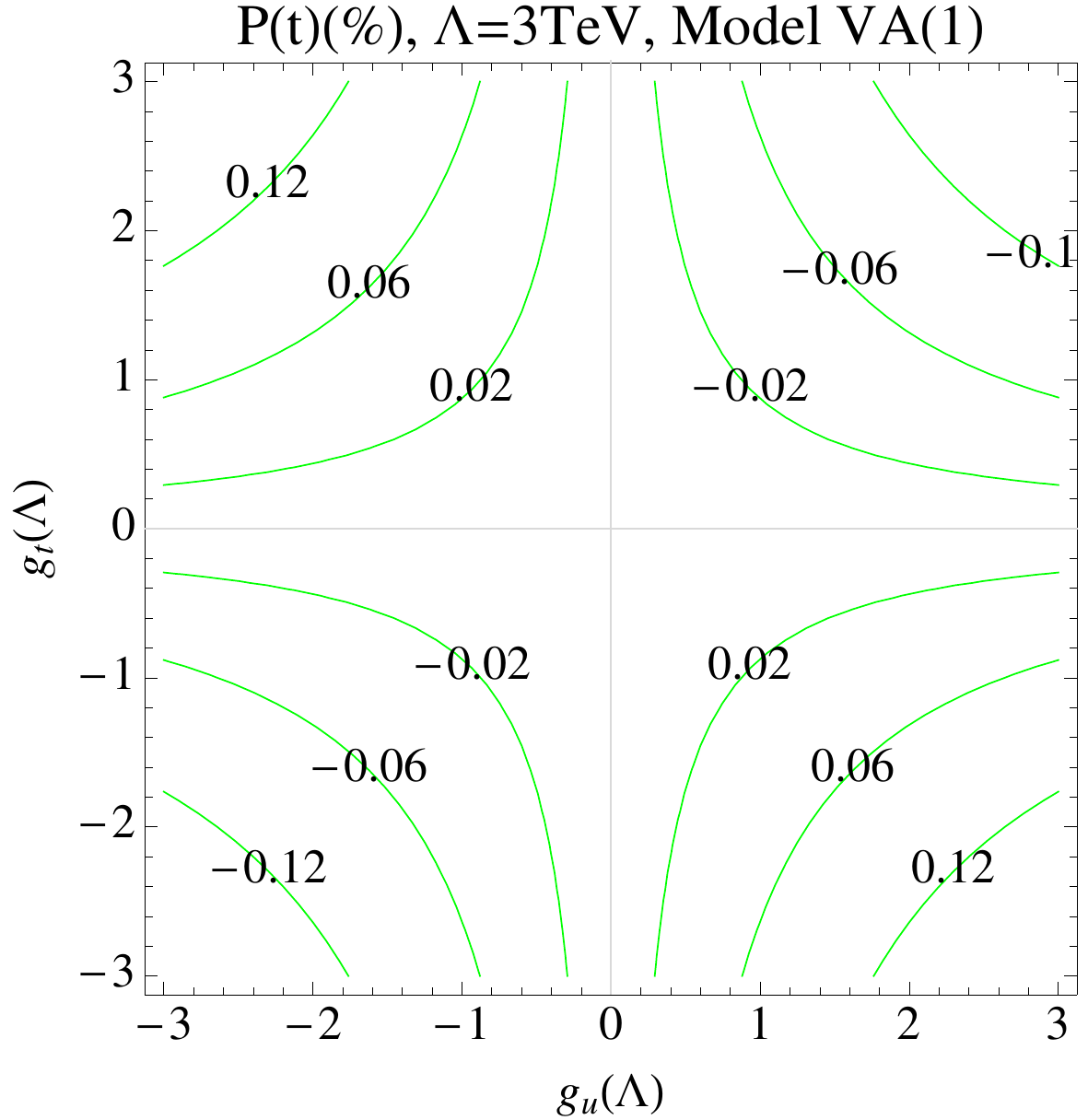}\\
\caption{RG-induced top polarizations in \%. Dashed lines(left) are polarizations with $m_{t\bar{t}}>800$GeV, and dotdashed lines(right) are tree-level results without RG effects added. All other details are as in \Fig{fig:span1}.}
\label{fig:span2}
\end{figure}

Next interesting observable is the top polarization. Although the polarization is currently measured only at the LHC with a low accuracy~\cite{Aad:2013ksa,Chatrchyan:2013wua}, the Tevatron has been measuring spin correlations based on related techniques. It is also known that many new physics can be efficiently measured and distinguished through the polarization measurements~\cite{Degrande:2010kt,Jung:2010yn,oai:arXiv.org:1010.1458,Krohn:2011tw}.

At tree-level, the top polarization is induced by ${\cal O}_{\rm VA}^{(8)}$ operator. All possible RG-induction of top polarizations is thus based on one of the following operator mixing patterns: ${\cal O}_{\rm AV}^{(8)}, {\cal O}_{\rm AV}^{(1)}, {\cal O}_{\rm VA}^{(1)} \, \to \, {\cal O}_{\rm VA}^{(8)}$. Three models, AV(8), AV(1) and VA(1), are considered to illustrate each pattern. They span a whole set of models that can induce top polarizations first at one-loop order. Similarly, the left- and right-handed top polarizations can mix under QCD RG evolution in the sense that one polarization can induce the other based on the operator mixing ${\cal O}_{\rm VR} \leftrightarrow {\cal O}_{\rm VL}$.

RG-induced Wilson coefficients are evaluated in \Fig{fig:coeff-pol}, the Tevatron top polarization is calculated in the plane of model parameters in \Fig{fig:span2} and model predictions (with and without RG effects taken into account) are compared in Table~\ref{tab:polpred}.

The model AV(1) can have largest RG effects on the polarization. With $g_{u,t} \sim 2.5$ at $\Lambda =3$TeV, the AV(1) model induces $\sim 17\%$ polarization with $m_{t\bar{t}}>800$GeV which may be big enough to be measured. Although no certain higher-order calculation of the top polarization is available, theoretical uncertainties of polarizations will not be much larger than that of the top asymmetry. With a weaker coupling about QCD coupling strength, the polarization falls down to 3\%. The top polarization can also be mixed by QCD RG evolution. The VR(8) model induces a positive right-handed top polarization at tree-level, but the right- and left-handed tops are mixed by QCD RG evolution and the polarization is enhanced slightly; see \Fig{fig:span2}. Numerically, the enhancement is small, however: ${\cal P}(t) = 23\% \to24\%$ with $g_{u,t}=2.5$ and $m_{t\bar{t}}>800$GeV as partly shown in Table~\ref{tab:polpred}.

We also observe from \Fig{fig:coeff-pol} that the AV(1) induces a large $C_{\rm VA}^{(8)}$ while the VA(1) does not (rather, it induces a large $C_{\rm AV}^{(8)}$). It is understood similarly as why the VV(1) induces a large $C_{\rm AA}^{(8)}$ while the AA(1) induces a large $C_{\rm VV}^{(8)}$ as discussed in \Sec{sec:rgafb}. The approximate solution of the RG equation in \Eq{eq:sol4-4by4} can again be used to understand it. Also, from the opposite signs of off-diagonal elements in the ADM \Eq{eq:4by4adm}, one can expect that $C_{\rm AV}^{(1)} = \frac{1}{4} (C_{\rm LL}^{(1)} + c_{\rm LR}^{(1)} - C_{\rm RR}^{(1)} - C_{\rm RL}^{(1)})$ induces a large $C_{\rm VA}^{(8)} = \frac{1}{4}( C_{\rm LL}^{(8)} - C_{\rm LR}^{(8)} - C_{\rm RR}^{(8)} + C_{\rm RL}^{(8)})$.

Another interesting feature in \Fig{fig:coeff-pol} that does not exist in the parity-even sector in \Fig{fig:coeff-afb} is that parity-even operators are RG-induced from parity-odd operators. This is due to penguin effects and will be discussed in later subsections.

\subsection{RG effects as subleading corrections} \label{sec:rgesubleading}

\begin{figure}[t] \centering
\includegraphics[width=0.49\textwidth]{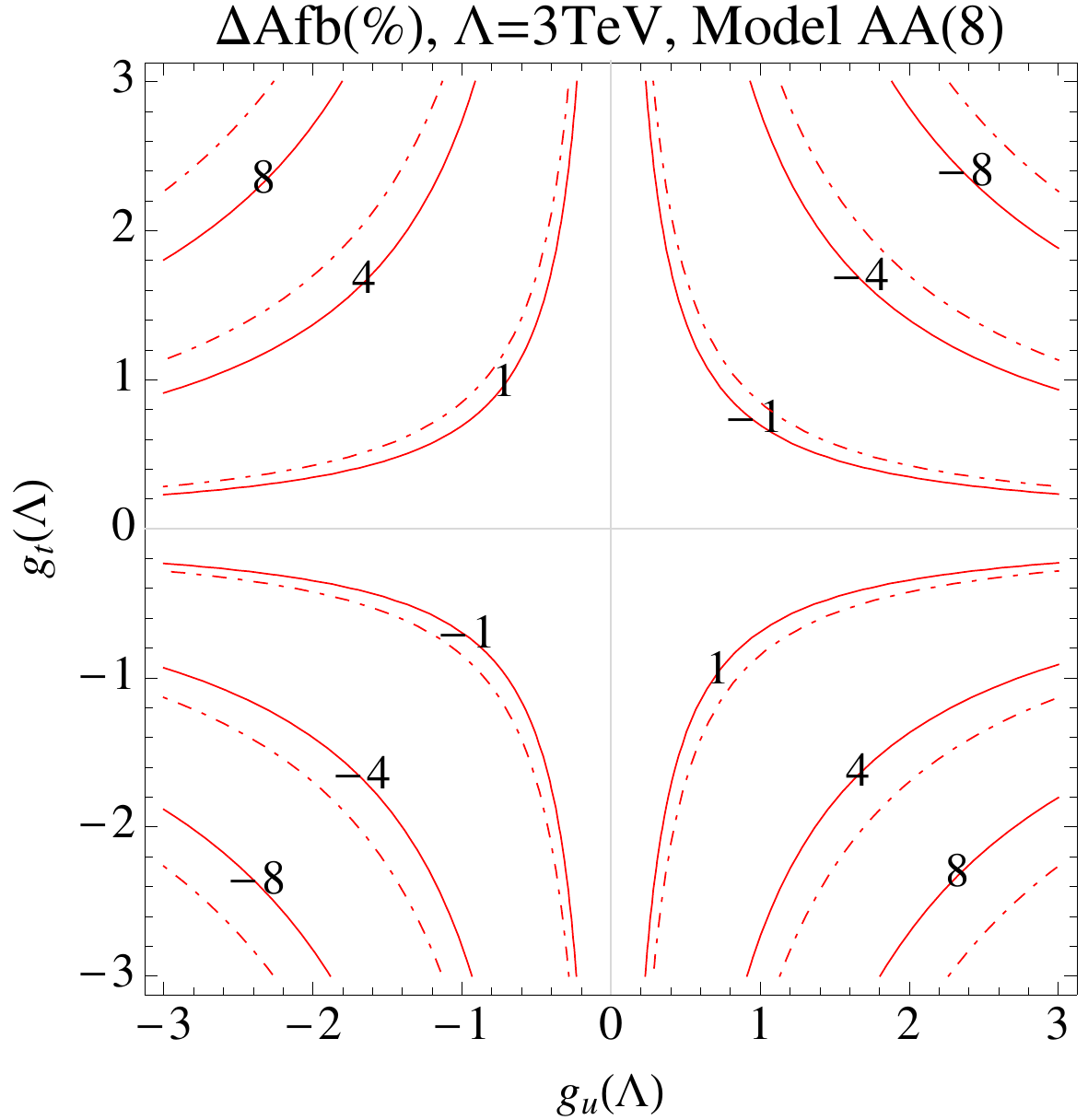}
\includegraphics[width=0.49\textwidth]{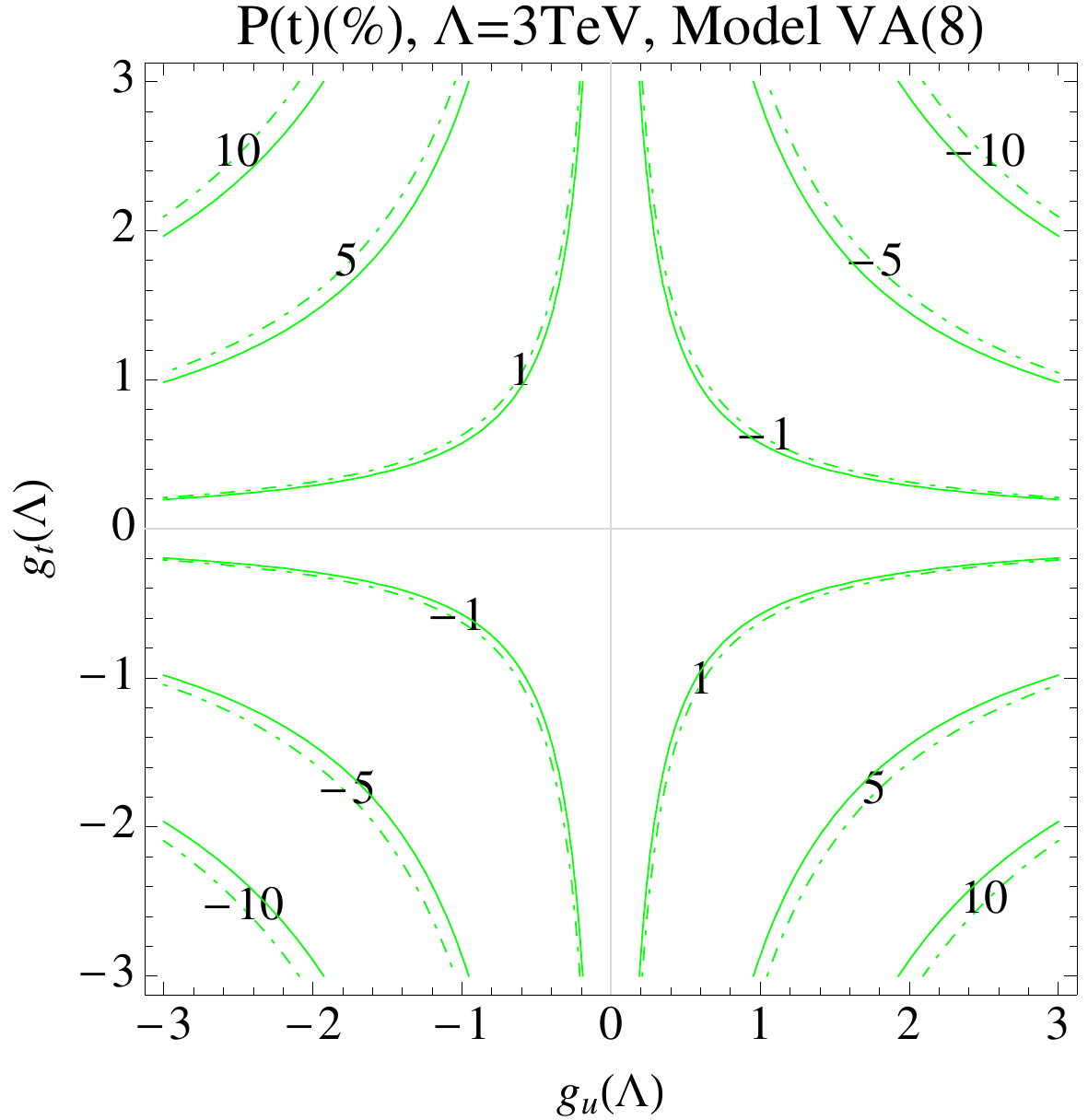}
\caption{Top asymmetries(left) and polarizations(right) are show with(solid) and without(dotdashed) RG effects are taken into account; solid lines are only leading-order tree-level predictions.}
\label{fig:rgecorr}
\end{figure}

RG effects as sub-leading corrections are also important results to discuss. Needless to say that measuring new physics parameters from low-energy data requires such RG effects to be taken into account. From approximate solutions in \Eq{eq:sol1-4by4}-\Eq{eq:sol4-4by4}, the fractional corrections for all operators in the V--A basis induced by themselves are
\beq
\frac{ c_i(m_t) - c_i(\Lambda) }{c_i(\Lambda)} \, \simeq \,   9 \frac{\alpha_s(\Lambda)}{4\pi} \ln \frac{\Lambda}{m_t}.
\eeq
This is numerically about 16\% for $\Lambda = 3$TeV. By taking the AA(8) and the VA(8) models which generate top asymmetries and polarizations at tree-level respectively, we show the relative sizes of sub-leading RG effects in \Fig{fig:rgecorr}. The full results in \Fig{fig:rgecorr} including penguin effects and cross-section shifts tend to slightly enhance the fractional correction to the top asymmetry and to suppress that to the polarization. Penguin effects are not same on the parity-even and -odd sectors, and they also modify the total cross-section which then normalizes the observables differently; see \Sec{sec:lambda} for related discussion. In any case, 5$\sim$ 20\% NLO corrections are expected for all observables which is a typical size of NLO corrections.

\subsection{RG-induced versus the square of tree-level operators}  \label{sec:rgvssq}

We have been interested in the RG-induction of operators and their interference with QCD in the top pair productions. Does the square of effective operators at tree-level contribute with similar sizes? Parametrically, both effects are of similar order as
\beq
\left( C\frac{m_t^2}{\Lambda^2} \right)^2 \, \sim \, \alpha_s \cdot \left( C \frac{m_t^2}{\Lambda^2} \right) \frac{\alpha_s}{4\pi} \ln \frac{m_t}{\Lambda} \, \sim  \,{\cal O}(0.01),
\eeq
where the left-hand side denotes the naive estimations of square of tree-level effective operators whereas the right-hand side denotes the interference between QCD and the RG-induced effective operator. We analytically calculate the contributions of ${\cal O}(1/\Lambda^4)$ in Appendix~\ref{app:helamp}. Notably, if one starts with only one operator in the V--A basis, no observables other than the total cross-section (and spin-correlation) are affected at ${\cal O}(1/\Lambda^4)$. Thus, all the RG-induced effects from the models, VV, AA, AV, VA, discussed in this paper are indeed leading effects. It is another reason to use the V--A basis for the operator mixing.

On the other hand, the total cross-section always receives the contributions from the square of effective operators. Thus, RG-induced cross-sections may not be the leading ones, and we do not further study them.

\subsection{Penguin effects in inferring the scale $\Lambda$} \label{sec:lambda}

\begin{figure}[t] \centering
\includegraphics[width=0.49\textwidth]{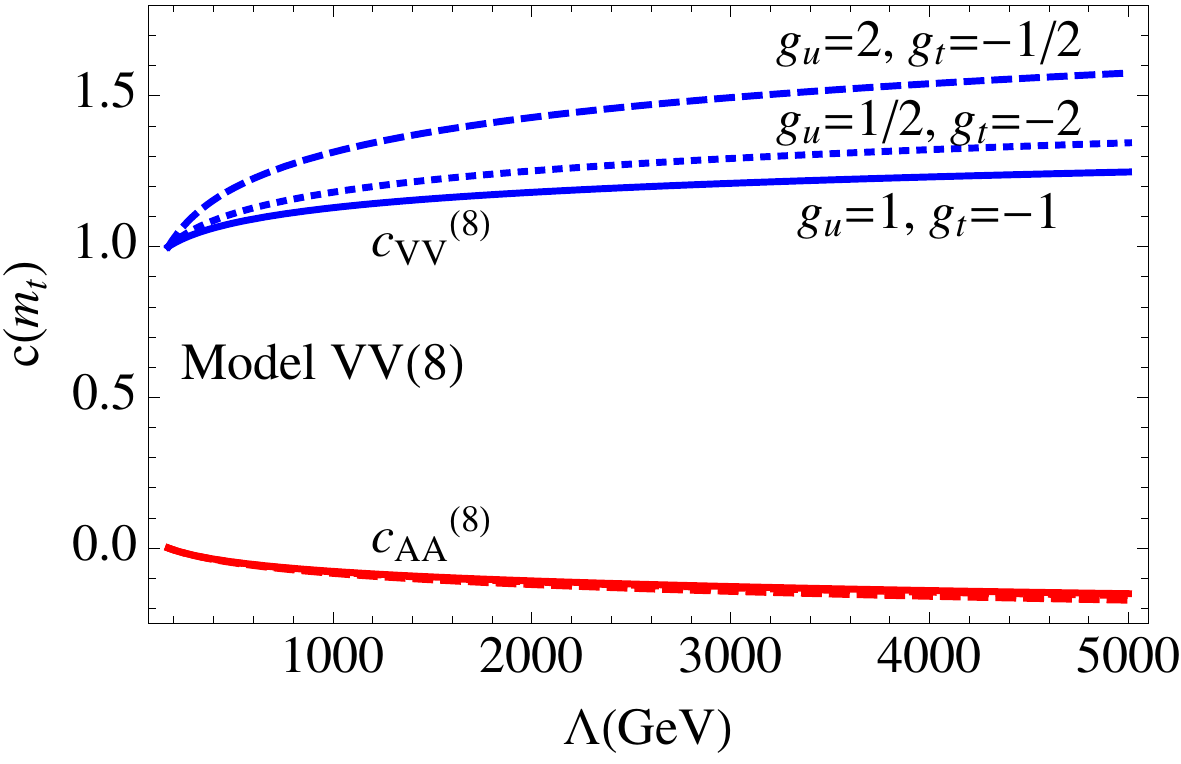}
\includegraphics[width=0.49\textwidth]{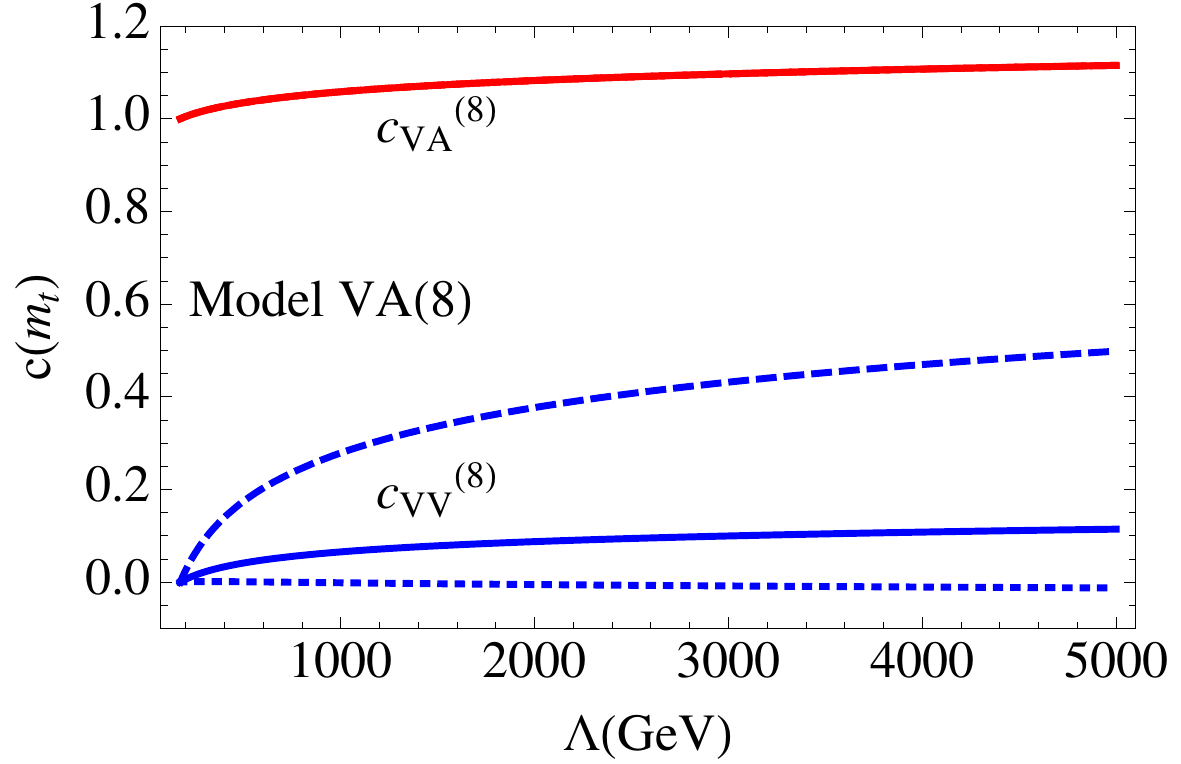}
\caption{Different RG trajectories from three different sets of parameters giving the same product $g_u g_t$. The VV(8)(left) and VA(8)(right) models are used for illustrations.}
\label{fig:lambda}
\end{figure}

At the interference level, the top pair production depends on the $C/\Lambda^2 \sim g_u g_t / \Lambda^2$ combination. At higher-order, however, the production depends on individual model parameters. For example, $g_u^2$ and $g_t^2$ become individually influential through penguin diagrams. In \Fig{fig:lambda}, we show how operators evolve with different choices of $g_u$ and $g_t$ restricted to have the same product $g_u g_t=-1$. Notably, $C_{\rm VV}^{(8)}$ varies significantly depending on individual parameters while other Wilson coefficients are not so sensitive.

Consider penguin effects converting the $(\bar{u} \gamma_\mu T^a u)_{\rm A} (\bar{u} \gamma^\mu T^a u)_{\rm B}$ operators to the $(\bar{u} \gamma_\mu T^a u)_{\rm C} (\bar{t} \gamma^\mu T^a t)_{\rm D}$ operators. The penguin conversion of the latter to the former operators has indirect and smaller effects on top pair productions, and therefore we ignore them in this approximate discussion. The ADM $\gamma$ for the relevant penguin conversion within the fully expanded basis described in Appendix~\ref{sec:app-op144} is given by
\beq
uuut \, \ni \,  \frac{\alpha_s}{4\pi} \, \left( \bmat \frac{8}{9} & \frac{8}{9} & 0 & 0 \\ \frac{2}{3} & \frac{2}{3} & \frac{2}{3} & \frac{2}{3} \\
 0 & 0 & \frac{8}{9} & \frac{8}{9} \emat \right),
\eeq
where the basis consists only of color-octet operators with $\rm (AB)=(LL,LR=RL,RR)$ and $\rm (CD)=(LL,LR,RL,RR)$. See Appendix~\ref{sec:app-op144} for the notation in the left-hand side. It is convenient to transform the ADM to the V--A basis if we define models to have either vectorial or axial-vectorial couplings. Now $\rm (AB) =(VV,AA)$\footnote{Note that AV- or VA-type currents $(\bar{u} \gamma^\mu T^a u)_{\rm A} (\bar{u} \gamma_\mu T^a u)_{\rm B}$ do not exist at $\Lambda$ if we define models to have either vectorial or axial-vectorial couplings to up quarks, for example.} and $\rm (CD)=(VV,AA,AV,VA)$, and the ADM becomes
\beq
uuut \, \ni \, \frac{\alpha_s}{4\pi} \, \left( \bmat \frac{80}{9} & 0 & 0 & 0 \\ -\frac{16}{9} &0 & 0 & 0 \emat \right).
\eeq
Interestingly, in this approximation, penguin diagrams induce only VV-type four-quark operators, ${\cal O}_{\rm VV}$, in the $(\bar{u} \gamma^\mu u) (\bar{t} \gamma_\mu t)$ sector. This explains why $C_{\rm VV}$ is most sensitive to the underlying model parameters through penguin effects. In other words, when the scale $\Lambda$ and model parameters are inferred, the total cross-section data should be especially used with care by taking into account the effects from full RG evolutions as well as the square of effective operators.

It is also worthwhile to summarize penguin effects discussed in several places in this paper. Penguin effects cannot be symmetric to the parity-even and -odd sectors because QCD is parity-conserving. They especially modify the total cross-section which may be measurable and then normalizes the observables differently. They make the low-energy prediction sensitive to the underlying model parameters even though those parameters would predict the same physics at tree-level.

\section{Validity check: RG as a proxy of one-loop effects} \label{sec:rgvalid}

In \Sec{sec:rgafb}, we have noticed that the sign of the RG-induced top asymmetry from effective theory calculation is opposite to that of the full theory one-loop results in Ref.~\cite{Jung:2014gfa}. In this section, we discuss two possible origins of the discrepancy: missing one-loop diagrams that have no overlap with the RG calculation and terms other than large logarithms in effective theory one-loop diagrams. We also discuss the validity of the four-quark effective theory.

First of all, our RG calculation does not include the interference between one-loop QCD box diagrams of the SM and tree-level effective operators that contribute to the forward-backward asymmetric cross-section. These interference diagrams are called \emph{missing} one-loop diagrams in this section; by missing diagrams, we mean that their contributions have no overlap with our RG calculation. On the other hand, the one-loop amplitudes of the effective theory contain large logarithms of ${\cal O}(\alpha_s \log \Lambda/m_t)$ as well as other non-large-logarithmic terms. The large logarithmic terms are overlapped with our RG contribution and are resummed to all order in $\alpha_s$. The non-large-logarithmic terms as well as the missing diagrams do not give rise to the operator running, and we have taken them to be sub-leading to the large logarithmic terms. Numerically, however, it turns out that the missing diagrams can be as important as the leading logarithms.

\begin{figure}[t] \centering
\includegraphics[width=0.75\textwidth]{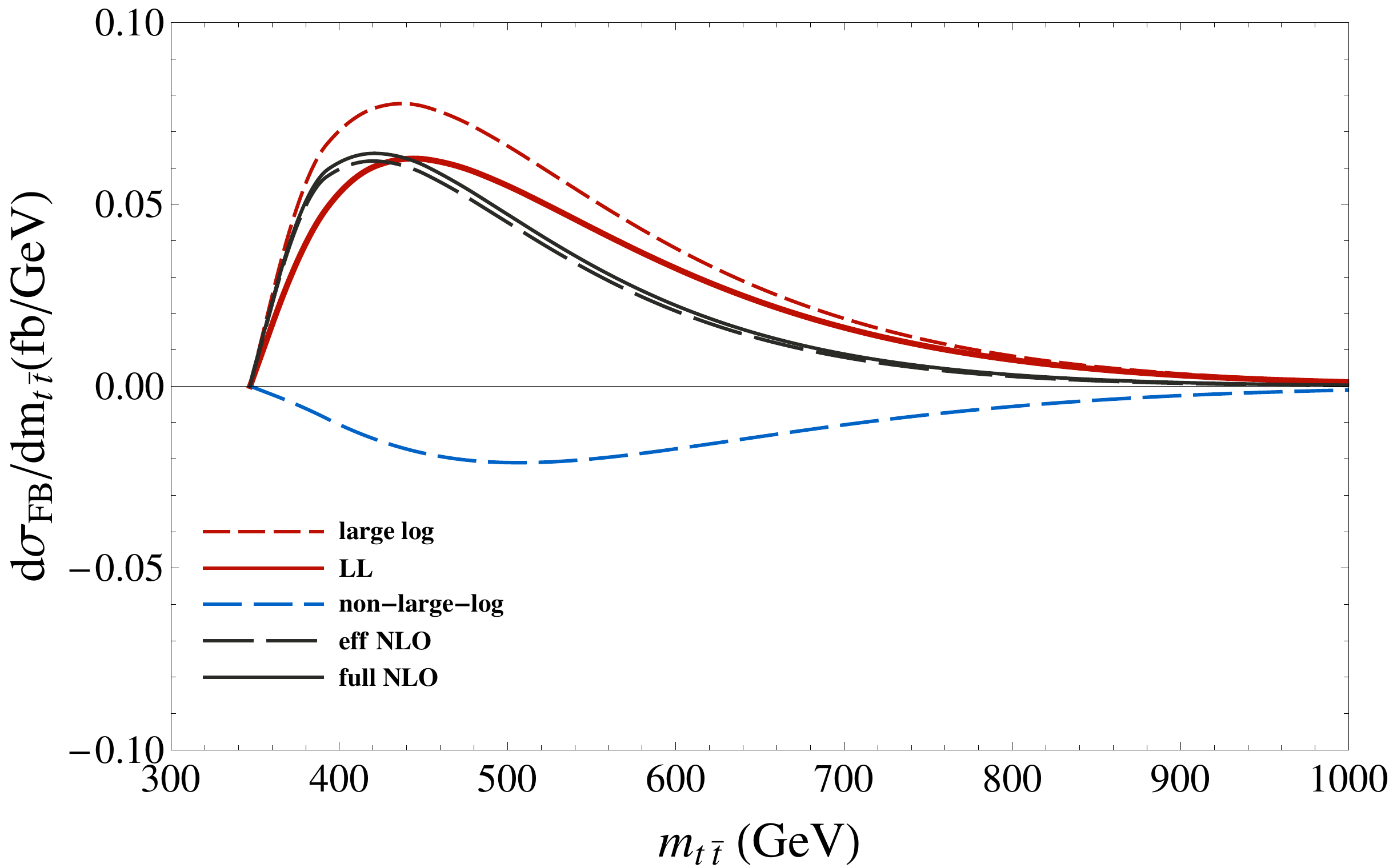}
\caption{The differential $\sigma_{\rm FB}$ with respect to $t \bar t$ invariant mass. The VV(1) effective theory box diagram (b) and (c) in \Fig{fig:ADMdiagrams} (black-dashed) are decomposed  into large logarithmic terms (red-dashed) and non-large-logarithmic terms (blue-dashed). The red-solid is the leading-log resummed result. The black-solid is the one-loop result of a VV(1) toy model with $M_X = 3\TeV$ and $g_u=g_t=1$. See text for more discussion.}
\label{fig:LL}
\end{figure}

In \Fig{fig:LL}, we compare various contributions of the box diagram (b) and (c) in \Fig{fig:ADMdiagrams} for the VV(1) effective theory and the VV(1) toy model of heavy $X_\mu$ gauge bosons. The corresponding box diagram for the toy model consists of one gluon and one $X^\mu$. Since the results of the effective theory(black-dashed) and the toy model(black-solid) agree well even to the high-mass region, the validity of the effective theory is not threatened. Rather, we observe that the non-logarithmic contributions(blue-dashed) are sizably negative compared to the logarithmic ones(red-dashed). In any case, the logarithmic contributions are larger\footnote{The leading expansion in $\alpha_s$ of the leading-log resummed result(red-solid in \Fig{fig:LL}) gives the large log terms in red-dashed. It turns out that the resummation effect is marginal -15\% to the large log terms.}. Thus, we conclude that the sign flip is largely due to the missing diagrams rather than the non-logarithmic contributions.

Do missing diagrams exist for other top pair observables? We can, at least, argue that similar missing diagrams exist for other observables. The main point in our argument is that there are two operator sets that can interfere and induce the same observable. For example, ${\cal O}_{\rm VV}^{(8)}$ and ${\cal O}_{\rm AA}^{(8)}$ as well as ${\cal O}_{\rm VV}^{(1)}$ and ${\cal O}_{\rm AA}^{(1)}$ can interfere and induce the top asymmetry. The latter is actually related to why the missing interference diagrams contribute. The one-loop QCD box can be effectively thought of as the QCD running of ${\cal O}_{\rm VV}^{(8)}$ which would induce ${\cal O}_{\rm AA}^{(1)}$ (among many) and interfere with tree-level ${\cal O}_{\rm VV}^{(1)}$ of the VV(1) model -- this is effectively the \emph{missing} contribution. Exactly in the same way, all top pair observables have two sets of operators that can interfere and induce them. For the top polarization from the AV(1) model, for another example, the ${\cal O}_{\rm AA}^{(1)}$ from the QCD box diagram interferes with the tree-level ${\cal O}_{\rm AV}^{(1)}$ to induce missing contributions to the polarization. The numerical sizes of such missing contributions are, however, \emph{a priori} not known until the full one-loop results of a model is available.

We find it useful to compare the RG effects on the top pair and the $h \to \gamma \gamma$ studied in Ref.~\cite{Grojean:2013kd}. Based on the similar argument made above, we conclude that missing diagrams do not exist for the $h \to \gamma \gamma$. It is because only one (CP-even) operator ${\cal O}_{FF} = H^\dagger H F_{\mu \nu} F^{\mu \nu}$ can contribute to the process. Suppose that, in the effective theory, the ${\cal O}_{FF}$ operator is RG-induced but absent at tree-level. Then, the one-loop SM diagrams inducing the ${\cal O}_{FF}$ interfere only with the RG-induced ${\cal O}_{FF}$ but not with the tree-level operators of the effective theory. This is the only contribution in this case (thus, no missing diagrams) whose logarithmic contributions are indeed resummed by the RG calculation. This contrast makes it clear why the missing diagrams exist in top pair observables.

In all, for top pair productions, one-loop corrections to the effective theory may be numerically as important as one-loop RG effects, and full one-loop calculation is motivated. One-loop effects can be consistently implemented by performing consistent one-loop calculation in the effective theory framework.
A few useful one-loop results of new physics models are already available, e.g. in Refs.~\cite{Xiao:2010hm,Bauer:2010iq,Jung:2014gfa}.
We emphasize that our qualitative discussions will still be valid regardless of whether missing diagrams exist or not. Although a full one-loop calculation would be numerically more accurate as it should be, it will not tell us much more about the higher-order physics.

\section{Summary and conclusions} \label{sec:conclusion}

Useful to the future top pair precision test is the categorization of possible one-loop phenomena of new physics. All top pair observables in the process $q\bar{q} \to t\bar{t}$ can be induced at one-loop order of new physics even when they are not induced at tree-level. We summarize which underlying models can induce which observables at one-loop and the related operator mixing patterns of four-quark effective operators:
\bei
\item VV(1), AA(1), VV(8) $\to \, A_{\rm FB}$ based on the mixing ${\cal O}_{\rm VV}^{(1)}, {\cal O}_{\rm AA}^{(1)}, {\cal O}_{\rm VV}^{(8)} \to {\cal O}_{\rm AA}^{(8)}$,
\item AV(1), VA(1), AV(8) $\to \, {\cal P}(t)$ based on the mixing ${\cal O}_{\rm AV}^{(1)}, {\cal O}_{VA}^{(1)}, {\cal O}_{\rm AV}^{(8)} \to {\cal O}_{\rm VA}^{(8)}$,
\item VA(1), AV(1), VA(8) $\to \, {\cal D}_{\rm FB}$ based on the mixing ${\cal O}_{\rm VA}^{(1)}, {\cal O}_{\rm AV}^{(1)}, {\cal O}_{\rm VA}^{(8)} \to {\cal O}_{\rm AV}^{(8)}$.
\eei
Models are introduced in Table~\ref{tab:initial}, and parity quantum numbers of operators and corresponding observables are summarized in \Eq{eq:corres}. Interestingly, color-singlet models (the first model in each item listed above) generally induce the largest RG effects even though they do not contribute at tree-level. The maximal effects on the top asymmetry (and perhaps also on the polarzation) from color-singlet models are larger than theoretical uncertainties on the observables, thus will be measurable in the future; see Table~\ref{tab:datapred} and \ref{tab:polpred}. One should keep in mind that our estimation is based solely on the RG evolution of effective operators which resums large leading logarithms, but there are other contributions at full one-loop order that can be numerically important as discussed in \Sec{sec:rgvalid}.
Motivated from these, we have carried out the one-loop calculation and related collider studies of certain color-singlet models in Ref.~\cite{Jung:2014gfa} which is useful on its own and for future one-loop effective theory calculation.
In the upcoming precision era, we hope that our discussions based on the global QCD RG evolution can insightfully motivate more dedicated model buildings, higher-order calculations and collider physics studies.

\section*{Acknowledgements}
We thank Chul Kim for useful discussions. SJ and YWY thank KIAS Center for Advanced Computation for providing computing resources.

SJ is supported in part by National Research Foundation (NRF) of Korea under grant 2013R1A1A2058449. PK is supported in part by Basic Science Research Program through the NRF under grant 2012R1A2A1A01006053 and by SRC program of NRF funded by MEST (20120001176) through Korea Neutrino Research Center at Seoul National University. CY is supported by Basic Science Research Program through the NRF funded by the Ministry of Education Science and Technology 2011-0022996.

\begin{landscape}

\begin{table}[t] \centering \footnotesize
\begin{tabular}{cc||cccc||cc}
\hline \hline
 & & $\Delta \sigma^{Tev}$ & $\Delta A_{FB}$ & $\frac{\Delta \sigma }{ \sigma_{SM}}|_{800\GeV}$ & $\Delta A_{FB}|_{650\GeV}$ & $\Delta \sigma^{LHC8}$ & $\frac{\Delta \sigma }{ \sigma_{SM}}|_{1\TeV}$ \\
\hline \hline
\multirow{2}{*}{}
& CDF data & $(7.62^{ \pm 0.42})$pb\cite{Aaltonen:2013wca} & $(16.4^{ \pm 4.7})\%$\cite{Aaltonen:2012it}   & $\frac{(40.8 \pm 21)\fb}{38\fb}$ \cite{Aaltonen:2009iz} & $(49.3^{ \pm 19.3})\%$ \cite{Aaltonen:2012it} & $(238^{\pm11.3})$pb\cite{TheATLAScollaboration:2013dja} & $\lesssim 0.2$\cite{Chatrchyan:2013lca} \\
& D0 data & $(7.62^{ \pm 0.42})$pb\cite{Aaltonen:2013wca} & $(10.6^{ \pm 3.0})\%$\cite{Abazov:2014cca}   & $\frac{(150 \pm 100)\fb}{150\fb}^*$\cite{Abazov:2014vga} & $(-12.3^{ \pm 29.6})\%^{**}$\cite{Abazov:2014cca} & -- & -- \\
& SM prediction & $(7.16^{+0.20}_{-0.23})$pb\cite{Czakon:2013goa} & $(8.8^{\pm 0.6})\%$\cite{Bernreuther:2012sx} & 0 & $(14.3^{\pm4.3})\%$ \cite{Aaltonen:2012it} & $(246^{+8.7}_{-10.6})$pb\cite{Czakon:2013goa} & 0 \\
\hline \hline
\multirow{2}{*}{VV(8)}
& $g_u = g_t = 1.7$ & $-0.44$\pb & 0.41\% & $-0.32$ & 2.0\% & $-3.5$\pb & $-0.14$ \\
& $g_u = -g_t = 1.7$ & 0.60\pb & $-0.39$\% & 0.43 & $-1.4$\% & 4.8\pb & 0.19 \\
\hline
\multirow{1}{*}{AA(8)}
& $g_u=-g_t=2.5$ & $-0.15$\pb & 9.3\% & $-0.11$ & 41\% & $-1.2$\pb & $-0.05$ \\
\hline
\multirow{1}{*}{VV(1)}
& $g_u =g_t = 2.5$ & 0.04\pb & 2.0\% & 0.03 & 8.5\% & 0.35\pb & 0.01 \\
\hline
\multirow{1}{*}{AA(1)}
& $g_u =g_t = 2.5$ & 0.31\pb & $-0.1$\% & 0.23 & $-0.5$\% & 2.5\pb & 0.10 \\
\hline \hline
\end{tabular}
\caption{Comparison of reference model predictions with data. Models are described in text and Table.\ref{tab:initial}. $\Lambda=3\TeV$ is taken for model predictions. (*): The D0 data is measured with $m_{t\bar{t}}>750$GeV. (**): Although it is negative, the D0 data in the $m_{t\bar{t}}=550-650$GeV region is large positive 37.6\% compared to the SM prediction 10.9\%.}
\label{tab:datapred}\end{table}

\begin{table}[t] \centering \footnotesize
\begin{tabular}{c ||c|c || c|c}
\hline \hline
 & \multicolumn{2}{ | c ||}{ ${\cal P}(t)$ at Tevatron}  & \multicolumn{2}{ | c}{ ${\cal P}(t)$ at LHC8}  \\
\cline{2-5}
Model & Total & $m_{t\bar{t}}>800$GeV & Total & $m_{t\bar{t}}>1000$GeV \\
\hline \hline
\multicolumn{1}{ c ||}{ ATLAS data } & -- & -- & $(-3.5^{\pm 1.4 \pm 3.7})\%^*$~\cite{Aad:2013ksa} & -- \\
\multicolumn{1}{ c ||}{ CMS data } & -- & -- & $(0.5^{\pm 1.3 \pm 1.6})\%^*$~\cite{Chatrchyan:2013wua} & -- \\
\multicolumn{1}{ c ||}{ SM prediction } & -- & -- & $\sim(0.0^{\pm 0.1 \pm 1.5})\%^{**}$~\cite{Chatrchyan:2013wua} & -- \\
\hline \hline
\multicolumn{1}{ c ||}{ AV(8), $g_u=g_t=2.5$ } & 1\% & 7\% & 0.4\% & 3\% \\
\hline
\multicolumn{1}{ c ||}{ VR(8), $g_u=-g_t=2.5$ } & 4.5 $\to$ 5.0\% & 23 $\to$ 24\%  & $-1.6 \to -1.8$\% & $-13 \to -14$\% \\
\hline \hline
\multicolumn{1}{ c ||}{ AV(1), $g_u=g_t=2.5$ } & 3.0\% & 17\%  & 0.9\% & 7\% \\
\hline
\multicolumn{1}{ c ||}{ VA(1), $g_u=g_t=2.5$ } &  $-0.1$\% & $-0.9$\%  & $-0.05$\% & $-0.4$\% \\
\hline \hline
\end{tabular}
\caption{Top polarization. Total and the high-$m_{t\bar{t}}$ polarizations are shown at Tevatron and LHC8. For the VR(8), we show two polarizations for each column before $\to$ after the RG effects are taken into account. $\Lambda=3\TeV$ is used. (*): The only currently available data are from 7TeV 5/fb, and uncertainties are statistical and systematic. (**): The systematic uncertainty of the SM prediction is guessed unofficially by authors by referring to those of other similar observables listed in Ref.\cite{Chatrchyan:2013wua}.}
\label{tab:polpred} \end{table}

\end{landscape}

\appendix
\section{ADM for the tree-penguin classified basis} \label{sec:app-adm68}

Leading order one loop diagrams for calculating ADM of effective four quark operators are shown in Figure \ref{fig:ADMdiagrams}. Diagrams in first line are tree diagrams while diagrams in second line are penguin diagrams.
 Here, we describe ADM of following ordered operators with arbitrary generation indices $i, j ~(i\neq j )$ and flavor index $q, q^\prime~(q\neq q^\prime)$. The RG equation of any operator from the full set can be inferred from this ADM.
\begin{eqnarray}
{\rm Tree:} &&
\op_{Q_iQ_j}^{(1)},~\op_{Q_iQ_j}^{(8)},~\op_{Q_iQ_i}^{(1)},~\op_{Q_iQ_i}^{(8)}
,~\op_{Q_jQ_j}^{(1)},~\op_{Q_jQ_j}^{(8)},\nn\\
&&\op_{qq^\prime}^{(1)},~\op_{qq^\prime}^{(8)},~\op_{qq}^{(8)},~\op_{q^\prime q^\prime}^{(8)},\nn\\
&&\op_{Q_i q}^{(1)},\op_{Q_i q}^{(8)},\op_{Q_i q^\prime}^{(1)},\op_{Q_i q^\prime}^{(8)},
  \op_{Q_j q}^{(1)},\op_{Q_j q}^{(8)},\op_{Q_j q^\prime}^{(1)},\op_{Q_j q^\prime}^{(8)},\\
{\rm Penguin:}&&
\op_{Q_i \Sigma_Q}^{(1)},~\op_{Q_i \Sigma_Q}^{(8)},~\op_{Q_j \Sigma_Q}^{(1)},~\op_{Q_j \Sigma_Q}^{(8)},\nn\\
&& \op_{q \Sigma_q}^{(1)}, \op_{q \Sigma_q}^{(8)},
\op_{q^\prime \Sigma_q}^{(1)}, \op_{q^\prime \Sigma_q}^{(8)}, \nn\\
&& \op_{Q_i \Sigma_q}^{(1)}, \op_{Q_i \Sigma_q}^{(8)},
\op_{Q_j \Sigma_q}^{(1)}, \op_{Q_j \Sigma_q}^{(8)}, \nn \\
&& \op_{q \Sigma_Q}^{(1)}, \op_{q \Sigma_Q}^{(8)},
\op_{q^\prime \Sigma_Q}^{(1)}, \op_{q^\prime \Sigma_Q}^{(8)}, \\
{\rm Double~Penguin:}
&& \op_{\Sigma_Q \Sigma_Q}^{(1)}, \op_{\Sigma_Q \Sigma_Q}^{(8)},
\op_{\Sigma_q \Sigma_q}^{(1)}, \op_{\Sigma_q \Sigma_q}^{(8)},
\op_{\Sigma_Q \Sigma_q}^{(1)}, \op_{\Sigma_Q \Sigma_q}^{(8)}\,.
\end{eqnarray}
We find that $40\times40$ anomalous dimension matrix ${\mathbf \gamma}$ has following block-triangular form
\begin{equation}
\label{eq:admTriangle}
\gamma = \left(\begin{array}{ccc}
 \gamma_{\rm tt} & \gamma_{\rm tp}  & 0   \\
0   & \gamma_{\rm pp} & \gamma_{\rm pd}  \\
0           & 0  & \gamma_{\rm dd}
\end{array} \right).
\end{equation}
Note that tree operators mix with only penguin operators and  themselves while penguin operators mix with only double penguin operators and themselves. Double-penguin operators do not mix with neither tree operators nor penguin operators. It should be emphasized that off-diagonal components  $\gamma_{\rm tp}$, $\gamma_{\rm pd}$ are quite smaller than diagonal components $\gamma_{\rm tt}$, $\gamma_{\rm pp}$, $\gamma_{\rm dd}$ since they are induced from penguin diagrams and therefore they do not have large Dirac contraction factor. This feature indicates that mixing effect between tree and penguin or penguin and double-penguin operators during RG evolution is much smaller than their own RG evolution.

The anomalous dimension matrix $\gamma$ of effective operators is expanded in $\alpha_s$ as
\begin{equation}
\gamma = \sum_{k=0}\gamma^{(k)} \Big(\frac{\alpha_s}{4\pi}\Big)^{k+1}\,.
\end{equation}
The leading order sub-matrices of Eq. (\ref{eq:admTriangle}) are as follows.
\begin{equation}
\gamma_{\rm tt }^{(0)}=
\left(
\begin{array}{cccccccccccccccccc}
 0 & 12 & 0 & 0 & 0 & 0 & 0 & 0 & 0 & 0 & 0 & 0 & 0 & 0 & 0 & 0  & 0 & 0  \\
 \frac{8}{3} & -4 & 0 & 0 & 0 & 0 & 0 & 0 & 0 & 0 & 0 & 0 & 0 & 0 & 0 & 0  & 0 & 0 \\
 0 & 0 & 0 & 12 & 0 & 0 & 0 & 0 & 0 & 0 & 0 & 0 & 0 & 0 & 0 & 0  & 0 & 0  \\
 0 & 0 & \frac{8}{3} & -4 & 0 & 0 & 0 & 0 & 0 & 0 & 0 & 0 & 0 & 0 & 0 & 0  & 0 & 0  \\
 0 & 0 & 0 & 0 & 0 & 12 & 0 & 0 & 0 & 0 & 0 & 0 & 0 & 0 & 0 & 0  & 0 & 0  \\
 0 & 0 & 0 & 0 & \frac{8}{3} & -4 & 0 & 0 & 0 & 0 & 0 & 0 & 0 & 0 & 0 & 0 & 0 & 0 \\
 0 & 0 & 0 & 0 & 0 & 0 & 0 & 12 & 0 & 0 & 0 & 0 & 0 & 0 & 0 & 0   & 0 & 0  \\
 0 & 0 & 0 & 0 & 0 & 0 & \frac{8}{3} & -4 & 0 & 0 & 0 & 0 & 0 & 0 & 0 & 0 & 0 & 0 \\
 0 & 0 & 0 & 0 & 0 & 0 & 0 & 0 & 4 & 0 & 0 & 0 & 0 & 0 & 0 & 0  & 0 & 0   \\
 0 & 0 & 0 & 0 & 0 & 0 & 0 & 0 & 0 & 4 & 0 & 0 & 0 & 0 & 0 & 0 & 0 & 0 \\
 0 & 0 & 0 & 0 & 0 & 0 & 0 & 0 & 0 & 0 & 0 & -12 & 0 & 0 & 0 & 0  & 0 & 0 \\
 0 & 0 & 0 & 0 & 0 & 0 & 0 & 0 & 0 & 0 & -\frac{8}{3} & -14 & 0 & 0 & 0 & 0 & 0 & 0  \\
 0 & 0 & 0 & 0 & 0 & 0 & 0 & 0 & 0 & 0 & 0 & 0 & 0 & -12 & 0 & 0  & 0 & 0 \\
 0 & 0 & 0 & 0 & 0 & 0 & 0 & 0 & 0 & 0 & 0 & 0 & -\frac{8}{3} & -14 & 0 & 0 & 0 & 0 \\
 0 & 0 & 0 & 0 & 0 & 0 & 0 & 0 & 0 & 0 & 0 & 0 & 0 & 0 & 0 & -12 & 0 & 0 \\
 0 & 0 & 0 & 0 & 0 & 0 & 0 & 0 & 0 & 0 & 0 & 0 & 0 & 0 & -\frac{8}{3} & -14 & 0 & 0  \\
  0 & 0 & 0 & 0 & 0 & 0 & 0 & 0 & 0 & 0 & 0 & 0 & 0 & 0 & 0 & 0 & 0 & -12 \\
 0 & 0 & 0 & 0 & 0 & 0 & 0 & 0 & 0 & 0 & 0 & 0 & 0 & 0 & 0 & 0 &-\frac{8}{3} & -14
\end{array}
\right)\,.
\end{equation}
\begin{equation}
\gamma_{\rm tp}^{(0)}=
\left(
\begin{array}{cccccccccccccccc}
 0 & 0 & 0 & 0 & 0 & 0 & 0 & 0 & 0 & 0 & 0 & 0 & 0 & 0 & 0 & 0 \\
 0 & \frac{4}{3} & 0 & \frac{4}{3} & 0 & 0 & 0 & 0 & 0 & \frac{4}{3} & 0 & \frac{4}{3} & 0 & 0 & 0 & 0\\
 0 & \frac{8}{3} & 0 & 0 & 0 & 0 & 0 & 0 & 0 & \frac{8}{3} & 0 & 0 & 0 & 0 & 0 & 0\\
 0 & \frac{14}{9} & 0 & 0 & 0 & 0 & 0 & 0 & 0 & \frac{14}{9} & 0 & 0 & 0 & 0 & 0 & 0\\
 0 & 0 & 0 & \frac{8}{3} & 0 & 0 & 0 & 0 & 0 & 0 & 0 & \frac{8}{3} & 0 & 0 & 0 & 0\\
 0 & 0 & 0 & \frac{14}{9} & 0 & 0 & 0 & 0 & 0 & 0 & 0 & \frac{14}{9} & 0 & 0 & 0 & 0\\
 0 & 0 & 0 & 0 & 0 & 0 & 0 & 0 & 0 & 0 & 0 & 0 & 0 & 0 & 0 & 0 \\
 0 & 0 & 0 & 0 & 0 & \frac{2}{3} & 0 & \frac{2}{3} & 0 & 0 & 0 & 0 & 0 & \frac{2}{3} & 0 & \frac{2}{3} \\
 0 & 0 & 0 & 0 & 0 & \frac{8}{9} & 0 & 0 & 0 & 0 & 0 & 0 & 0 & \frac{8}{9} & 0 & 0 \\
 0 & 0 & 0 & 0 & 0 & 0 & 0 & \frac{8}{9} & 0 & 0 & 0 & 0 & 0 & 0 & 0 & \frac{8}{9} \\
 0 & 0 & 0 & 0 & 0 & 0 & 0 & 0 & 0 & 0 & 0 & 0 & 0 & 0 & 0 & 0 \\
 0 & \frac{2}{3} & 0 & 0 & 0 & \frac{4}{3} & 0 & 0 & 0 & \frac{2}{3} & 0 & 0 & 0 & \frac{4}{3} & 0 & 0 \\
 0 & 0 & 0 & 0 & 0 & 0 & 0 & 0 & 0 & 0 & 0 & 0 & 0 & 0 & 0 & 0 \\
 0 & \frac{2}{3} & 0 & 0 & 0 & 0 & 0 & \frac{4}{3} & 0 & \frac{2}{3} & 0 & 0 & 0 & 0 & 0 & \frac{4}{3} \\
 0 & 0 & 0 & 0 & 0 & 0 & 0 & 0 & 0 & 0 & 0 & 0 & 0 & 0 & 0 & 0 \\
 0 & 0 & 0 & \frac{2}{3} & 0 & \frac{4}{3} & 0 & 0 & 0 & 0 & 0 & \frac{2}{3} & 0 & \frac{4}{3} & 0 & 0 \\
 0 & 0 & 0 & 0 & 0 & 0 & 0 & 0 & 0 & 0 & 0 & 0 & 0 & 0 & 0 & 0 \\
 0 & 0 & 0 & \frac{2}{3} & 0 & 0 & 0 & \frac{4}{3} & 0 & 0 & 0 & \frac{2}{3} & 0 & 0 & 0 & \frac{4}{3}
\end{array}
\right)\,.
\end{equation}

\begin{equation}
\gamma_{\rm pp}^{(0)}=
\left(
\begin{array}{cccccccccccccccc}
 0 & \frac{44}{3} & 0 & 0 & 0 & 0 & 0 & 0 & 0  & \frac{8}{3} & 0 & 0 & 0 & 0 & 0 & 0 \\
 \frac{8}{3} & \frac{6n_f-40}{9} & 0 & 0 & 0 & 0 & 0 & 0 & 0 & \frac{6n_f-4}{9} & 0 & 0 & 0 & 0 & 0 & 0 \\
 0 & 0 & 0  & \frac{44}{3} & 0 & 0 & 0 & 0 & 0 & 0 & 0 & \frac{8}{3} & 0 & 0 & 0 & 0 \\
 0 & 0 & \frac{8}{3} & \frac{6n_f-40}{9} & 0 & 0 & 0 & 0 & 0 & 0 & 0 & \frac{6n_f-4}{9} & 0 & 0 & 0 & 0 \\
 0 & 0 & 0 & 0 & 0 & \frac{44}{3} & 0 & 0 & 0 & 0 & 0 & 0 & 0 & \frac{8}{3} & 0 & 0 \\
 0 & 0 & 0 & 0 & \frac{8}{3} & \frac{6n_f-40}{9} & 0 & 0 & 0 & 0 & 0 & 0 & 0 & \frac{6n_f-4}{9} & 0 & 0 \\
 0 & 0 & 0 & 0 & 0 & 0 & 0 & \frac{44}{3} & 0 & 0 & 0 & 0 & 0 & 0 & 0 & \frac{8}{3} \\
 0 & 0 & 0 & 0 & 0 & 0 & \frac{8}{3} & \frac{6n_f-40}{9} & 0 & 0 & 0 & 0 & 0 & 0 & 0 & \frac{6n_f-4}{9} \\
 0 & 0 & 0 & 0 & 0 & 0 & 0 & 0 & 0  & -12 & 0 & 0 & 0 & 0 & 0 & 0 \\
 0 & \frac{2n_f}{3} & 0 & 0 & 0 & 0 & 0 & 0 & -\frac{8}{3} & \frac{2n_f-42}{3} & 0 & 0 & 0 & 0 & 0 & 0 \\
 0 & 0 & 0 & 0 & 0 & 0 & 0 & 0 & 0 & 0 & 0  & -12 & 0 & 0 & 0 & 0 \\
 0 & 0 & 0 & \frac{2n_f}{3} & 0 & 0 & 0 & 0 & 0 & 0 & -\frac{8}{3} & \frac{2n_f-42}{3} & 0 & 0 & 0 & 0 \\
 0 & 0 & 0 & 0 & 0 & 0 & 0 & 0 & 0 & 0 & 0 & 0 & 0  & -12 & 0 & 0 \\
 0 & 0 & 0 & 0 & 0 & \frac{2n_f}{3} & 0 & 0 & 0 & 0 & 0 & 0 & -\frac{8}{3} & \frac{2n_f-42}{3} & 0 & 0 \\
 0 & 0 & 0 & 0 & 0 & 0 & 0 & 0 & 0 & 0 & 0 & 0 & 0 & 0 & 0  & -12 \\
 0 & 0 & 0 & 0 & 0 & 0 & 0 & \frac{2n_f}{3} & 0 & 0 & 0 & 0 & 0 & 0 & -\frac{8}{3} & \frac{2n_f-42}{3}
 \end{array}
\right)\,.
\end{equation}
\begin{equation}
\gamma_{\rm pd}^{(0)} = \left(\begin{array}{cccccccccccccccc}
0  &  0  &   0  &  0  &         0  &  0  &  0  &  0  &      0  &  0  &   0  &  0  &         0  &  0  &  0  &  0 \\
0  &  \frac{4}{3}  &   0  &  \frac{4}{3}&         0  &  0  &  0  &  0  &      0  &  0  &   0  &  0  &  0  &  \frac{2}{3}  &  0  &  \frac{2}{3} \\
0  &  0  &   0  &  0  &         0  &  0  &  0  &  0  &      0  &  0  &   0  &  0  &         0  &  0  &  0  &  0 \\
0  &  0  &   0  &  0  &         0  &  \frac{2}{3}  &  0  &  \frac{2}{3}  &      0  &  \frac{4}{3}  &   0  &  \frac{4}{3}  &         0  &  0  &  0  &  0 \\
0  &  0  &   0  &  0  &         0  &  0  &  0  &  0  &      0  &  0  &   0  &  0  &         0  &  0  &  0  &  0 \\
0  &  \frac{4}{3}  &   0  &  \frac{4}{3}  &         0  &  \frac{2}{3}  &  0  &  \frac{2}{3}  &      0  &  \frac{4}{3}  &   0  &  \frac{4}{3}  &         0  &  \frac{2}{3}  &  0  &  \frac{2}{3}
\end{array} \right)^T.
\end{equation}
\begin{equation}
\gamma_{\rm dd}^{(0)} =\left(
\begin{array}{cccccc}
 0 & \frac{44}{3} & 0 & 0 & 0 & \frac{8}{3} \\
 \frac{8}{3} & \frac{12 n_f-40}{9} & 0 & 0 & 0 & \frac{12 n_f-4}{9} \\
 0 & 0 & 0 & \frac{44}{3} & 0 & \frac{8}{3} \\
 0 & 0 & \frac{8}{3} & \frac{12 n_f-40}{9} & 0 & \frac{12 n_f-4}{9} \\
 0 & 0 & 0 & 0 & 0 & -12 \\
 0 & \frac{2 n_f}{3} & 0 & \frac{2 n_f}{3} & -\frac{8}{3} & \frac{4 n_f-42}{3}
\end{array}
\right).
\end{equation}

\section{144$\times$144 ADM for the fully expanded basis} \label{sec:app-op144}

The basis is introduced in \Sec{sec:expop}. Including all six flavors, independent 144 operators are counted as follows. For $6\times 5 \times 1/2=15$ combinations of $(q,q^\prime)$, where $q \neq q^\prime$ for the current-current $j^\mu_q j_{q^\prime \mu}$, four chiralities $\rm (LL,LR,RL,RR)$ and two colors (octet or singlet) are possible. For six combinations of $q=q^\prime$, only four combinations, $\rm LL(8),LR(1),LR(8),RR(8)$ are possible since color-singlet operators of $\rm LL$ or $\rm RR$ are reduced to their octet counterparts. In total, $15 \times 8 + 6 \times 4 = 144$. However, if the $\rm SU(2)_L$ gauge symmetry is imposed, not all operators are independent. In this basis, $\rm SU(2)_L$ symmetry can be imposed by $\rm SU(2)_L$ invariant boundary conditions at the matching scale, i.e., $g_{u_{\rm L}} = g_{d_{\rm L}}$, etc. QCD RG evolution will then preserve the $\rm SU(2)_L$.

All information of our full $144 \times 144$ ADM is contained in the following subset
\beq
\gamma \=  \frac{\alpha_s}{4\pi} \, \left(
\begin{array}{c|c c c| c c c}
 & 1 & 5 & 13 & 17 & 21 & 29 \\
 & (uu) & (ut) & (tt) & (dd) & (ud) & (td) \\
\hline
(uu) & uuuu & uuut & 0 & 0 & uuut & 0 \\
(ut) & utuu & utut & uttt & 0 & udut & tddb \\
(tt) & 0 & ttut & uuuu & 0 & 0 & uuut \\
\hline
(dd) & 0 & 0 & 0 & uuuu & ttut & ttut \\
(ud) & utuu & udut & 0 & uttt & utut & udtd \\
(td) & 0 & tdut & utuu & uttt & udtd & utut \\
\end{array}
\right).
\label{eq:dZ36} \eeq
Each row and column is labeled as $(qq^\prime)$ which means operators of $(\bar{q}q)(\bar{q}^\prime q^\prime)$ type. Each label $(qq^\prime)$ is 4(8)-dimension if $q=q^\prime(q\ne q^\prime)$ for the chirality and color indices. On the top of the matrix, we explicitly show the numbering of the first operator in each $(qq^\prime)$ sector. For the 8-dimensional $(qq^\prime)$ sector, the order of operators is ${\cal O}_{\rm LL}^{(1,8)},{\cal O}_{\rm LR}^{(1,8)},{\cal O}_{\rm RL}^{(1,8)},{\cal O}_{\rm RR}^{(1,8)}$. For the 4-dimensional $(qq^\prime)$ sector, the order is ${\cal O}_{\rm LL}^{(8)},{\cal O}_{\rm LR}^{(1,8)},{\cal O}_{\rm RR}^{(8)}$. The sub-matrix $utuu$ (where ($ut$)-row and ($uu$)-column meet), for example, describes operator mixing from $(\bar{u}u)(\bar{t}t)$-type into $(\bar{u}u)(\bar{u}u)$-type.
Since there are 8 independent $(\bar{u}u)(\bar{t}t)$ operators and 4 independent $(\bar{u}u)(\bar{u}u)$ operators, the sub-matrix $utuu$ is $8 \times 4$ dimension.

The full 144$\times$144 ADM contains many repeated sub-matrices; only 10 different sub-matrices appear. One can use our results to construct the ADM for a theory with any number of flavors. For example, the ADM for the $(ut)$ row and the $(bt)$ column (which means the induction of the $(bt)$ from the $(ut)$) is given by the $(utbt)=(udtd)$ above; the equality is obtained simply by renaming the flavors.

We list all 10 sub-matrices appearing in \Eq{eq:dZ36}.
\beq
uuuu \= \left(
\begin{array}{cccc}
 \frac{44}{9} & 0 & \frac{8}{9} & 0 \\
 0 & 0 & -12 & 0 \\
 \frac{2}{3} & -\frac{8}{3} & -\frac{38}{3} & \frac{2}{3} \\
 0 & 0 & \frac{8}{9} & \frac{44}{9}
\end{array}
\right),
\qquad
uuut \= \left(
\begin{array}{cccccccc}
 0 & \frac{8}{9} & 0 & \frac{8}{9} & 0 & 0 & 0 & 0 \\
 0 & 0 & 0 & 0 & 0 & 0 & 0 & 0 \\
 0 & \frac{2}{3} & 0 & \frac{2}{3} & 0 & \frac{2}{3} & 0 & \frac{2}{3} \\
 0 & 0 & 0 & 0 & 0 & \frac{8}{9} & 0 & \frac{8}{9}
\end{array}
\right)
\eeq
\beq
utuu \= \left(
\begin{array}{cccc}
 0 & 0 & 0 & 0 \\
 \frac{2}{3} & 0 & \frac{2}{3} & 0 \\
 0 & 0 & 0 & 0 \\
 \frac{2}{3} & 0 & \frac{2}{3} & 0 \\
 0 & 0 & 0 & 0 \\
 0 & 0 & \frac{2}{3} & \frac{2}{3} \\
 0 & 0 & 0 & 0 \\
 0 & 0 & \frac{2}{3} & \frac{2}{3}
\end{array}
\right),
\qquad
utut \= \left(
\begin{array}{cccccccc}
 0 & 12 & 0 & 0 & 0 & 0 & 0 & 0 \\
 \frac{8}{3} & -\frac{8}{3} & 0 & \frac{2}{3} & 0 & \frac{2}{3} & 0 & 0 \\
 0 & 0 & 0 & -12 & 0 & 0 & 0 & 0 \\
 0 & \frac{2}{3} & -\frac{8}{3} & -\frac{38}{3} & 0 & 0 & 0 & \frac{2}{3} \\
 0 & 0 & 0 & 0 & 0 & -12 & 0 & 0 \\
 0 & \frac{2}{3} & 0 & 0 & -\frac{8}{3} & -\frac{38}{3} & 0 & \frac{2}{3} \\
 0 & 0 & 0 & 0 & 0 & 0 & 0 & 12 \\
 0 & 0 & 0 & \frac{2}{3} & 0 & \frac{2}{3} & \frac{8}{3} & -\frac{8}{3}
\end{array}
\right)
\eeq
\beq
uttt \= \left(
\begin{array}{cccc}
 0 & 0 & 0 & 0 \\
 \frac{2}{3} & 0 & \frac{2}{3} & 0 \\
 0 & 0 & 0 & 0 \\
 0 & 0 & \frac{2}{3} & \frac{2}{3} \\
 0 & 0 & 0 & 0 \\
 \frac{2}{3} & 0 & \frac{2}{3} & 0 \\
 0 & 0 & 0 & 0 \\
 0 & 0 & \frac{2}{3} & \frac{2}{3}
\end{array}
\right),
\qquad
ttut \= \left(
\begin{array}{cccccccc}
 0 & \frac{8}{9} & 0 & 0 & 0 & \frac{8}{9} & 0 & 0 \\
 0 & 0 & 0 & 0 & 0 & 0 & 0 & 0 \\
 0 & \frac{2}{3} & 0 & \frac{2}{3} & 0 & \frac{2}{3} & 0 & \frac{2}{3} \\
 0 & 0 & 0 & \frac{8}{9} & 0 & 0 & 0 & \frac{8}{9}
\end{array}
\right)
\eeq
\beq
udtd \= \left(
\begin{array}{cccccccc}
 0 & 0 & 0 & 0 & 0 & 0 & 0 & 0 \\
 0 & \frac{2}{3} & 0 & 0 & 0 & \frac{2}{3} & 0 & 0 \\
 0 & 0 & 0 & 0 & 0 & 0 & 0 & 0 \\
 0 & 0 & 0 & \frac{2}{3} & 0 & 0 & 0 & \frac{2}{3} \\
 0 & 0 & 0 & 0 & 0 & 0 & 0 & 0 \\
 0 & \frac{2}{3} & 0 & 0 & 0 & \frac{2}{3} & 0 & 0 \\
 0 & 0 & 0 & 0 & 0 & 0 & 0 & 0 \\
 0 & 0 & 0 & \frac{2}{3} & 0 & 0 & 0 & \frac{2}{3}
\end{array}
\right),
\qquad
udut \= \left(
\begin{array}{cccccccc}
 0 & 0 & 0 & 0 & 0 & 0 & 0 & 0 \\
 0 & \frac{2}{3} & 0 & \frac{2}{3} & 0 & 0 & 0 & 0 \\
 0 & 0 & 0 & 0 & 0 & 0 & 0 & 0 \\
 0 & \frac{2}{3} & 0 & \frac{2}{3} & 0 & 0 & 0 & 0 \\
 0 & 0 & 0 & 0 & 0 & 0 & 0 & 0 \\
 0 & 0 & 0 & 0 & 0 & \frac{2}{3} & 0 & \frac{2}{3} \\
 0 & 0 & 0 & 0 & 0 & 0 & 0 & 0 \\
 0 & 0 & 0 & 0 & 0 & \frac{2}{3} & 0 & \frac{2}{3}
\end{array}
\right)
\eeq
\beq
tdut \= \left(
\begin{array}{cccccccc}
 0 & 0 & 0 & 0 & 0 & 0 & 0 & 0 \\
 0 & \frac{2}{3} & 0 & 0 & 0 & \frac{2}{3} & 0 & 0 \\
 0 & 0 & 0 & 0 & 0 & 0 & 0 & 0 \\
 0 & \frac{2}{3} & 0 & 0 & 0 & \frac{2}{3} & 0 & 0 \\
 0 & 0 & 0 & 0 & 0 & 0 & 0 & 0 \\
 0 & 0 & 0 & \frac{2}{3} & 0 & 0 & 0 & \frac{2}{3} \\
 0 & 0 & 0 & 0 & 0 & 0 & 0 & 0 \\
 0 & 0 & 0 & \frac{2}{3} & 0 & 0 & 0 & \frac{2}{3}
\end{array}
\right),
\qquad
tddb \= \left(
\begin{array}{cccccccc}
 0 & 0 & 0 & 0 & 0 & 0 & 0 & 0 \\
 0 & \frac{2}{3} & 0 & \frac{2}{3} & 0 & 0 & 0 & 0 \\
 0 & 0 & 0 & 0 & 0 & 0 & 0 & 0 \\
 0 & 0 & 0 & 0 & 0 & 0 \frac{2}{3} & 0 & \frac{2}{3} \\
 0 & 0 & 0 & 0 & 0 & 0 & 0 & 0 \\
 0 & \frac{2}{3} & 0 & \frac{2}{3} & 0 & 0 & 0 & 0 \\
 0 & 0 & 0 & 0 & 0 & 0 & 0 & 0 \\
 0 & 0 & 0 & 0 & 0 & \frac{2}{3} & 0 & \frac{2}{3}
\end{array}
\right)
\eeq

\section{${\cal O}(1/\Lambda^4)$ contributions} \label{app:helamp}

We list the helicity cross-sections and top pair observables including ${\cal O}(1/\Lambda^{4})$ terms. We aim to answer whether the contributions from the square of effective operators are as important as the RG-induced effects; see \Sec{sec:lambda} for the related discussion.

The helicity cross-sections from the square of color-octet effective operators are
\bea
\sigma_{++} = \sigma_{--} &\ni& \frac{1}{432\pi \hat s} \frac{m_t^2 \beta_t}{\hat s} \left(\frac{\hat s}{\Lambda^2}\right)^2
\bigg[(C_{\rm LL}^{(8)}+C_{\rm LR}^{(8)})^2+(C_{\rm RL}^{(8)}+C_{\rm RR}^{(8)})^2\bigg]\,, \\
\sigma_{+-} &\ni &   \frac{1}{864 \pi \hat s}\beta_t \left(\frac{\hat s}{\Lambda^2}\right)^2
\bigg[\Big((1-\beta_t)C_{\rm LL}^{(8)}+(1+\beta_t) C_{\rm LR}^{(8)}\Big)^2 \nn \\
&& ~~~~~~~~~~~~~~~~~~~
\qquad  \qquad + \Big((1-\beta_t)C_{\rm RL}^{(8)}+(1+\beta)C_{\rm RR}^{(8)}\Big)^2\bigg]\,, \\
\sigma_{-+} &\ni &   \frac{1}{864 \pi \hat s}\beta_t \left(\frac{\hat s}{\Lambda^2}\right)^2
\bigg[\Big((1+\beta_t)C_{\rm LL}^{(8)}+(1-\beta_t) C_{\rm LR}^{(8)}\Big)^2 \nn \\
&& ~~~~~~~~~~~~~~~~~~~
\qquad \qquad  + \Big((1+\beta_t)C_{\rm RL}^{(8)}+(1-\beta)C_{\rm RR}^{(8)}\Big)^2\bigg]\,.
\eea
The forward-backward asymmetric helicity cross-sections are (defined as in \Eq{eq:afbhelcrx})
\beq
a_{++} \= a_{--} \, \ni \,  0,
\eeq
\beq
a_{+-} \, \ni \, \frac{1}{1152 \pi \hat{s}} \beta_t \left( \frac{\hat{s}}{\Lambda^2} \right)^2  \Big[ - \left( (1-\beta_t)C_{\rm LL}^{(8)} + (1+\beta_t) C^{(8)}_{\rm LR} \right)^2 \+ \left( (1-\beta_t) C^{(8)}_{\rm RL} + (1+\beta_t) C^{(8)}_{\rm RR} \right)^2 \Big],
\eeq
\beq
a_{-+} \, \ni \, \frac{1}{1152 \pi \hat{s}} \beta_t \left( \frac{\hat{s}}{\Lambda^2} \right)^2  \Big[ - \left( (1-\beta_t)C_{\rm RR}^{(8)} + (1+\beta_t) C^{(8)}_{\rm RL} \right)^2 \+ \left( (1-\beta_t) C^{(8)}_{\rm LR} + (1+\beta_t) C^{(8)}_{\rm LL} \right)^2 \Big].
\eeq

The ${\cal O}(1/\Lambda^4)$ terms contribute to the observables as
\bea
\sigma_{\rm tot} & \ni &
\frac{\beta_t}{2304\pi \hat{s}} \, \left( \frac{\hat{s}}{\Lambda^2 } \right)^2 \cdot \Big[ \, \frac{32}{3} \frac{m_t^2}{\hat{s}}  \big( \big(C^{(8)}_{\rm LL}+C^{(8)}_{\rm LR} \big)^2 + \big(C^{(8)}_{\rm RR}+C^{(8)}_{\rm RL}\big)^2 \big) \Big. \nonumber\\
&& \qquad \+ \frac{8}{3} \big( (1-\beta_t)^2+(1+\beta_t)^2 \big) \big( {C^{(8)}_{\rm LL}}^2 + {C^{(8)}_{\rm LR}}^2 + {C^{(8)}_{\rm RL}}^2 + {C^{(8)}_{\rm RR}}^2 \big) \\
&& \qquad \+ \frac{32}{3}  (1+\beta_t)(1-\beta_t) \big( C^{(8)}_{\rm LL} C^{(8)}_{\rm LR} + C^{(8)}_{\rm RL} C^{(8)}_{\rm RR} \big)  \, \Big]. \nonumber
\eea
\beq
\sigma_{\rm FB} \, \ni \,
\frac{\beta_t}{1152\pi \hat{s}} \, \left( \frac{\hat{s}}{\Lambda^2 } \right)^2 \cdot \Big[ \, \big( (1+\beta_t)^2-(1-\beta_t)^2 \big) \big( {C^{(8)}_{\rm LL}}^2 + {C^{(8)}_{\rm RR}}^2 - {C^{(8)}_{\rm LR}}^2 - {C^{(8)}_{\rm RL}}^2 \big) \, \Big].
\eeq
\beq
\sigma_{{\cal P}_t} \, \ni \,
\frac{\beta_t}{2304\pi \hat{s}} \, \left( \frac{\hat{s}}{\Lambda^2 } \right)^2 \cdot \Big[ \, \frac{8}{3} \big( (1+\beta_t)^2-(1-\beta_t)^2 \big) \big( {C^{(8)}_{\rm RR}}^2 + {C^{(8)}_{\rm LR}}^2 - {C^{(8)}_{\rm RL}}^2 - {C^{(8)}_{\rm LL}}^2 \big) \, \Big].
\eeq
\bea
\sigma_{{\cal D}_{\rm FB}} & \ni &
\frac{\beta_t}{1152 \pi \hat{s}} \, \left( \frac{\hat{s}}{\Lambda^2 } \right)^2 \cdot \Big[ \, \big( (1-\beta_t)^2 + (1+\beta_t)^2 \big) \big( {C^{(8)}_{\rm RL}}^2 + {C^{(8)}_{\rm RR}}^2 - {C^{(8)}_{\rm LL}}^2 - {C^{(8)}_{\rm LR}}^2 \big) \Big. \nonumber\\ && \qquad \Big. \+ 4(1+\beta_t)(1-\beta_t) ( C^{(8)}_{\rm RL} C^{(8)}_{\rm RR} - C^{(8)}_{\rm LL} C^{(8)}_{\rm LR} ) \, \Big].
\eea
\bea
\sigma_{{\cal C}_{t\bar{t}}} & \ni &
\frac{\beta_t}{2304\pi \hat{s}} \, \left( \frac{\hat{s}}{\Lambda^2 } \right)^2 \cdot \Big[ \, -\frac{32}{3} \frac{m_t^2}{\hat{s}}  \big( \big( C^{(8)}_{\rm LL}+ C^{(8)}_{\rm LR} \big)^2 + \big( C^{(8)}_{\rm RR}+ C^{(8)}_{\rm RL} \big)^2 \big) \Big. \nonumber\\
&& \qquad \+ \frac{8}{3} \big( (1-\beta_t)^2+(1+\beta_t)^2 \big) \big( {C^{(8)}_{\rm LL}}^2 + {C^{(8)}_{\rm LR}}^2 + {C^{(8)}_{\rm RL}}^2 + {C^{(8)}_{\rm RR}}^2 \big) \\
&& \qquad \+ \frac{32}{3}  (1+\beta_t)(1-\beta_t) \big( C^{(8)}_{\rm LL} C^{(8)}_{\rm LR} + C^{(8)}_{\rm RL} C^{(8)}_{\rm RR} \big)  \, \Big]. \nonumber
\eea

Among all observables, the total cross-section and spin-correlation are modified by all ${\cal O}_{\rm VV, AA, AV, VA}$ operators at the ${\cal O}(1/\Lambda^4)$. All other observables are not contributed from any of these operators. This is another reason to use the V--A basis to discuss operator mixing effects.

\section{Conversion between the L--R and V--A basis} \label{app:lrva}

Operators in the V--A basis can be decomposed into operators in the L--R basis
\bea
{\cal O}_{\rm VV} &=& {\cal O}_{\rm RR} + {\cal O}_{\rm LL} + {\cal O}_{\rm RL}+ {\cal O}_{\rm LR}, \quad \quad {\cal O}_{\rm AA} = {\cal O}_{\rm LL} + {\cal O}_{\rm RR} - {\cal O}_{\rm LR} - {\cal O}_{\rm RL},  \nonumber\\
{\cal O}_{\rm VA} &=& {\cal O}_{\rm RR} - {\cal O}_{\rm LL} - {\cal O}_{\rm RL} + {\cal O}_{\rm LR}, \quad \quad {\cal O}_{\rm AV} = {\cal O}_{\rm RR} - {\cal O}_{\rm LL} + {\cal O}_{\rm RL} - {\cal O}_{\rm LR}.
\eea
Likewise, operators in the L--R basis can be transformed into the V--A basis
\bea
{\cal O}_{\rm RR} &=& \frac{1}{4} \left( {\cal O}_{\rm VV} + {\cal O}_{\rm AA} + {\cal O}_{\rm VA} + {\cal O}_{\rm AV} \right), \nn \\
\qquad {\cal O}_{\rm LL} &\=& \frac{1}{4} \left( {\cal O}_{\rm VV} + {\cal O}_{\rm AA} - {\cal O}_{\rm VA} - {\cal O}_{\rm AV} \right),  \nonumber\\
{\cal O}_{\rm RL} &=& \frac{1}{4} \left( {\cal O}_{\rm VV} - {\cal O}_{\rm AA} - {\cal O}_{\rm VA} + {\cal O}_{\rm AV} \right), \nn \\
\qquad {\cal O}_{\rm LR} &\=& \frac{1}{4} \left( {\cal O}_{\rm VV} - {\cal O}_{\rm AA} + {\cal O}_{\rm VA} - {\cal O}_{\rm AV} \right).
\eea
Thus, operator coefficients are related as
\bea
C_{\rm VV} &=& \frac{1}{4} \left( C_{\rm RR} + C_{\rm LL} + C_{\rm RL}+ C_{\rm LR} \right), \quad C_{\rm AA} = \frac{1}{4} \left( C_{\rm RR} + C_{\rm LL} - C_{\rm RL} - C_{\rm LR} \right),  \nonumber\\
C_{\rm VA} &=& \frac{1}{4} \left( C_{\rm RR} - C_{\rm LL} - C_{\rm RL} + C_{\rm LR} \right), \quad C_{\rm AV} = \frac{1}{4} \left( C_{\rm RR} - C_{\rm LL} + C_{\rm RL} - C_{\rm LR} \right),
\label{eq:cvaclr} \eea
and
\bea
C_{\rm RR} &=&  C_{\rm VV} + C_{\rm AA} + C_{\rm VA}+ C_{\rm AV}, \quad
C_{\rm LL} =  C_{\rm VV} + C_{\rm AA} - C_{\rm VA} - C_{\rm AV},  \nonumber\\
C_{\rm RL} &=&  C_{\rm VV} - C_{\rm AA} - C_{\rm VA} + C_{\rm AV}, \quad
C_{\rm LR} = C_{\rm VV} - C_{\rm AA} + C_{\rm VA} - C_{\rm AV} .
\eea
Relations apply both to color-octet and -singlet operators.



\begin{thebibliography}{99}

\bibitem{Almeida:2008ug}
  L.~G.~Almeida, G.~F.~Sterman and W.~Vogelsang,
  ``Threshold Resummation for the Top Quark Charge Asymmetry,''
  Phys.\ Rev.\ D {\bf 78}, 014008 (2008)
  [arXiv:0805.1885 [hep-ph]].
  V.~Ahrens, A.~Ferroglia, M.~Neubert, B.~D.~Pecjak and L.~L.~Yang,
  ``The top-pair forward-backward asymmetry beyond NLO,''
  Phys.\ Rev.\ D {\bf 84}, 074004 (2011)
  [arXiv:1106.6051 [hep-ph]].
  W.~Bernreuther, M.~Fuecker and Z.~G.~Si,
  ``Mixed QCD and weak corrections to top quark pair production at hadron colliders,''
  Phys.\ Lett.\ B {\bf 633}, 54 (2006)
  [hep-ph/0508091].
  J.~H.~Kuhn, A.~Scharf and P.~Uwer,
  ``Electroweak corrections to top-quark pair production in quark-antiquark annihilation,''
  Eur.\ Phys.\ J.\ C {\bf 45}, 139 (2006)
  [hep-ph/0508092].
  A.~V.~Manohar and M.~Trott,
  ``Electroweak Sudakov Corrections and the Top Quark Forward-Backward Asymmetry,''
  Phys.\ Lett.\ B {\bf 711}, 313 (2012)
  [arXiv:1201.3926 [hep-ph]].
  W.~Hollik and D.~Pagani,
  ``The electroweak contribution to the top quark forward-backward asymmetry at the Tevatron,''
  Phys.\ Rev.\ D {\bf 84}, 093003 (2011)
  [arXiv:1107.2606 [hep-ph]].
  J.~H.~Kuhn and G.~Rodrigo,
  ``Charge asymmetries of top quarks at hadron colliders revisited,''
  JHEP {\bf 1201}, 063 (2012)
  [arXiv:1109.6830 [hep-ph]].

\bibitem{Bernreuther:2012sx}
  W.~Bernreuther and Z.~-G.~Si,
  ``Top quark and leptonic charge asymmetries for the Tevatron and LHC,''
  Phys.\ Rev.\ D {\bf 86}, 034026 (2012)
  [arXiv:1205.6580 [hep-ph]].

\bibitem{Czakon:2013goa}
  M.~Czakon, P.~Fiedler and A.~Mitov,
  ``The total top quark pair production cross-section at hadron colliders through O($\alpha_S^4$),''
  Phys.\ Rev.\ Lett.\  {\bf 110}, 252004 (2013)
  [arXiv:1303.6254 [hep-ph]].

\bibitem{Xiao:2010hm}
  B.~Xiao, Y.~-k.~Wang and S.~-h.~Zhu,
  ``Forward-backward Asymmetry and Differential Cross Section of Top Quark in Flavor Violating Z' model at ${\cal O}(\alpha_s^2 \alpha_X)$,''
  Phys.\ Rev.\ D {\bf 82}, 034026 (2010)
  [arXiv:1006.2510 [hep-ph]].
  H.~X.~Zhu, C.~S.~Li, L.~Dai, J.~Gao, J.~Wang and C.~-P.~Yuan,
  ``One-loop Helicity Amplitudes for Top Quark Pair Production in Randall-Sundrum Model,''
  JHEP {\bf 1109}, 043 (2011)
  [arXiv:1106.2243 [hep-ph]].
  K.~Yan, J.~Wang, D.~Y.~Shao and C.~S.~Li,
  ``Next-to-leading order QCD effect of $W'$ on top quark Forward-Backward Asymmetry,''
  Phys.\ Rev.\ D {\bf 85}, 034020 (2012)
  [arXiv:1110.6684 [hep-ph]].

\bibitem{Bauer:2010iq}
  M.~Bauer, F.~Goertz, U.~Haisch, T.~Pfoh and S.~Westhoff,
  ``Top-Quark Forward-Backward Asymmetry in Randall-Sundrum Models Beyond the Leading Order,''
  JHEP {\bf 1011}, 039 (2010)
  [arXiv:1008.0742 [hep-ph]].

\bibitem{oai:arXiv.org:hep-ph/9606222}
  See for example:
  A.~V.~Manohar,
  ``Effective field theories,''
  In *Schladming 1996, Perturbative and nonperturbative aspects of quantum field theory* 311-362
  [hep-ph/9606222].

 \bibitem{oai:arXiv.org:hep-ph/9512380}
  G.~Buchalla, A.~J.~Buras and M.~E.~Lautenbacher,
  ``Weak decays beyond leading logarithms,''
  Rev.\ Mod.\ Phys.\  {\bf 68}, 1125 (1996)
  [hep-ph/9512380].

\bibitem{Grojean:2013kd}
  C.~Grojean, E.~E.~Jenkins, A.~V.~Manohar and M.~Trott,
  ``Renormalization Group Scaling of Higgs Operators and $\Gamma(h -> \gamma \gamma)$,''
  JHEP {\bf 1304}, 016 (2013)
  [arXiv:1301.2588 [hep-ph]].

\bibitem{Elias-Miro:2013gya}
  J.~Elias-Miró, J.~R.~Espinosa, E.~Masso and A.~Pomarol,
  ``Renormalization of dimension-six operators relevant for the Higgs decays $h\rightarrow \gamma\gamma,\gamma Z$,''
  JHEP {\bf 1308}, 033 (2013)
  [arXiv:1302.5661 [hep-ph]].
  J.~Elias-Miro, J.~R.~Espinosa, E.~Masso and A.~Pomarol,
  ``Higgs windows to new physics through d=6 operators: constraints and one-loop anomalous dimensions,''
  JHEP {\bf 1311}, 066 (2013)
  [arXiv:1308.1879 [hep-ph]].
  J.~Elias-Miró, C.~Grojean, R.~S.~Gupta and D.~Marzocca,
  ``Scaling and tuning of EW and Higgs observables,''
  JHEP {\bf 1405}, 019 (2014)
  [arXiv:1312.2928 [hep-ph], arXiv:1312.2928].

\bibitem{oai:arXiv.org:1107.4012}
  D.~Y.~Shao, C.~S.~Li, J.~Wang, J.~Gao, H.~Zhang and H.~X.~Zhu,
  ``Model independent analysis of top quark forward-backward asymmetry at the Tevatron up to ${\cal O}(as^2/\Lambda^2)$,''
  Phys.\ Rev.\ D {\bf 84}, 054016 (2011)
  [arXiv:1107.4012 [hep-ph]].

\bibitem{oai:arXiv.org:1204.4773}
  J.~Gao, C.~S.~Li and C.~P.~Yuan,
  ``NLO QCD Corrections to dijet Production via Quark Contact Interactions,''
  JHEP {\bf 1207}, 037 (2012)
  [arXiv:1204.4773 [hep-ph]].

\bibitem{Jenkins:2013zja}
  E.~E.~Jenkins, A.~V.~Manohar and M.~Trott,
  ``Renormalization Group Evolution of the Standard Model Dimension Six Operators I: Formalism and lambda Dependence,''
  JHEP {\bf 1310}, 087 (2013)
  [arXiv:1308.2627 [hep-ph]].

\bibitem{Alonso:2013hga}
  R.~Alonso, E.~E.~Jenkins, A.~V.~Manohar and M.~Trott,
  ``Renormalization Group Evolution of the Standard Model Dimension Six Operators III: Gauge Coupling Dependence and Phenomenology,''
  JHEP {\bf 1404}, 159 (2014)
  [arXiv:1312.2014 [hep-ph]].

\bibitem{Zhang:2014rja}
  C.~Zhang,
  ``Effective field theory approach to top-quark decay at next-to-leading order in QCD,''
  arXiv:1404.1264 [hep-ph].

\bibitem{Antunano:2007da}
  O.~Antunano, J.~H.~Kuhn and G.~Rodrigo,
  ``Top quarks, axigluons and charge asymmetries at hadron colliders,''
  Phys.\ Rev.\ D {\bf 77}, 014003 (2008)
  [arXiv:0709.1652 [hep-ph]].
  Y.~-k.~Wang, B.~Xiao and S.~-h.~Zhu,
  ``One-side Forward-backward Asymmetry in Top Quark Pair Production at CERN Large Hadron Collider,''
  Phys.\ Rev.\ D {\bf 82}, 094011 (2010)
  [arXiv:1008.2685 [hep-ph]].
  S.~Jung, A.~Pierce and J.~D.~Wells,
  ``Top quark asymmetry from a non-Abelian horizontal symmetry,''
  Phys.\ Rev.\ D {\bf 83}, 114039 (2011)
  [arXiv:1103.4835 [hep-ph]].
  J.~A.~Aguilar-Saavedra, A.~Juste and F.~Rubbo,
  ``Boosting the t tbar charge asymmetry,''
  Phys.\ Lett.\ B {\bf 707}, 92 (2012)
  [arXiv:1109.3710 [hep-ph]].

\bibitem{Aad:2013cea}
  G.~Aad {\it et al.}  [ATLAS Collaboration],
  ``Measurement of the top quark pair production charge asymmetry in proton-proton collisions at $\sqrt{s}$ = 7 TeV using the ATLAS detector,''
  JHEP {\bf 1402}, 107 (2014)
  [arXiv:1311.6724 [hep-ex]].
  CMS Collaboration [CMS Collaboration],
  ``Measurement of the ttbar charge asymmetry with lepton+jets events at 8 TeV,''
  CMS-PAS-TOP-12-033.


\bibitem{Degrande:2010kt}
  C.~Degrande, J.~-M.~Gerard, C.~Grojean, F.~Maltoni and G.~Servant,
  ``Non-resonant New Physics in Top Pair Production at Hadron Colliders,''
  JHEP {\bf 1103}, 125 (2011)
  [arXiv:1010.6304 [hep-ph]].

\bibitem{Jung:2010yn}
  D.~-W.~Jung, P.~Ko and J.~S.~Lee,
  ``Longitudinal top polarization as a probe of a possible origin of forward-backward asymmetry of the top quark at the Tevatron,''
  Phys.\ Lett.\ B {\bf 701}, 248 (2011)
  [arXiv:1011.5976 [hep-ph]].

\bibitem{Mahlon:1995zn}
  G.~Mahlon and S.~J.~Parke,
  ``Angular correlations in top quark pair production and decay at hadron colliders,''
  Phys.\ Rev.\ D {\bf 53}, 4886 (1996)
  [hep-ph/9512264].

\bibitem{Jung:2009pi}
  D.~-W.~Jung, P.~Ko, J.~S.~Lee and S.~-h.~Nam,
  ``Model independent analysis of the forward-backward asymmetry of top quark production at the Tevatron,''
  Phys.\ Lett.\ B {\bf 691}, 238 (2010)
  [arXiv:0912.1105 [hep-ph]].

\bibitem{Krohn:2011tw}
  D.~Krohn, T.~Liu, J.~Shelton and L.~-T.~Wang,
  ``A Polarized View of the Top Asymmetry,''
  Phys.\ Rev.\ D {\bf 84}, 074034 (2011)
  [arXiv:1105.3743 [hep-ph]].

\bibitem{Bowen:2005ap}
  M.~T.~Bowen, S.~D.~Ellis and D.~Rainwater,
  ``Standard model top quark asymmetry at the Fermilab Tevatron,''
  Phys.\ Rev.\ D {\bf 73}, 014008 (2006)
  [hep-ph/0509267].

\bibitem{oai:arXiv.org:1201.1790}
  E.~L.~Berger, Q.~-H.~Cao, C.~-R.~Chen, J.~-H.~Yu and H.~Zhang,
  ``The Top Quark Production Asymmetries $A_{FB}^t$ and $A_{FB}^{\ell}$,''
  Phys.\ Rev.\ Lett.\  {\bf 108}, 072002 (2012)
  [arXiv:1201.1790 [hep-ph]].

\bibitem{Aguilar-Saavedra:2014yea}
  J.~A.~Aguilar-Saavedra,
  arXiv:1405.1412 [hep-ph].

\bibitem{Falkowski:2011zr}
  A.~Falkowski, G.~Perez and M.~Schmaltz,
  ``Spinning the Top,''
  Phys.\ Rev.\ D {\bf 87}, 034041 (2013)
  [arXiv:1110.3796 [hep-ph]].

\bibitem{Fajfer:2012si}
  S.~Fajfer, J.~F.~Kamenik and B.~Melic,
  ``Discerning New Physics in Top-Antitop Production using Top Spin Observables at Hadron Colliders,''
  JHEP {\bf 1208}, 114 (2012)
  [arXiv:1205.0264 [hep-ph]].

\bibitem{oai:arXiv.org:0802.0007}
  P.~M.~Nadolsky, H.~-L.~Lai, Q.~-H.~Cao, J.~Huston, J.~Pumplin, D.~Stump, W.~-K.~Tung and C.~-P.~Yuan,
  ``Implications of CTEQ global analysis for collider observables,''
  Phys.\ Rev.\ D {\bf 78}, 013004 (2008)
  [arXiv:0802.0007 [hep-ph]].

\bibitem{Zhang:2010dr}
  C.~Zhang and S.~Willenbrock,
  ``Effective-Field-Theory Approach to Top-Quark Production and Decay,''
  Phys.\ Rev.\ D {\bf 83}, 034006 (2011)
  [arXiv:1008.3869 [hep-ph]].

\bibitem{Ciuchini:1993ks}
  M.~Ciuchini, E.~Franco, G.~Martinelli, L.~Reina and L.~Silvestrini,
  ``Scheme independence of the effective Hamiltonian for b ---> s gamma and b ---> s g decays,''
  Phys.\ Lett.\ B {\bf 316}, 127 (1993)
  [hep-ph/9307364].

\bibitem{Ciuchini:1993fk}
  M.~Ciuchini, E.~Franco, L.~Reina and L.~Silvestrini,
  ``Leading order QCD corrections to b ---> s gamma and b ---> s g decays in three regularization schemes,''
  Nucl.\ Phys.\ B {\bf 421}, 41 (1994)
  [hep-ph/9311357].

\bibitem{Grzadkowski:2010es}
  B.~Grzadkowski, M.~Iskrzynski, M.~Misiak and J.~Rosiek,
  ``Dimension-Six Terms in the Standard Model Lagrangian,''
  JHEP {\bf 1010}, 085 (2010)
  [arXiv:1008.4884 [hep-ph]].

\bibitem{Jung:2009jz}
  S.~Jung, H.~Murayama, A.~Pierce and J.~D.~Wells,
  ``Top quark forward-backward asymmetry from new t-channel physics,''
  Phys.\ Rev.\ D {\bf 81}, 015004 (2010)
  [arXiv:0907.4112 [hep-ph]].

\bibitem{Jenkins:2013sda}
  E.~E.~Jenkins, A.~V.~Manohar and M.~Trott,
  ``Naive Dimensional Analysis Counting of Gauge Theory Amplitudes and Anomalous Dimensions,''
  Phys.\ Lett.\ B {\bf 726}, 697 (2013)
  [arXiv:1309.0819 [hep-ph]].

\bibitem{Berends:QED}
  F.~A.~Berends, K.~J.~F.~Gaemers and R.~Gastmans,
  ``alpha**3 Contribution to the angular asymmetry in e+ e- ---> mu+ mu-,''
  Nucl.\ Phys.\ B {\bf 63}, 381 (1973).

\bibitem{oai:arXiv.org:hep-ph/9802268}
  J.~H.~Kuhn and G.~Rodrigo,
  ``Charge asymmetry in hadroproduction of heavy quarks,''
  Phys.\ Rev.\ Lett.\  {\bf 81}, 49 (1998)
  [hep-ph/9802268].
  J.~H.~Kuhn and G.~Rodrigo,
  ``Charge asymmetry of heavy quarks at hadron colliders,''
  Phys.\ Rev.\ D {\bf 59}, 054017 (1999)
  [hep-ph/9807420].

\bibitem{Jung:2014gfa}
  S.~Jung, P.~Ko, Y.~W.~Yoon and C.~Yu,
  ``Correlation of top asymmetries: loop versus tree origins,''
  arXiv:1405.5313 [hep-ph].

\bibitem{Zhu:2012ts}
  H.~X.~Zhu, C.~S.~Li, H.~T.~Li, D.~Y.~Shao and L.~L.~Yang,
  ``Transverse-momentum resummation for top-quark pairs at hadron colliders,''
  Phys.\ Rev.\ Lett.\  {\bf 110}, 082001 (2013)
  [arXiv:1208.5774 [hep-ph]].

\bibitem{Li:2013mia}
  H.~T.~Li, C.~S.~Li, D.~Y.~Shao, L.~L.~Yang and H.~X.~Zhu,
  ``Top quark pair production at small transverse momentum in hadronic collisions,''
  Phys.\ Rev.\ D {\bf 88}, 074004 (2013)
  [arXiv:1307.2464].

\bibitem{Skands:2012mm}
  P.~Skands, B.~Webber and J.~Winter,
  ``QCD Coherence and the Top Quark Asymmetry,''
  JHEP {\bf 1207}, 151 (2012)
  [arXiv:1205.1466 [hep-ph]].

\bibitem{Gripaios:2013rda}
  B.~Gripaios, A.~Papaefstathiou and B.~Webber,
  ``Probing the Colour Structure of the Top Quark Forward-Backward Asymmetry,''
  JHEP {\bf 1311}, 105 (2013)
  [arXiv:1309.0810 [hep-ph]].

\bibitem{lightaxi1}
  G.~Marques Tavares and M.~Schmaltz,
  ``Explaining the t-tbar asymmetry with a light axigluon,''
  Phys.\ Rev.\ D {\bf 84}, 054008 (2011)
  [arXiv:1107.0978 [hep-ph]].
  C.~Gross, G.~Marques Tavares, M.~Schmaltz and C.~Spethmann,
  ``Light axigluon explanation of the Tevatron ttbar asymmetry and multijet signals at the LHC,''
  Phys.\ Rev.\ D {\bf 87}, 014004 (2013)
  [arXiv:1209.6375 [hep-ph]].

\bibitem{Krnjaic:2011ub}
  G.~Z.~Krnjaic,
  ``Very Light Axigluons and the Top Asymmetry,''
  Phys.\ Rev.\ D {\bf 85}, 014030 (2012)
  [arXiv:1109.0648 [hep-ph]].

\bibitem{AguilarSaavedra:2011ci}
  J.~A.~Aguilar-Saavedra and M.~Perez-Victoria,
  Phys.\ Lett.\ B {\bf 705}, 228 (2011)
  [arXiv:1107.2120 [hep-ph]].
  

\bibitem{Aad:2013ksa}
  G.~Aad {\it et al.}  [ATLAS Collaboration],
  ``Measurement of top quark polarization in top-antitop events from proton-proton collisions at $\sqrt{s}$ = 7 TeV using the ATLAS detector,''
  Phys.\ Rev.\ Lett.\  {\bf 111}, 232002 (2013)
  [arXiv:1307.6511 [hep-ex]].

\bibitem{Chatrchyan:2013wua}
  S.~Chatrchyan {\it et al.}  [CMS Collaboration],
  ``Measurements of $t\bar{t}$ spin correlations and top-quark polarization using dilepton final states in pp collisions at $\sqrt{s}$ = 7 TeV,''
  Phys.\ Rev.\ Lett.\  {\bf 112}, 182001 (2014)
  [arXiv:1311.3924 [hep-ex]].

\bibitem{oai:arXiv.org:1010.1458}
  R.~M.~Godbole, K.~Rao, S.~D.~Rindani and R.~K.~Singh,
  ``On measurement of top polarization as a probe of $t \bar t$ production mechanisms at the LHC,''
  JHEP {\bf 1011}, 144 (2010)
  [arXiv:1010.1458 [hep-ph]].
  J.~Cao, L.~Wu and J.~M.~Yang,
  ``New physics effects on top quark spin correlation and polarization at the LHC: a comparative study in different models,''
  Phys.\ Rev.\ D {\bf 83}, 034024 (2011)
  [arXiv:1011.5564 [hep-ph]].
  D.~Choudhury, R.~M.~Godbole, S.~D.~Rindani and P.~Saha,
  ``Top polarization, forward-backward asymmetry and new physics,''
  Phys.\ Rev.\ D {\bf 84}, 014023 (2011)
  [arXiv:1012.4750 [hep-ph]].

\bibitem{Aaltonen:2013wca}
  T.~A.~Aaltonen {\it et al.}  [ CDF and  D0 Collaborations],
  ``Combination of measurements of the top-quark pair production cross section from the Tevatron Collider,''
  arXiv:1309.7570 [hep-ex].

\bibitem{Aaltonen:2012it}
  T.~Aaltonen {\it et al.}  [CDF Collaboration],
  ``Measurement of the top quark forward-backward production asymmetry and its dependence on event kinematic properties,''
  Phys.\ Rev.\ D {\bf 87}, 092002 (2013)
  [arXiv:1211.1003 [hep-ex]].

\bibitem{Aaltonen:2009iz}
  T.~Aaltonen {\it et al.}  [CDF Collaboration],
  ``First Measurement of the t anti-t Differential Cross Section d sigma/dM(t anti-t) in p anti-p Collisions at s**(1/2)=1.96-TeV,''
  Phys.\ Rev.\ Lett.\  {\bf 102}, 222003 (2009)
  [arXiv:0903.2850 [hep-ex]].

\bibitem{TheATLAScollaboration:2013dja}
  The ATLAS collaboration,
  ``Measurement of the $t\bar{t}$ production cross-section in $pp$ collisions at $\sqrt{s}=8$ TeV using $e\mu$ events with $b$-tagged jets,''
  ATLAS-CONF-2013-097.

\bibitem{Chatrchyan:2013lca}
  S.~Chatrchyan {\it et al.}  [CMS Collaboration],
  ``Searches for new physics using the $t\bar{t}$ invariant mass distribution in pp collisions at $\sqrt{s}$=8TeV,''
  Phys.\ Rev.\ Lett.\  {\bf 111}, 211804 (2013)
  [arXiv:1309.2030 [hep-ex]].

\bibitem{Abazov:2014vga}
  V.~M.~Abazov {\it et al.}  [D0 Collaboration],
  ``Measurement of differential $t\bar{t}$ production cross sections in $p\bar{p}$ collisions,''
  arXiv:1401.5785 [hep-ex].

\bibitem{Abazov:2014cca}
  V.~M.~Abazov {\it et al.}  [ D0 Collaboration],
  ``Measurement of the forward-backward asymmetry in top quark-antiquark production in ppbar collisions using the lepton+jets channel,''
  arXiv:1405.0421 [hep-ex].





\end{thebibliography}
\end{document}